\documentclass[journal]{IEEEtran}

\IEEEoverridecommandlockouts

\ifCLASSINFOpdf
  \usepackage[pdftex]{graphicx}
  \graphicspath{{../pdf/}{../jpeg/,./figures}}
  \DeclareGraphicsExtensions{.pdf,.jpeg,.png}
\else
  \usepackage[dvips]{graphicx}
  \graphicspath{{../eps/,./figures}}
  \DeclareGraphicsExtensions{.eps}
\fi

\usepackage{amsmath}
\usepackage{amsthm}
\usepackage{amssymb}
\usepackage[tight,footnotesize]{subfigure}
\usepackage{xcolor}
\usepackage[noadjust]{cite}
\usepackage{multirow}
\usepackage{epstopdf}
\usepackage[keeplastbox]{flushend}
\usepackage[printonlyused]{acronym}
\usepackage[mathscr]{euscript}
\usepackage{pifont}
\usepackage{tablefootnote}
\usepackage{framed}
\usepackage{boldline}
\acrodefplural{ID}[IDs]{identities}

\newcommand{\rev}[1]{\textcolor{blue}{#1}}
\renewcommand{\rev}[1]{#1}

\newcommand{\revv}[1]{\textcolor{blue}{#1}}
\newcommand{\tabcolor}{\color{blue}}
\renewcommand{\revv}[1]{#1}
\renewcommand{\tabcolor}{}

\newcommand{\field}[1]{\mathbb{#1}}

\newcommand{\set}[1]{\mathcal{#1}}
\newcommand{\mat}[1]{\boldsymbol{\mathbf{#1}}}
\newcommand{\ve}[1]{\boldsymbol{\mathbf{#1}}}

\newcommand{\operator}[1]{\mathrm{#1}}

\newcommand{\arma}{\operator{ARMA}}
\newcommand{\arima}{\operator{ARIMA}}

\newcommand{\col}{\operator{col}}
\begin{document}
\title{\rev{A Survey of Anticipatory Mobile Networking: Context-Based Classification, Prediction Methodologies, and Optimization Techniques}}

\author{Nicola Bui,~\IEEEmembership{Student Member,~IEEE,}
Matteo Cesana,~\IEEEmembership{Member,~IEEE,}
S. Amir Hosseini,~\IEEEmembership{Student Member,~IEEE,}
Qi~Liao,~\IEEEmembership{Member,~IEEE,} 
Ilaria Malanchini,~\IEEEmembership{Member,~IEEE,} 
and Joerg Widmer,~\IEEEmembership{Senior Member,~IEEE}
\thanks{Nicola Bui and Joerg Widmer are with IMDEA Networks Institute, Madrid, Spain. email:\{nicola.bui,~joerg.widmer\}@imdea.org. Matteo Cesana is with Politecnico di Milano, Italy. email:matteo.cesana@polimi.it. S. Amir Hosseini is with NYU Tandon School of Engineering, US. email:amirhs.hosseini@nyu.edu. Qi Liao and Ilaria Malanchini are with Nokia Bell Labs, Stuttgart, Germany. email:\{qi.liao,~ilaria.malanchini\}@nokia-bell-labs.com. This work has been has been supported in part by the European Union H2020-ICT grant 644399 (MONROE), European Union H2020-MSCA-ITN grant 643002 (ACT5G), by the Madrid Regional Government through the TIGRE5-CM program (S2013/ICE-2919), the Ramon y Cajal grant from the Spanish Ministry of Economy and Competitiveness RYC-2012-10788 and grant TEC2014-55713-R.}}

\maketitle

\begin{abstract}
A growing trend for information technology is to not just react to changes, but anticipate them as much as possible. This paradigm made modern solutions, such as recommendation systems, a ubiquitous presence in today's digital transactions. Anticipatory networking extends the idea to communication technologies by studying patterns and periodicity in human behavior and network dynamics to optimize network performance. This survey collects and analyzes recent papers leveraging context information to forecast the evolution of network conditions and, in turn, to improve network performance. In particular, we identify the main prediction and optimization tools adopted in this body of work and link them with objectives and constraints of the typical applications and scenarios. Finally, we consider open challenges and research directions to make anticipatory networking part of next generation networks.
\end{abstract}

\begin{IEEEkeywords}
Anticipatory, Prediction, Optimization, 5G, Mobile Networks.
\end{IEEEkeywords}

\IEEEpeerreviewmaketitle

\newcommand{\tabrelated}{
\begin{table*}
	\tabcolor
	\centering
	\caption{Related works}
	\label{tab:related}
	\begin{tabular}{|lV{2.5}p{12cm}|}
	\hline
	{\bf Topic} & {\bf Content} \\
	\hlineB{2.5}
	Big Data & \cite{zheng2016big}~studies big data analytics for network optimization. \\
	\hline
	\multirow{1}{*}{Context Information} & \cite{makris2013survey, pejovic2015anticipatory} discuss acquisition, modeling, exchange and usage of contextual information for different scenarios. \\
	\hline
	Data Classification & \cite{boucheron2005theory} surveys a variety of classifiers and uses them to predict unknown data. \\
	\hline
	\multirow{2}{*}{Traffic \& Throughput} & \cite{liu2015empirical} uses trace-driven simulation to compare prediction errors obtained using different techniques. \\
	    & \cite{nguyen2008survey} uses real network traffic to evaluate prediction techniques and to discuss their practical challenges. \\
	\hline
	\multirow{2}{*}{Social Patterns} & \cite{jin2013understanding} uses social network information to study traffic patterns. \\
	 & \cite{barakovic2013survey} investigates the impact of prediction on \ac{QoE} \\
	\hline
	\multirow{2}{*}{Cognitive Radios} & \cite{hoyhtya2016spectrum} investigates spectrum occupancy models and their reliability. \\
	 & \cite{chen2016survey} focus on spectrum occupancy and channel status prediction. \\ 
	\hline
	\end{tabular}
\end{table*}
}

\newcommand{\overalltable}{
\begin{table*}
	\centering
	\caption{Survey classification and structure}
	\label{tab:class}
	\begin{tabular}{c|l|l|l|}
		\multicolumn{2}{c}{} & \multicolumn{1}{c}{{\bf Prediction} (Section~\ref{sec:prediction})} & \multicolumn{1}{c}{{\bf Optimization} (Section~\ref{sec:optimization})} \\
		\cline{2-4}
		\multirow{16}{*}{ \rotatebox[origin=c]{90}{ {\bf Context} (Section~\ref{sec:classification}) } } 
		& \multirow{4}{*}{ \rotatebox[origin=c]{90}{ Geo } } & \emph{Ideal:} \cite{lu2013optimizing, abouzeid2013optimal, siris2013enhancing, abouzeid2015evaluating} & \emph{ConvOpt$^a$:} \cite{abouzeid2013optimal} \\
		\cline{3-4}
		& & \emph{Time series:} \cite{scellato2011nextplace, de2013interdependence, jiang2013tracking, bui2014model, liao2015channel, riiser2012video,  yang2013broadcasting} & \emph{\acs{MDP}$^b$/\acs{MPC}$^c$:} \cite{chon2014smartdc, chon2011mobility} \\
		\cline{3-4}
		& & \emph{Regression, classification:} \cite{chen2013predicting, chon2013understanding, froehlich2008route, monreale2009wherenext,ghouti2013mobility, hao2014gtube, margolies2014exploiting, sridaran2013location} & \emph{Game theory:} \cite{namvar2014context} \\
		\cline{3-4}
		& & \emph{Probabilistic:} \cite{song2010limits, lee2006modeling, abu2010application, barth2011mobility, barth2012combining, chon2014smartdc, chon2012evaluating, chon2011mobility, gidofalvi2012and,  lu2013approaching, xiong2014mpaas, chon2014adaptive} & \emph{Heuristic:} \cite{lu2013optimizing, hao2014gtube, riiser2012video, siris2013enhancing, margolies2014exploiting, yang2013broadcasting, chon2014adaptive} \\
		\cline{2-4}
		& \multirow{4}{*}{ \rotatebox[origin=c]{90}{ Link } } & \emph{Ideal:} \cite{abouzeid2014energy, abouzeid2014efficient, bui2015anticipatory, bui2015anticipatoryb, draxler2013cross, draxler2015smarterphones,  valentin2014anticipatory, zou2015can, muppirisetty2015proactive, muppirisetty2016channel, blobel2015anticipatory, tsilimantos2016anticipatory, atawia2014robust, mangla2016video, atawia2015chance, atawia2016joint} & \emph{ConvOpt:} \cite{abouzeid2014energy, abouzeid2014efficient, bui2015anticipatoryb, bui2015anticipatory, draxler2015smarterphones, draxler2013cross, zou2015can, valentin2014anticipatory, liu2016hop, blobel2015anticipatory, tsilimantos2016anticipatory, atawia2014robust, mangla2016video, atawia2015chance, atawia2016joint} \\
		\cline{3-4}
		& & \emph{Time series:} \cite{bui2015mobile, naimi2014anticipation, wang2013ames, kurdoglu2016realtime} & \emph{\acs{MDP}/\acs{MPC}:} \cite{bao2015bitrate, bianchi2013networked, yin2011prediction, hosseini2015not} \\
		\cline{3-4} 
		& & \emph{Regression, classification:} \cite{dallanese2011channel, kasparick2015kernel, muppirisetty2015spatial, piacentini2010path, tarsa2015taming, tie2011anticipatory, liu2016hop} & \emph{Game theory:} \cite{semiari2015context} \\
		\cline{3-4}
		& & \emph{Probabilistic:} \cite{bao2015bitrate, bianchi2013networked, nicholson2008breadcrumbs, seetharam2015managing, yin2011prediction, hosseini2015not, fazio2016pattern} & \emph{Heuristic:} \cite{bui2015mobile, wang2013ames, seetharam2015managing, tie2011anticipatory, naimi2014anticipation, tarsa2015taming, kurdoglu2016realtime, fazio2016pattern} \\
		\cline{2-4}
		& \multirow{4}{*}{ \rotatebox[origin=c]{90}{ Traffic } } & \emph{Ideal:} \cite{abedini2014content, abouzeid2013predictive, huang2014backpressure, proebster2011context, tadrous2013proactive, yao2012improving, yin2015control, zahran2016oscar} & \emph{ConvOpt:} \cite{abouzeid2013predictive, proebster2011context, pollakis2016anticipatory, zahran2016oscar, wang2016squad, yu2016power, yu2014predictive, du2016traffic, du2016resource, miller2015control} \\
		\cline{3-4}
		& & \emph{Time series:} \cite{lee2013generalized, okutani1984dynamic, papagiannaki2003long, riiser2013commute, sadek2004multi, zhou2005network, fu2010systematic, wang2016squad} & \emph{\acs{MDP}/\acs{MPC}:} \cite{lee2013generalized, fu2010systematic, yin2015control, yi2016cs2p} \\
		\cline{3-4}
		& & \emph{Regression, classification:} \cite{chen2015rate, samulevicius2015most, paul2011understanding, sayeed2015cloud, sekar2013developing, shafiq2011characterizing, xu2013proteus, jiang2016cfa, millan2015tracking, yu2016power, yu2014predictive, du2016traffic, du2016resource} & \emph{Game theory:} \cite{semiari2016context} \\
		\cline{3-4}
		& & \emph{Probabilistic:} \cite{beister2014predicting, shafiq2011characterizing, yi2016cs2p} & \emph{Heuristic:} \cite{abedini2014content, huang2014backpressure, samulevicius2015most, yao2012improving, sekar2013developing, xu2013proteus, jiang2016cfa} \\
		\cline{2-4}
		& \multirow{4}{*}{ \rotatebox[origin=c]{90}{ Social } } & \emph{Ideal:} \cite{bastug2013proactive, proebster2012context, tsiropoulos2011impact, siris2016exploiting} & \emph{ConvOpt:} \cite{proebster2012context, tsiropoulos2011impact, witheephanich2014min, tadrous2015optimal, tadrous2015joint} \\
		\cline{3-4}
		& & \emph{Time series:} \cite{wanalertlak2011behavior} & \emph{\acs{MDP}/\acs{MPC}:} \cite{chen2013markov} \\
		\cline{3-4}
		& & \emph{Regression, classification:} \cite{bastug2014think, bastug2014living, bastug2014anticipatory, dutta2015predictive, mozer2000predicting, noulas2012mining, yi2016spatial} & \emph{Game theory:} \cite{semiari2015context, semiari2016context, hamidouche2014many, namvar2014context, gu2015matching} \\
		\cline{3-4}
		& & \emph{Probabilistic:} \cite{bapierre2011variable, chen2013markov, witheephanich2014min, calabrese2010human, golrezaei2012femtocaching, zhang2015social, semiari2015context, semiari2016context, tadrous2015optimal, tadrous2015joint} & \emph{Heuristic:} \cite{golrezaei2012femtocaching, bastug2014think, bastug2013proactive, bastug2014anticipatory, bastug2014living, dutta2015predictive, wanalertlak2011behavior, zhang2015social, siris2016exploiting} \\
		\cline{2-4} 
		\multicolumn{2}{c}{} & \multicolumn{2}{l}{{\bf Surveys}: \cite{zheng2016big, makris2013survey, pejovic2015anticipatory, boucheron2005theory, liu2015empirical, nguyen2008survey, jin2013understanding, barakovic2013survey, jackson2008social, harvey1990forecasting, lee2012comparative, xu2005survey, murthy1998automatic, ramsay1991some, ramsay2006functional, boyd2004convex, schrijver1998theory, qin2003survey, puterman2014markov, sutton1998reinforcement} } \\
		\multicolumn{2}{c}{} & \multicolumn{2}{l}{{\bf Projects}: \cite{momentum, geopkdd, TelecomItalia} } \\
		\multicolumn{2}{c}{} & \multicolumn{2}{l}{\tiny{$^a$convex optimization $^b$Markov decision process $^c$model predictive control}} \\		
	\end{tabular}
\end{table*}
}

\newcommand{\overalltablehc}{
\begin{table*}
	\tabcolor	
	\centering
	\caption{Survey classification and structure}
	\label{tab:class}
	\begin{tabular}{c|lV{2.5}lV{2.5}l|}
		\multicolumn{2}{c}{} & \multicolumn{1}{c}{{\bf Prediction} (Section~\ref{sec:prediction})} & \multicolumn{1}{c}{{\bf Optimization} (Section~\ref{sec:optimization})} \\
		\cline{2-4}
		\multirow{16}{*}{ \rotatebox[origin=c]{90}{ {\bf Context} (Section~\ref{sec:classification}) } } 
		& \multirow{4}{*}{ \rotatebox[origin=c]{90}{ Geo } } & \emph{Ideal:} [31, 42, 43, 45] & \emph{ConvOpt$^a$:} [43] \\
		\cline{3-4}
		& & \emph{Time series:} [13, 28, 29, 32, 37, 38, 41] & \emph{\acs{MDP}$^b$/\acs{MPC}$^c$:} [24, 26] \\
		\cline{3-4}
		& & \emph{Regression, classification:} [14, 15, 22, 33-35, 44, 46] & \emph{Game theory:} [131] \\
		\cline{3-4}
		& & \emph{Probabilistic:} [11, 12, 16-21, 23-26] & \emph{Heuristic:} [25, 32, 41, 42, 44-46] \\
		\clineB{2-4}{2.5}
		& \multirow{4}{*}{ \rotatebox[origin=c]{90}{ Link } } & \emph{Ideal:} [56, 57, 65-70, 72-79] & \emph{ConvOpt:} [64-70 72-79] \\
		\cline{3-4}
		& & \emph{Time series:} [54, 58, 59, 63] & \emph{\acs{MDP}/\acs{MPC}:} [50, 60, 62, 158] \\
		\cline{3-4} 
		& & \emph{Regression, classification:} [47-49, 51, 52, 55, 64] & \emph{Game theory:} [129] \\
		\cline{3-4}
		& & \emph{Probabilistic:} [30, 50, 53, 60-62, 158] & \emph{Heuristic:} [30, 47, 51, 54, 58, 59, 61, 63] \\
		\clineB{2-4}{2.5}
		& \multirow{4}{*}{ \rotatebox[origin=c]{90}{ Traffic } } & \emph{Ideal:} [95-97, 111, 112, 115, 118, 138] & \emph{ConvOpt:} [103-107, 111, 118-120, 138] \\
		\cline{3-4}
		& & \emph{Time series:} [100, 108-110, 113, 119, 145, 165] & \emph{\acs{MDP}/\acs{MPC}:} [100, 115, 116, 165] \\
		\cline{3-4}
		& & \emph{Regression, classification:} [92-94, 98, 99, 101, 104-107, 114, 117, 156] & \emph{Game theory:} [117] \\
		\cline{3-4}
		& & \emph{Probabilistic:} [93, 102, 116] & \emph{Heuristic:} [96-99, 101, 112, 117] \\
		\clineB{2-4}{2.5}
		& \multirow{4}{*}{ \rotatebox[origin=c]{90}{ Social } } & \emph{Ideal:} [121, 124, 137, 140] & \emph{ConvOpt:} [126, 127, 137, 140, 159] \\
		\cline{3-4}
		& & \emph{Time series:} [40] & \emph{\acs{MDP}/\acs{MPC}:} [157] \\
		\cline{3-4}
		& & \emph{Regression, classification:} [122, 123, 134, 139, 148, 149, 154] & \emph{Game theory:} [128-131, 133] \\
		\cline{3-4}
		& & \emph{Probabilistic:} [125-127, 129, 130, 132, 135, 136, 157, 159] & \emph{Heuristic:} [40, 121-125, 132, 148, 149] \\
		\cline{2-4} 
		\multicolumn{2}{c}{} & \multicolumn{2}{l}{\tiny{$^a$convex optimization $^b$Markov decision process $^c$model predictive control}} \\
	\end{tabular}	
\end{table*}
}

\newcommand{\classtable}{
\begin{table*}[t]
	\begin{minipage}{2\columnwidth}
	\centering
	\tabcolor	
	\caption{Context Classification Summary: each context is associated to its most popular applications, prediction techniques, optimization methods and main notable characteristics.}
	\label{tab:class_opt_eval}
	\begin{tabular}{|p{1.6cm}V{2.5}p{2.55cm}|p{2.05cm}|p{3.8cm}|p{5.9cm}|}
		\hline
		{\bf Context} & {\bf Applications} & {\bf Prediction\footnote{Ranking based on the number of papers reviewed in this survey using the predictor.}} & {\bf Optimization} & {\bf Remarks} \\
		\hlineB{2.5}
		{\bf Geographic}\newline[11-26, 28, 29, 31-35, 37, 38, 41-46, 131] & 
		Mobility prediction\newline Multimedia streaming \newline Broadcast\newline  Resource allocation\newline  Duty cycling& 
		1$^{\rm st}$ Probabilistic\newline 2$^{\rm nd}$ Regression\newline 3$^{\rm rd}$ Time series\newline 4$^{\rm th}$ Classification & 
		1) Prediction to define convex optimization problems \newline 2) Prediction as the optimization objective &
		1) Prediction accuracy is inversely proportional to the time scale and granularity\newline 2) High prediction accuracy can be obtained on long time scales if periodicity and/or trends are present\newline 3) Prediction is more effectively used in delay tolerant applications \\
		\hline
		{\bf Link}\newline[30, 47-70, 72-79, 129, 158] & 
		Channel forecast\newline Resource allocation\newline Network mapping\newline Routing\newline Multimedia streaming &
		1$^{\rm st}$ Regression\newline 2$^{\rm nd}$ Time series\newline 3$^{\rm rd}$ Probabilistic\newline 4$^{\rm th}$ Classification & 
		1) Markov decision process is used when statistical knowledge of the system is available\newline 2) Convex optimization is preferred when it is possible to perform accurate forecast & 
		1) Channel quality maps can be effectively used to improve networking\newline 2) Mobility dynamics affect the prediction effectiveness\newline 3) Channel is most often predicted by means of functional regression or Markovian models\\
		\hline
		{\bf Traffic}\newline[92-102, 104-120, 138, 145, 156, 165] &
		Traffic analysis\newline Resource allocation\newline Multimedia streaming &
		1$^{\rm st}$ Regression\newline 2$^{\rm nd}$ Classification\newline 3$^{\rm rd}$ Probabilistic &
		1) Maps are used to deterministically guide the optimization\newline 2) Convex optimization problems can be formulated to obtain bounds &
		1) Improved long-term network optimization and reconfiguration\newline 2) Traffic distribution is skewed both with regards to users and locations\newline 3) Traffic has a strong time periodicity\newline 4) Geo-localized information can be used as inputs for optimization\\
		\hline
		{\bf Social}\newline[40, 121-140, 148, 149, 154, 157, 159] & 
		Network caching\newline Mobility prediction\newline Resource allocation\newline Multimedia streaming  &
		1$^{\rm st}$ Classification\newline 2$^{\rm nd}$ Regression\newline 3$^{\rm rd}$ Time series\newline 4$^{\rm th}$ Probabilistic & 
		1) Formal optimization problems can be defined, but they are usually impractical to be solved\newline 2) Game theory and heuristics are the preferable online solutions &
		1) A fraction of social information can be accurately predicted\newline 2) Prediction obtained from social information is usually coarse\newline 3) Social information prediction can effectively improve application performance \\
		\hline
	\end{tabular}
	\label{tab:prediction_class}
	\end{minipage}
\end{table*}
}

\newcommand{\figtrajectory}{
\begin{figure}
\centering
\includegraphics[width=1\columnwidth]{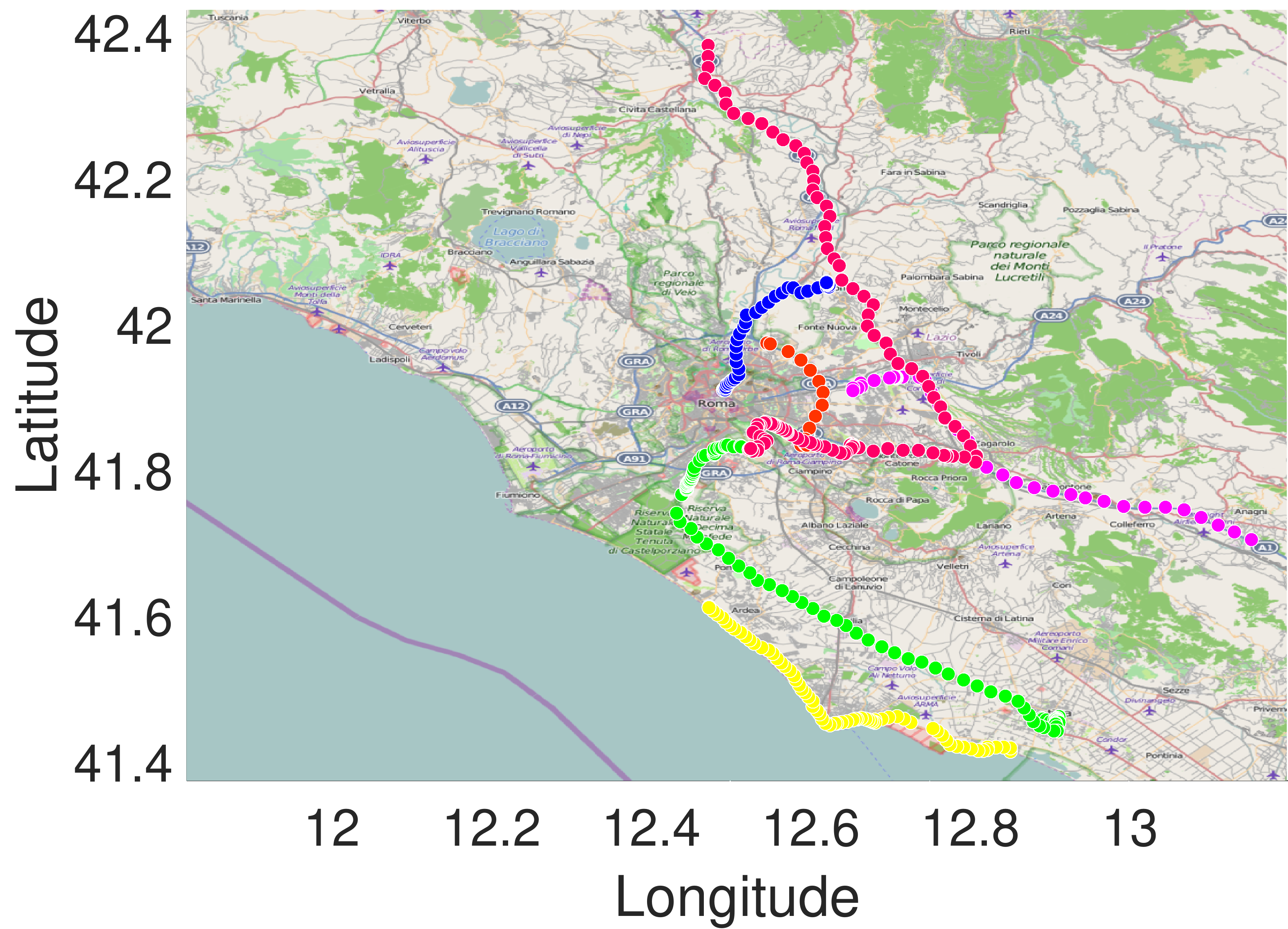}
\vspace{-0.7cm}\caption{Geographic context example: an example of estimated trajectories of 6 mobile users.}
\label{fig:trajectory}
\end{figure}
}

\newcommand{\figpathlossmap}{
\begin{figure}
  \centering
  \ifCLASSINFOpdf
  \subfigure{\includegraphics[trim={1cm 0 0 0},clip, width=.9\columnwidth]{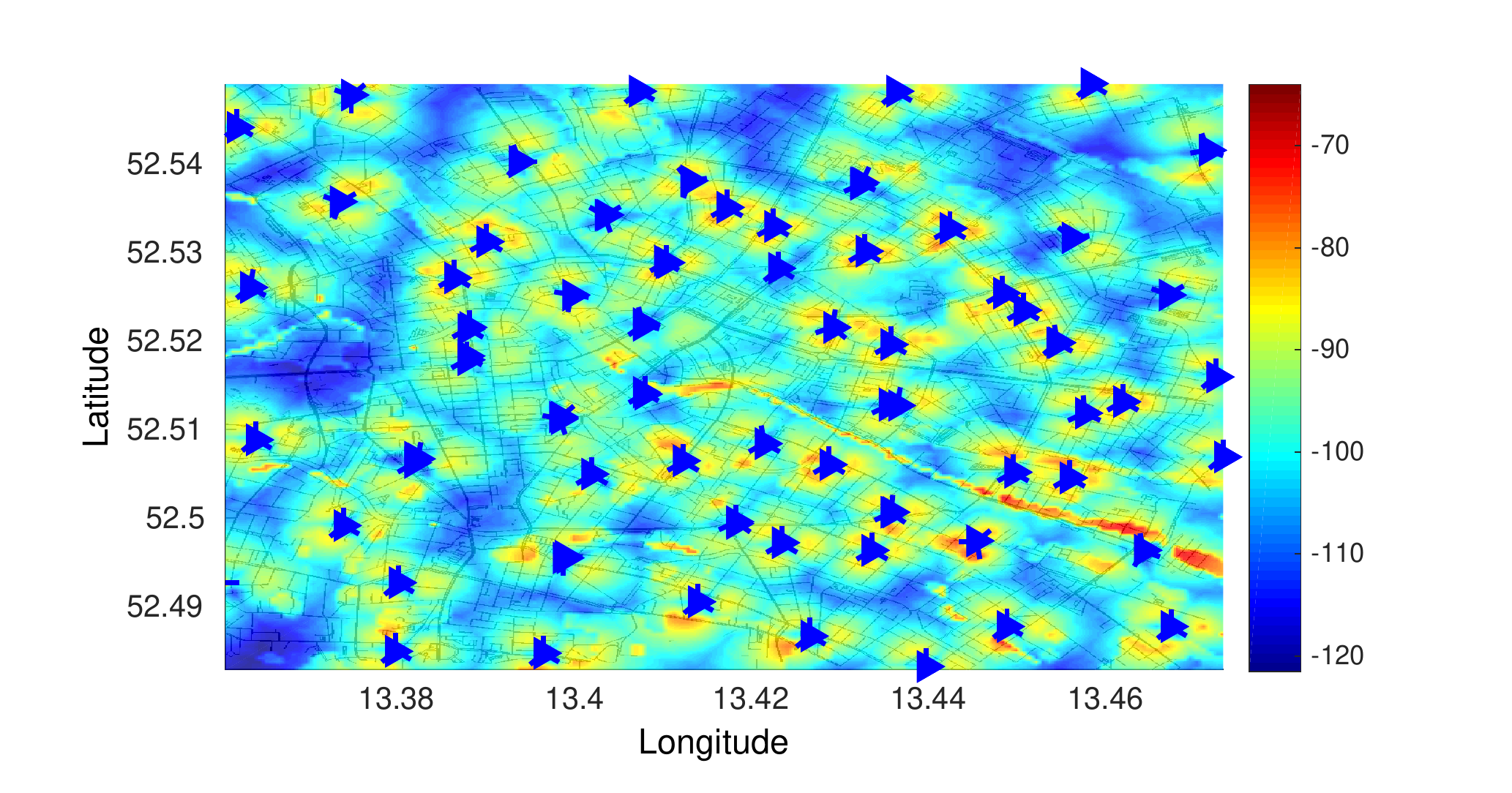}}
  \else
  \subfigure{\includegraphics[width=0.9\columnwidth]{./fig/BerlinPathlossMap}}
  \fi
  \vspace{-0.2cm}\caption{Link context example: a pathloss map of Berlin downtown obtained from the data of the MOMENTUM project~\cite{momentum}, where the triangles represent base stations. Pathloss maps are frequently used to predict the evolution of the connection quality in mobile networks.}
  \label{fig:map}
\end{figure}
}

\newcommand{\figtimeseries}{
\begin{figure}
\centering
\subfigure[Uplink and downlink traffic load in a cell grid in Rome, Italy.]{\label{fig:TS_CellLoad}\includegraphics[width=1\columnwidth]{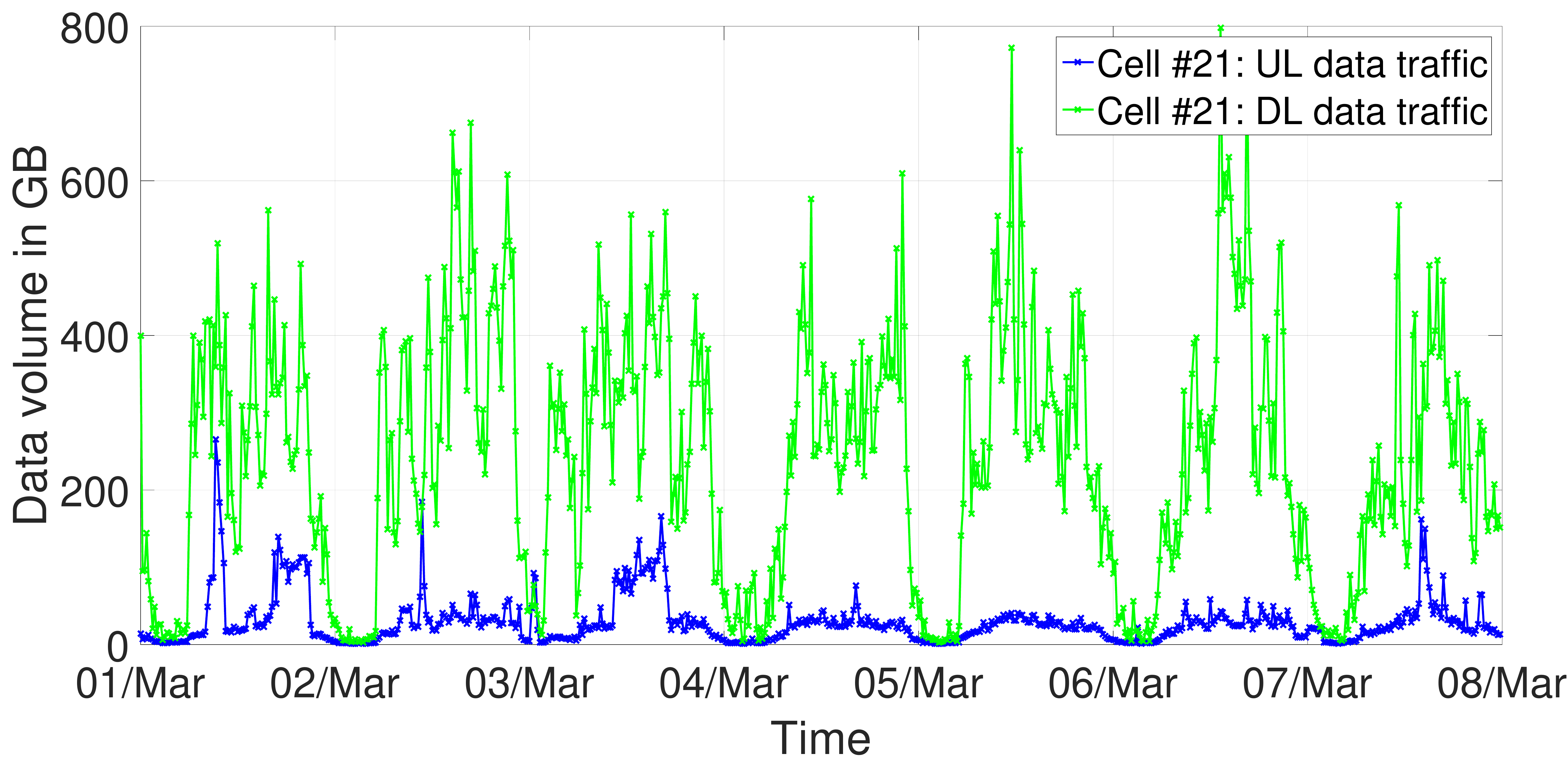}}
\subfigure[Aggregated uplink and downlink traffic load in Rome, Italy.]{\label{fig:TS_AggrLoad}\includegraphics[width=1\columnwidth]{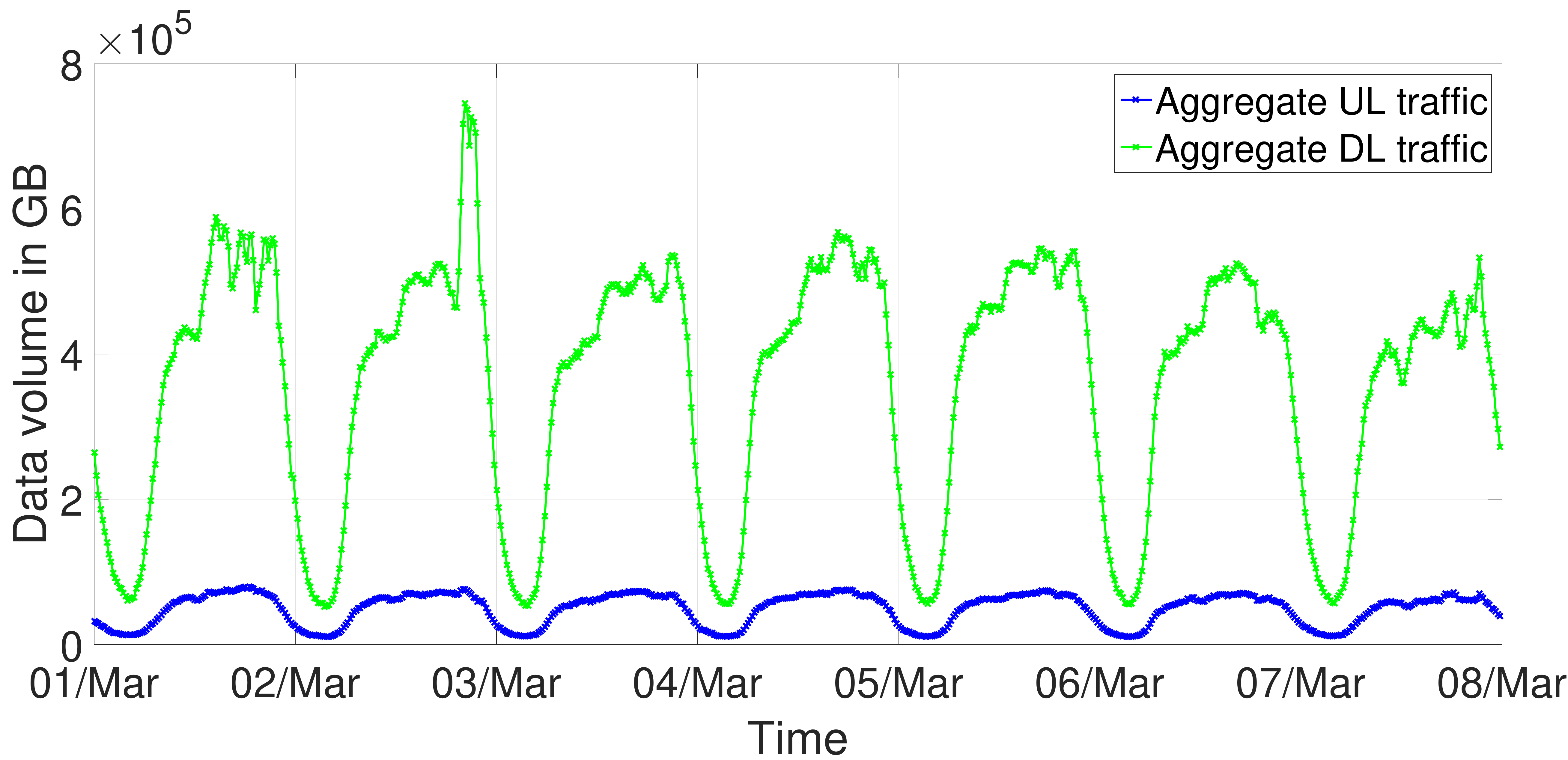}}
\caption{Example of time series: Traffic load (aggregated every 15 minutes) for a week in March 2015 in Rome, Italy. Data source from Telecom Italia's Big Data Challenge \cite{TelecomItalia}.}
\label{fig:TS_Load}
\end{figure}
}

\newcommand{\figfunctional}{
\begin{figure}
\centering
\includegraphics[width=.9\columnwidth]{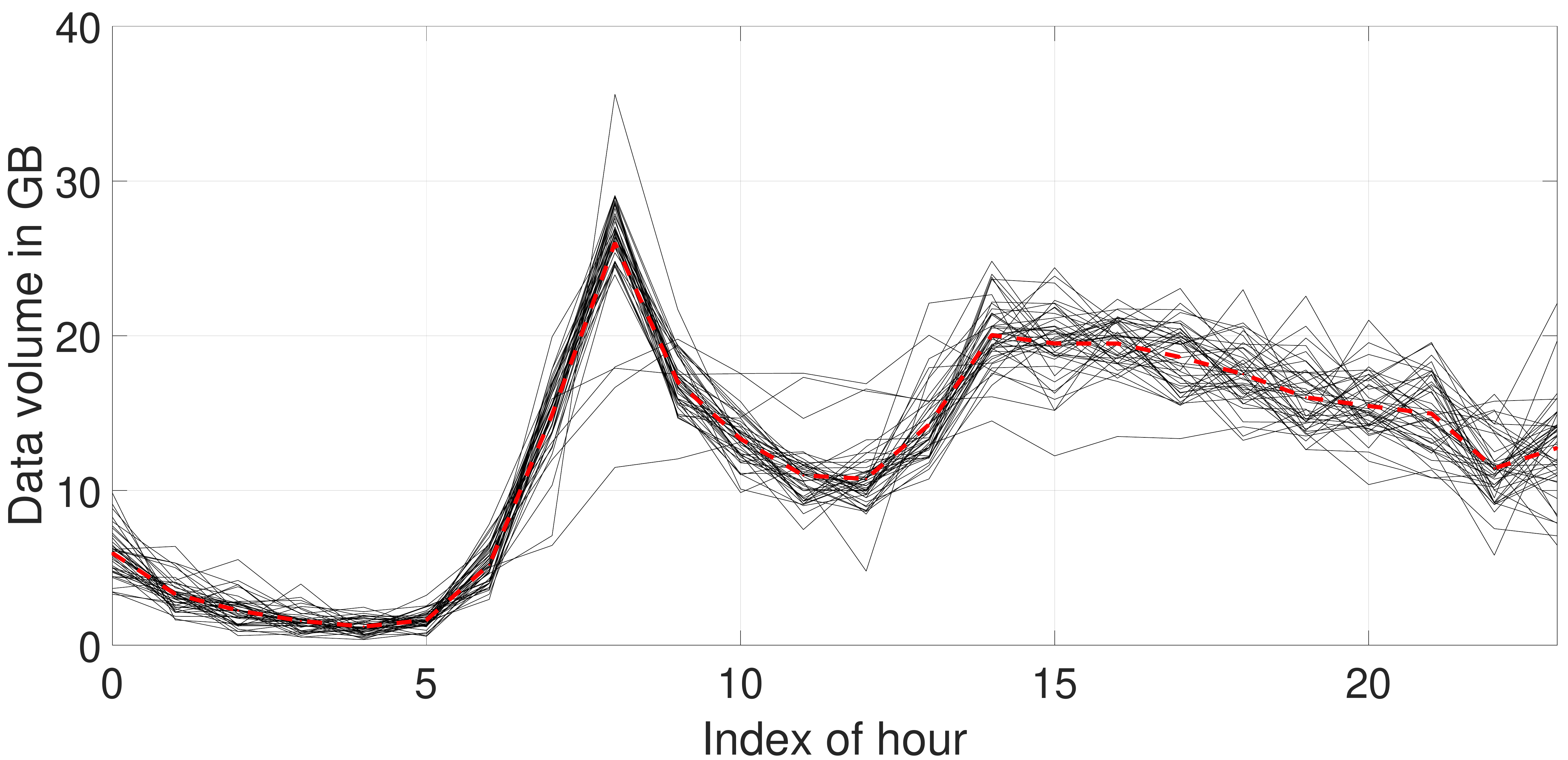}
\caption{Example of a functional dataset: WiFi traffic in Rome depending on hour of the day. Data source from Telecom Italia's Big Data Challenge~\cite{TelecomItalia}.}
\label{fig:FDA_wifiload}
\end{figure}
}

\newcommand{\figsvm}{
\begin{figure}
\centering
\subfigure[Linear]{\label{fig:svm:lin}\includegraphics[width=.51\columnwidth]{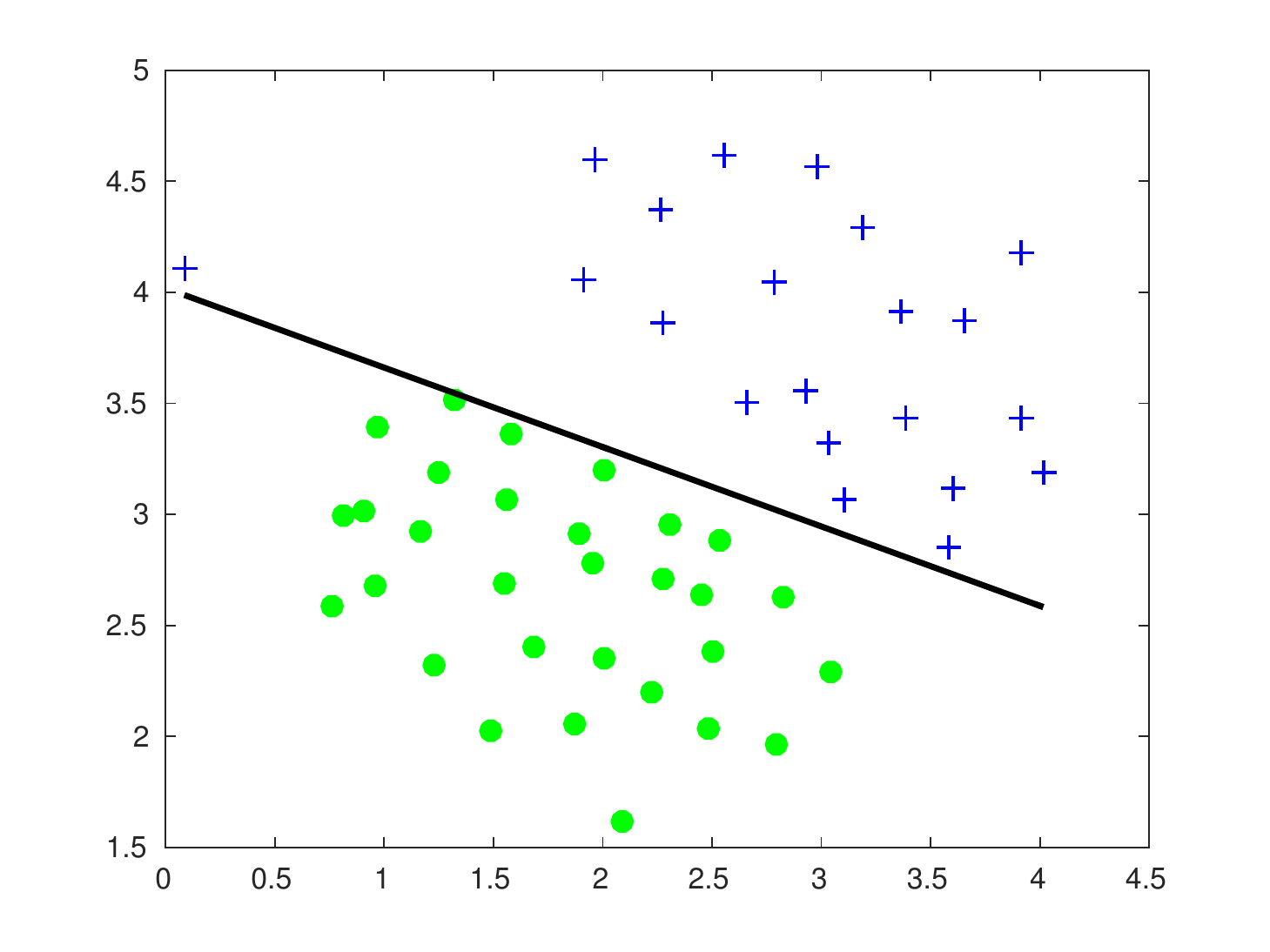}}
\hspace{-0.4cm}
\subfigure[Gaussian]{\label{fig:svm:gau}\includegraphics[width=.51\columnwidth]{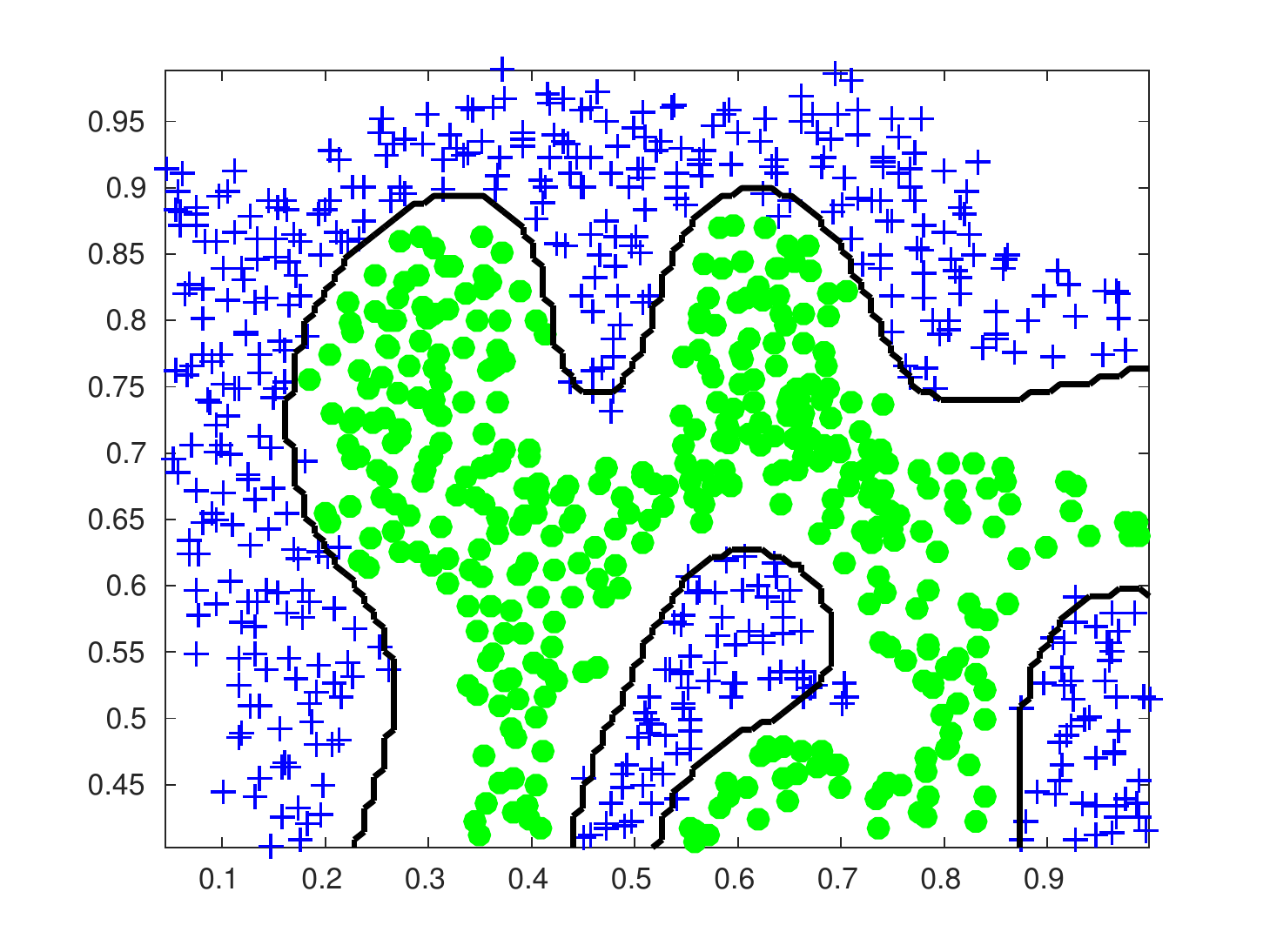}}
\caption{Examples of SVM, where different datasets are analyzed according to a linear (left) and a Gaussian (right) kernel.}
\label{fig:svm}
\end{figure}
}

\newcommand{\tableprediction}{
\begin{table*}[t]
\centering
\caption{Selected Prediction Methods: Variables of interest and constraints of modeling.}
\begin{tabular}{|p{1.8cm}|p{2cm}V{2.5}p{1.65cm}|p{1.65cm}|p{1.65cm}V{2.5}p{1.1cm}|p{1.4cm}|p{1.2cm}|p{1.2cm}|}
\hline
\multicolumn{2}{|cV{2.5}}{{\bf Prediction Method}} & \multicolumn{3}{cV{2.5}}{{\bf Properties of the Context}} & \multicolumn{4}{c|}{{\bf Constraints}} \\
\hline
Class & Methodology & Dimension & Granularity &  Range & Type & Linearity & Side Info. & Quality \\
\hlineB{2.5}
\multirow{3}{*}{Time series} & ARIMA & univariate & M/L & S & data  & Y & N & weak \\
\cline{2-9}
& Kalman filter &  multivariate  & M/L & S & data  & Y & N & weak \\
\cline{2-9}
& References & \multicolumn{7}{l|}{
\emph{ARIMA}: \cite{hao2014gtube, jiang2013tracking, liao2015channel, tie2011anticipatory, wang2013ames, naimi2014anticipation, bui2015mobile, lee2013generalized, wanalertlak2011behavior, kurdoglu2016realtime, wang2016squad} 
\emph{Kalman}: \cite{yang2013broadcasting, dallanese2011channel} } \\
\hlineB{2.5}
\multirow{4}{*}{Classification} & \ac{CF} & multivariate & L & M/L & data & Y & both & robust \\
\cline{2-9}
& Clustering & multivariate & L & M/L & data & both & both & robust \\
\cline{2-9}
& Decision trees & multivariate & L & any & data & both & Y & robust \\
\cline{2-9}
& References & \multicolumn{7}{l|}{
\emph{CF}: \cite{xiong2014mpaas, noulas2012mining, dutta2015predictive} 
\emph{Cluster}: \cite{chen2013predicting, froehlich2008route, tarsa2015taming, chen2015rate, bastug2014think, bastug2014living, bastug2014anticipatory, jiang2016cfa} 
\emph{Decision trees}: \cite{monreale2009wherenext, sekar2013developing, xu2013proteus}} \\
\hlineB{2.5}
\multirow{4}{*}{Regression} & Functional &  multivariate & any & M/L & models & both & Y & robust \\
\cline{2-9}
& \ac{SVM} & multivariate & any & any & both & both & both & weak \\
\cline{2-9}
& \ac{ANN} & multivariate & any & any & data & both & both & weak \\
\cline{2-9}
& References & \multicolumn{7}{l|}{
\emph{Functional}: \cite{de2013interdependence, scellato2011nextplace, liao2015channel, samulevicius2015most, liu2016hop, yu2016power, yu2014predictive} 
\emph{SVM}: \cite{tarsa2015taming, yi2016spatial, millan2015tracking} 
\emph{ANN}: \cite{ghouti2013mobility, piacentini2010path, du2016traffic, du2016resource}} \\
\hlineB{2.5}
\multirow{4}{*}{Probabilistic} & Markovian & multivariate & M/L & any & both & both & both & weak \\
\cline{2-9}
& Bayesian & multivariate & any & any & both & both & Y & weak \\
\cline{2-9}
& \multirow{2}{*}{References} & \multicolumn{7}{l|}{
\emph{Probabilistic}: \cite{xiong2014mpaas, barth2012combining, gidofalvi2012and, lu2013approaching, chon2014smartdc, chon2011mobility, abu2010application, barth2011mobility, chon2012evaluating, lee2006modeling, bapierre2011variable, bao2015bitrate, nicholson2008breadcrumbs, seetharam2015managing, yin2011prediction, beister2014predicting, chen2013markov, shafiq2011characterizing, fazio2016pattern, chon2014adaptive, yi2016cs2p} } \\
& & \multicolumn{7}{l|}{ \emph{Bayesian}: \cite{sridaran2013location, bui2014model, bui2015mobile, bianchi2013networked, witheephanich2014min, calabrese2010human, zhang2015social, semiari2015context, semiari2016context, tadrous2015optimal, tadrous2015joint} } \\
\hline
\end{tabular}
\label{tab:ObjectiveAndConstaints_predict}
\end{table*}
}

\newcommand{\taboptimization}{
\begin{table*}[t]
	\centering
	\caption{Optimization Methods Summary}
	\label{tab:class_opt_summary}
	\begin{tabular}{|p{2.5cm}V{2.5}p{7cm}V{2.5}p{7cm}|}
		\hline
		{\bf Methodology} & {\bf Properties of context} & {\bf Modeling constraints} \\
		\hlineB{2.5}
		ConvOpt & Can support any context property, but larger system states slow the solver performance. The solution accuracy is linked to the context precision. & Linearity can be exploited to improve the solver efficiency, while data reliability impacts the solution optimality. \\
		\hline
		MPC & Usually offers the highest precision by coupling prediction and optimization. & The most computationally intensive technique. \\
		\hline
		MDP & Limited range and precision. & The most robust approach to low data reliability. Although the system setup can be computationally intensive, it allows for lightweight policies to be implemented. \\
		\hline
		Game theory & Limited granularity to allow the system to converge to an equilibrium. & Very low computational complexity. Fast dynamics hinder the system convergence. \\
		\hline
	\end{tabular}
	\label{Tab:Optimization_Class}
\end{table*}
}

\newcommand{\tabnetwork}{
\begin{table*}[t]
	\tabcolor
	\centering
	\caption{Anticipatory Networking applicability to different Network Types}
	\label{tab:class_opt_summary}
	\begin{tabular}{|p{2cm}V{2.5}p{4.5cm}|p{4.5cm}|p{4.5cm}|}
		\hline
		{\bf Type} & {\bf Features} & {\bf Advantages} & {\bf Challenges} \\
		\hlineB{2.5}
		\emph{5G Cellular} & mm-waves\newline Massive MIMO\newline Cloud-RAN & Localization and tracking prediction\newline Load space-time distribution\newline Resource management & Channel models\newline Amount of data \\
		\hline
		\emph{\acs{MANET}} & Variable topology\newline Multi-hop communication\newline Self-management & Routing improvement\newline Load balancing & Infrastructure absence\newline Distributed optimization\newline Variable topology \\
		\hline
		\emph{Cognitive} & Primary/Secondary users\newline Sensing capabilities & Spectrum availability prediction\newline Load prediction and management\newline Transmission/Sensing ratio & Impact on models \\
		\hline
		\emph{\acs{D2D}} & Complex topology\newline Multi-RAN & Interference management\newline Resource allocation & Models complexity\newline Interference \\
		\hline
		\emph{\acs{IoT}} & Mostly deterministic traffic\newline High overhead\newline Sparse communication\newline Low-latency control loops & Prediction for compression\newline Models for anomaly detection\newline Overhead decrease & Amount of data and devices\newline Scalability\newline Constrained devices\\
		\hline		
	\end{tabular}
	\label{tab:network}
\end{table*}
}

\section{Introduction}
\label{sec:introduction}

Evolving from one generation to the next, wireless networks have been constantly increasing their performance in many different ways and for diverse purposes. Among them, communication efficiency has always been paramount to increase the network capabilities without updating the entire infrastructure. This survey investigates anticipatory networking, a recent research direction that supports network optimization through system state prediction.

The core concept of anticipatory networking is that, nowadays, tools exist to make reliable prediction about network status and performance.
Moreover, information availability is increasing every day as human behavior is becoming more socially and digitally interconnected. 
In addition, data centers are becoming more and more important in providing services and tools to access and analyze huge amounts of data.

As a consequence, not only can researchers tailor their solutions to specific places and users, but also they can anticipate the sequence of locations a user is going to visit or to forecast whether connectivity might be worsening, and to exploit the forecast information to take action before the event happens. This enables the possibility to take full advantage of good future conditions (such as getting closer to a base station or entering a less loaded cell) and to mitigate the impact of negative events (e.g., entering a tunnel).

This survey covers a body of recent works on anticipatory networking, which share two common aspects: 
\begin{itemize}
 \item {\it Anticipation}: they either explore prediction techniques directly or consider some future knowledge as given.
 \item {\it Networking}: they aim to optimize communications in mobile networks.
\end{itemize}
In addition, this survey delves into the following questions: How can prediction support wireless networks? Which type of information is possible to predict and which applications can take advantage of it? Which tools are the best for a given scenario or application? Which scenarios, among the ones envisioned for 5G networks, can benefit the most from anticipatory networking? What is yet to be studied in order for anticipatory networking to be implemented in 5G networks?

\rev{The main contributions of this survey are the following:
\begin{itemize}
 \item A thorough {\bf context-based analysis} of the literature classified according to the information exploited in the predictive framework.
 \item Two {\bf handbooks on the prediction and optimization} techniques used in the literature, which allow the reader to get familiar with them and critically assess the different approaches.
 \item \revv{An analysis of the applicability of anticipatory networking techniques to different {\bf types of wireless networks} and at different layers of the {\bf protocol stack}.} 
 \item Summaries of all the main parts of the survey, highlighting {\bf most popular choices and best practices}.
 \item A final section analyzing {\bf open challenges and potential issues} to the adoption of anticipatory networking solutions in future generation mobile networks.
\end{itemize}
}

\subsection{Background and Guidelines}
\label{sec:guidelines}

\overalltablehc

Anticipatory networking is the engineering branch that focuses on communication solutions that leverage the knowledge of the future evolution of a system to improve its operation. For instance, while a standard networking solution would answer the question \textit{``which is the best user to be served?''}, an anticipatory equivalent would answer \textit{``which are the best users to be served in the next time frames given the predicted evolution of their channel condition and service requirements?''}

A typical anticipatory networking solution is usually characterized by the following three attributes, which also determine the structure of this survey: 
\begin{itemize}
 \item {\it Context} defines the type of information considered to forecast the system evolution.
 \item {\it Prediction} specifies how the system evolution is forecast from the current and past context.
 \item {\it Optimization} describes how prediction is exploited to meet the application objectives.
\end{itemize}

To continue with the access selection example, the anticipatory networking solution might exploit the history of \ac{GPS} information (the \textit{context}) to train an \ac{AR} model (the \textit{prediction}) to predict the future positions of the users and their  channel conditions to solve an \ac{ILP} problem (the \textit{optimization}) that maximizes their \ac{QoE}.

The main body of the anticipatory networking literature can be split into four categories based on the context used to characterize the system state and to determine its evolution: {\it geographic}, such as human mobility patterns derived from location-based information; {\it link}, such as channel gain, noise and interference levels obtained from reference signal feedback; {\it traffic}, such as network load, throughput, and occupied physical resource blocks based on higher-layer performance indicators; {\it social}, such as user's behavior, profile, and information derived from user-generated contents and social networks. 

In order to determine which techniques are the most suitable to solve a given problem, it is important to analyze the following:
\begin{itemize}
 \item {\it Properties} of the context: \\ 
 1) {\it Dimension} describes the number of variables predicted by the model, which can be uni- or multivariate. \\ 
 2) {\it Granularity and precision} define the smallest variation of the parameter considered by the context and the accuracy of the data: the lower the granularity, the higher the precision and vice versa. Temporal and spatial granularities are crucial to strike a balance between efficiency and accuracy. \\ 
 3) {\it Range} characterizes the distance (usually time or space) between known data samples and the farthest predicted sample. It is also known as prediction (or optimization) horizon. 
 \item {\it Constraints} of the prediction or optimization model: \\ 
 1) {\it Availability of physical model} states whether a closed-form expression exists to describe the phenomenon. \\ 
 2) {\it Linearity} expresses the quality of the functions linking inputs and outputs of a problem. \\ 
 3) {\it Side information} determines whether the main context can be supported by auxiliary information. \\ 
 4) {\it Reliability and validity of information} specifies the noisiness of the data set, depending on which the prediction robustness should be calibrated.
\end{itemize}

\tabrelated

The  classification section will help the reader to understand the link between the different contexts and the solutions adopted to satisfy the given application requirements. Also, it is meant to provide a complete panorama of anticipatory networking. The two handbooks have the twofold objective of providing the reader with a short overview of the tools adopted in the literature and to analyze them in terms of variables of interest and constraints of the models. 
\rev{We believe that not only will this survey help researchers studying anticipatory networking, but also it will ease its adoption in future generation networks by providing a comprehensive overview of research directions, available solutions and application scenarios.}

Table~\ref{tab:class} provides a mapping between the techniques described in Section~\ref{sec:prediction} and~\ref{sec:optimization} (columns) and the context discussed in Section~\ref{sec:classification} (rows). Each main category is further split into subcategories according to its internal structure. Namely, the prediction category is subdivided into ideal (perfect prediction is assumed to be available), time series predictive modeling,  similarity-based classification and regression analysis, and probabilistic methods. The optimization category is split into \ac{ConvOpt}, \ac{MDP} and \ac{MPC}, game theoretic and, heuristic approaches. 

\rev{The rest of the survey consists of a quick overview of other surveys on related topics in Section~\ref{sec:related_work}, a context-based classification of the anticipatory networking literature in Section~\ref{sec:classification}, two handbooks on prediction and optimization techniques in Section~\ref{sec:prediction} and Section~\ref{sec:optimization}, respectively. \revv{Section~\ref{sec:network} and~\ref{sec:protocol} discuss how the anticipatory networking paradigm can be applied in a variety of network types and at different layers of the protocol stack.}
Section~\ref{sec:challenges} and~\ref{sec:conclusions} conclude the survey reporting the impact of anticipatory networking on future networks, the envisioned hindrances to its implementation and the open challenges.}

\rev{\section{Related Work}\label{sec:related_work}}

This section discusses a few recent survey on topics close to anticipatory networking and is summarized in Table~\ref{tab:related}.


\rev{Applying big data analytics for network optimization is studied in~\cite{zheng2016big}. Based on the papers they reviewed, the authors propose a generic framework to support big data based optimization of mobile networks. Using traffic patterns derived from case studies, they argue that their framework can be used to optimize resource allocation, base station deployment, and interference coordination in such networks. In~\cite{makris2013survey,pejovic2015anticipatory}, the ability to extract and process contextual information by entities in a network is identified as a key factor in improving network performance. In~\cite{makris2013survey}, the procedure of using context information in wireless networks is broken down into acquisition, modeling, exchanging and evaluating stages, where the first two deal with gathering information and predicting the future behavior, and the latter two perform self-optimization and decision making. A similar taxonomy is provided in~\cite{pejovic2015anticipatory} and various examples of different techniques are reviewed for each phase. In addition to that, the authors provide a thorough survey on potential use cases of anticipatory networks and their respective challenges.}

\rev{Predicting future states of network attributes is an essential task in designing anticipatory networks. Data classification, a popular prediction technique, has been thoroughly surveyed in~\cite{boucheron2005theory}. Among other attributes, the prediction of data traffic and throughput has been the subject of~\cite{liu2015empirical,nguyen2008survey}. In~\cite{liu2015empirical}, the authors consider seven algorithms for throughput prediction, ranging from mean-based and linear regression methods to \acp{ANN} and \acp{SVM} and compare their performance using a trace-driven simulator. Furthermore, they develop an information theoretic lower bound for the prediction error. In a similar attempt,~\cite{nguyen2008survey} reviews real time Internet traffic classification. Here, the authors not only review prediction algorithms, but also try to shed light on practical challenges in deploying different kinds of techniques under different network scenarios. For instance, they argue that algorithms that require packet inspection either in the form of port number or payload, might have limited applicability due to potential encryption compared to methods that rely on statistical traffic properties.  }

\rev{The capability to extract user behavior in online social networks and use it to learn the evolution of traffic patterns in mobile networks is the subject of another survey~\cite{jin2013understanding}. The general approach of the papers included in that review is to use social graphs and classify different types of interactions between users on social networks in order to monitor the corresponding network traffic. Another important attribute for network performance is modeling the Quality of Experience (QoE) or how the service is perceived by the user. The authors of~\cite{barakovic2013survey} provide a thorough survey including various methods for modeling QoE for different applications and also discuss tools for estimating and predicting QoE values by probing network parameters.}

\revv{\ac{CR} and \ac{REM} are two very important technologies to measure, estimate and predict spectrum availability and occupancy. For instance, \cite{hoyhtya2016spectrum, chen2016survey}~provide two independent taxonomies of methodologies, campaigns and models. In addition, they review the reliability of these types of measurements~\cite{hoyhtya2016spectrum} and they illustrate how to predict the system evolution thanks to available information and regression analysis~\cite{chen2016survey}.}

\rev{To the best of our knowledge, this survey is the first to specifically address anticipatory techniques for mobile networks. We believe that, while the topic is undeniably hot, an overarching review of the body of work is still missing and greatly needed to facilitate the adoption of such a promising direction.}
 
\rev{
\section{A Context-Based Classification of Anticipatory Networking Solutions}
\label{sec:classification}}

In this section, we classify the different types of context that can be predicted and exploited. For each one, we highlight the most popular prediction techniques as well as the applications for which an anticipatory optimization is performed. 

\subsection{Geographic Context}

Geographic context refers to the geographic area associated to a specific event or information. In wireless communications, it refers to the location of the mobile users, often enriched with speed information as well as past and future trajectories. Understanding human mobility is an emergent research field that especially in the last few years has significantly benefited from the rapid proliferation of wireless devices that frequently report status and location updates. Fig.~\ref{fig:trajectory} illustrates an example of estimated trajectories of 6 mobile users.

The potential predictability in user mobility can be as high as~$93\%$~\cite{song2010limits}\footnote{Value obtained for a high-income country with stable social conditions. The percentage can decrease for different countries, e.g., low-income country or natural disaster situation.}. Along the same line,~\cite{lu2013approaching} investigates both the maximal predictability and how close to this value practical algorithms can come when applied to a large mobile phone dataset. Those results indicate that human mobility is very far from being random. Therefore, collecting, predicting and exploiting geographic context is of crucial importance. 

In the rest of this section we organize the papers dealing with geographic context according to their main focus: the majority of them deals with pure geographical prediction 
and differs on secondary aspects such as whether they predict a single future location, a sequence of places or a trajectory. The second largest group of papers deals with multimedia streaming optimization.

\subsubsection{Next location prediction}
The simplest approach is to forecast where a given user will be at a predetermined instant of time in the future.
The authors of~\cite{jiang2013tracking} propose to track mobile nodes using topological coordinates and topology preserving maps. Nodes' location is identified with a vector of distances (in hops) from a set of nodes called anchors and a linear predictor is used to estimate the mobile nodes' future positions. Evaluation is performed on synthetic data and nodes are assumed to move at constant speed. Results show that the proposed method approaches an accuracy above $90\%$ for a prediction horizon of some tens of seconds. 

\figtrajectory

A more general approach that exploits \acp{ANN} is discussed in~\cite{ghouti2013mobility}. \acp{ELM}, which do not require any parameter tuning, are used to speed up the learning process. The method is evaluated using synthetic data over different mobility models. 

To extend the prediction horizon~\cite{chen2013predicting} exploits users' locations and short-term trajectories to predict the next handover. The authors use \ac{CSI} and handover history to solve a classification problem via supervised learning, i.e., employing a multi-class \ac{SVM}. In particular, each classifier corresponds to a possible previous cell and predicts the next cell. A real-time prediction scheme is proposed and the feedback is used to improve the accuracy over time. Simulation results have been derived using both synthetic and real datasets. The longer moves along a given path, the higher the accuracy of forecasting the rest. 

Location information can be extracted from cellular network records. In this way the granularity of the prediction is coarser, but positioning can be obtained with little extra energy. In particular, \cite{xiong2014mpaas}~aims at predicting a given user location from those of similar users. \textit{Collective behavioral patterns} and a Markovian predictor are used to compute the next six locations of a user with a one-hour granularity, i.e., a six-hour prediction horizon. Evaluation is done using a real dataset and shows that an accuracy of about $70\%$ can be achieved in the first hour, decreasing to $40-50\%$  for the sixth hour of prediction. 

\subsubsection{Space and time prediction}
Prediction of mobility in a combined space-time domain is often modeled using statistical methods. In~\cite{lee2006modeling}, the idea is to predict not only the future location a user will reach, but also \textit{when} and for \textit{how long} the user will stay there. To incorporate the \textit{sojourn} time during which a user remains in a certain location, mobility is modeled as a semi-Markov process. In particular, the transition probability matrix and the sojourn time distribution are derived  from the previous association history. Evaluation is done on a real dataset and shows approximately  $80\%$ accuracy. A similar approach is presented in~\cite{abu2010application}, where the prediction is extended from single to multi-transitions (estimating the likelihood of the future event after an arbitrary number of transitions). Both papers provide also some preliminary results on the benefits of the prediction on resource allocation and balancing. 

In~\cite{barth2011mobility}, the authors represent the network coverage and movements using graph theory. The user mobility is modeled using a \ac{CTM} process where the prediction of the next node to be visited depends not only on the current node but also on the previous one  (i.e., second-order Markovian predictor). Considering both local as well as global users' profiles,~\cite{barth2012combining} extends the previous Markovian predictor and improves accuracy by about $30\%$. 
As pointed out in~\cite{gidofalvi2012and}, sojourn times and transition probabilities are inhomogeneous. 
Thus, an inhomogeneous \ac{CTM} process is exploited to predict user mobility. Evaluation on a real dataset shows an accuracy of $67\%$ for long time scale prediction. 

The interdependence between time and space is investigated also in~\cite{chon2013understanding} by examining real data collected from smartphones during a two-month deployment. Furthermore, \cite{chon2012evaluating} shows the benefit of using a location-dependent Markov predictor with respect to a location-independent model based on nonlinear time series analysis. Additionally, it is shown that information on  arrival times and periodicity of location visits is needed to provide accurate prediction. A system design, named SmartDC, is presented in~\cite{chon2014smartdc, chon2014adaptive, chon2011mobility}. SmartDC comprises a mobility learner, a mobility predictor and an adaptive duty cycling. \rev{The proposed location monitoring scheme optimizes the sensing interval for a given energy budget.} The system has been implemented and tested in a real environment. Notably, this is also one of the few papers that takes into account the \textit{cost} of prediction, which in this case is evaluated in terms of energy. Namely, the authors detect approximately $90\%$ of location changes, while reducing energy consumption at the expense of higher detection delay. 

\subsubsection{Location sequences and trajectories}
A natural extension of the spatio-temporal perspective is the prediction of the location patterns and trajectories of the users.  
\revv{User mobility profiles have been introduced in~\cite{akyildiz2004predictive} to optimize call admission control, resource management and location updates. Statistical predictors are used to forecast the next cell to which a mobile phone is going to connect. The validation of the solution is done via  simulation.}
In~\cite{scellato2011nextplace}, an approach for location prediction based on nonlinear time series analysis is presented. The framework focuses on the \textit{temporal} predictability of users' location, considering their arrival and dwell time in relevant places. The evaluation is done considering four different real datasets. The authors evaluate first the predictability of the considered data and then show that the proposed nonlinear predictor outperforms both linear and Markov-based predictors. Precision approaches $70-90\% $ for medium scale prediction ($5$ minutes) and decreases to $20-40\%$ for long scale (up to $8$ hours). 

In order to improve the accuracy of time series techniques, in~\cite{de2013interdependence} the authors exploit the movement of friends, people, and, in general, entities, with  correlated mobility patterns. By means of multivariate nonlinear time series prediction techniques, they show that forecasting accuracy approaches $95\% $ for medium time scale prediction ($5$ to $10$ minutes) and is approximately $50 \%$ for $3$ hour prediction. Confidence bands show a significant improvement when prediction exploits patterns with high correlation.  Evaluation is done considering two different real datasets. 

\rev{Trajectory analysis and prediction also benefit from exploiting specific constraints such as streets, roads, traffic lights and public transportation routes. In~\cite{fazio2016pattern} the authors 
adapt the local Markovian prediction model for a specific coverage area in terms of a set of roads, moving directions, and traffic densities. When applying Markov prediction schemes, the authors consider a road compression approach to avoid dealing with a large number of locations, reduce the size of the state space, and minimize the approximation error. A more attractive candidate for trajectory prediction is the public transportation system, because of known routes and stops, and the large amount of generated mobile data traffic. In~\cite{abouzeid2015evaluating}, the authors investigate the predictability of mobility and signal variations along public transportation routes, to examine the viability of predictive content delivery. The analysis on a real dataset of a bus route, covering both urban and sub-urban areas, shows that modeling prediction uncertainty is paramount due to the high variability observed, which depends on combined effects of geographical area, time, forecasting window and contextual factors such as signal lights and bus stops.}

Moving from discrete to continuous trajectories, Kalman filtering is used to predict the future velocity and moving trends of vehicles and to improve the performance of broadcasting~\cite{yang2013broadcasting}. The main idea is that each node should send the message to be broadcast to the fastest candidate based on its neighbors' future mobility. Simulation results show modest gains, in terms of percentage of packet delivery and end-to-end delay, with respect to non-predictive methods. 

An alternative to Kalman filters is the use of regression techniques~\cite{sridaran2013location}, which analyze \ac{GPS} observations of past trips.
A systematic methodology, based on geometrical structures and data-mining techniques, is proposed to extract meaningful information for location patterns.
This work characterizes the location patterns, i.e., the set of locations visited, for several millions of users using nationwide call data records. The analysis highlights statistical properties of the typical covered area and route, such as its size, average length and spatial correlation. 

Along the same line, \cite{froehlich2008route} shows how the regularity of driver's behavior can be exploited to predict the current end-to-end route. The prediction is done by exploiting clustering techniques and is evaluated on a real dataset. 
A similar approach,  named \textit{WhereNext}, is proposed in~\cite{monreale2009wherenext}. This method predicts the next location of a moving object using past movement patterns that are based on both spatial and temporal information. The prediction is done by building a decision tree, whose nodes are the regions frequently visited. It is then used to predict the future location of a moving object. 
Results are shown using a real dataset provided by the GeoPKDD project~\cite{geopkdd}. The authors show the trade-off between the fraction of predicted trajectories and the accuracy. Both~\cite{froehlich2008route} and~\cite{monreale2009wherenext} show similar performance with an accuracy of approximately  $40 \%$  and medium time scale prediction (order of minutes). 

\subsubsection{Dealing with errors}
The  impact of estimation and prediction errors is modeled in~\cite{bui2014model}. The authors propose a comprehensive overview of several mobility predictors and associated errors and investigate the main error sources and their impact on prediction. Based on this, they propose a stochastic model to predict user throughput that accounts for uncertainty. The method is evaluated using synthetic data while  assuming that prediction's errors have a truncated Gaussian distribution. \rev{The joint analysis on the predictability of location and signal strength, which in this case is simply quantified by the standard deviation of the random variable, shown in~\cite{abouzeid2015evaluating} indicates that location-awareness is a key factor to enable accurate signal strength predictions.} Location errors are also considered in~\cite{liao2015channel} where both temporal and spatial correlation are exploited to predict the average channel gain. The proposed method combines an \ac{AR} model with functional linear regression and relies on location information. Results are derived using real data taken from the MOMENTUM project~\cite{momentum} and show that the proposed method outperforms \ac{SVM} and \ac{AR} processes.
\rev{\subsubsection{Mobility-assisted handover optimization}
Seamless mobility requires efficient resource reservation and context transfer procedures during handover, which should not be sensitive to randomness in user movement patterns. To guarantee the service continuity for mobile users, the conventional in-advance resource reservation schemes make a bandwidth reservation over all the cells that a mobile host will visit during its active connection. With mobility pattern prediction, it is possible to prepare resources in the most probable cells for the moving users. Using a Markov chain-based pattern prediction scheme, the authors in~\cite{fazio2016pattern} propose a statistical bandwidth management algorithm to handle proactive resource reservations to reduce bandwidth waste. Along similar lines, \cite{barth2011mobility, wanalertlak2011behavior} investigate mobility prediction schemes, considering not only location information but also user profiles, time-of-day, and duration characteristics, to improve the handover performance in terms of resource utilization, handover accuracy, call dropping and call blocking probabilities. 
}

\subsubsection{Geographically-assisted video optimization}

One of the main applications that has been used to show the benefits of geographic context is video streaming. A pioneer work showing the benefit of a long-term location-based scheduling for streaming is~\cite{riiser2012video}. The authors propose a system for bandwidth prediction based on geographic location and past network conditions.  Specifically, the streaming device can use a \ac{GPS}-based bandwidth-lookup service in order to predict the expected bandwidth availability and to optimally schedule the video playout. The authors present simulation as well as experimental results, where the prediction is performed for the upcoming $100$ meters. The predictive algorithm reduces the number of buffer underruns and provides  stable video quality. 

Application-layer video optimization based on prediction of user's mobility and expected capacity, is proposed also in~\cite{lu2013optimizing, abouzeid2013optimal, margolies2014exploiting}.
In~\cite{lu2013optimizing}, the authors minimize a utility function based on system utilization and rebuffering time. For the single user case they propose an online scheme based on partial knowledge, whereas the multiuser case is studied assuming complete future knowledge. In~\cite{abouzeid2013optimal}, different types of traffic are considered: full buffer, file download and buffered video. Prediction is assumed to be available and accurate over a limited time window. Three different utility functions are compared: maximization of the network throughput, maximization of the minimum user throughput, and minimization of the degradations of buffered video streams. Both works show results using synthetic data and assuming perfect prediction of the future wireless capacity variations over a time window with size ranging from tens to hundreds of seconds. 
In contrast,~\cite{margolies2014exploiting} introduces a data rate prediction mechanism that exploits mobility information and is used by an enhanced \ac{PF} scheduler. The performance gain is evaluated using a real dataset and shows a throughput increase of $15$\%-$55$\%.

Delay tolerant traffic can also benefit from offloading and prefetching as shown in~\cite{siris2013enhancing}. The authors propose methods to minimize the data transfer over a mobile network by increasing the traffic offloaded to WiFi hotspots. Three different algorithms are proposed for both delay tolerant and delay sensitive traffic. They are evaluated using empirical measurements and assuming errors in the prediction. Results show that offloaded traffic is maximized when using prediction, even when this is affected by errors. 

A \textit{geo-predictive streaming system} called GTube, is presented in~\cite{hao2014gtube}. The application obtains the user's \ac{GPS} locations and informs a server which provides the expected connection quality for future locations. The streaming parameters are adjusted accordingly. In particular, two quality adaptation algorithms are presented, where the video quality level is adapted for the upcoming 1 and $n$ steps, respectively, based on the estimated bandwidth. The system is tested using a real dataset and shows that accuracy reaches  almost $90\%$ for very short time scale prediction (few seconds), but it decreases very fast approaching zero for medium time scale prediction (few minutes). However, the proposed $n$-step algorithm improves the stability of the video quality and increases bandwidth utilization.

\subsection{Link Context}

Link context refers to the prediction of the evolution of the physical wireless channel, i.e., the channel quality and its specific parameters, so that it is possible either to take advantage of future link improvements or to counter bad conditions before they impact the system. As an example of link context, Fig.~\ref{fig:map} shows a pathloss map of the center of Berlin realized with the data of the MOMENTUM~\cite{momentum} project.


\subsubsection{Channel parameter prediction}
One possible approach to anticipate the evolution of the physical channel state is to predict the specific parameters that characterize it. In general, the variations of the physical channel can be caused by large-scale and small-scale fading. While predicting small-scale fading is quite challenging, if not impossible, several papers focuses on predicting path loss and shadowing effects. In~\cite{tie2011anticipatory}, the time-varying nonlinear wireless channel model is adopted to predict the channel quality variation anticipating distance and pathloss exponent. The performance evaluation is done using both an indoor and an outdoor testbed. The goodput obtained with the proposed bitrate control scheme can be almost doubled compared to other approaches.

Pathloss prediction in urban environments is investigated in~\cite{piacentini2010path}. The authors propose a two-step approach that combines machine learning and dimensional reduction techniques.  Specifically, they propose a new model for generating the input vector, the dimension of which is reduced by applying linear and nonlinear principal component analysis. The reduced vector is then given to a trained learning machine. The authors compare \acp{ANN} and \acp{SVM} using real measurements and conclude that slightly better results can be achieved using the \ac{ANN} regressors. 

Supporting the temporal prediction with spatial information is proposed in, e.g.,~\cite{dallanese2011channel} to study the evolution of shadow fading. The authors suggest to implement a \ac{KKF} to track the time varying shadowing using a network of \acp{CR}. The prediction is used to anticipate the position of the primary users and the expected interference and, consequently,  to maximize the transmission rate of \ac{CR} networks. Errors with the proposed model approach $2$ dB (compared to $10$ dB obtained with the pathloss based model). 
Targeting the same objective, but using a different methodology,~\cite{yin2011prediction} formulates the \ac{CR} throughput optimization problem as an \ac{MDP}. In particular, the predicted channel availability is used to maximize the throughput and to reduce the time overhead of channel sensing. 
Predictors robust to channel variations are investigated also in~\cite{tarsa2015taming}. A clustering method with supervised \ac{SVM} classification is proposed. The performance is shown for  bulk data transport via \ac{TCP} and it is also shown that the predictive approach outperforms non-predictive ones. 

\figpathlossmap


Finally, maps can be used to summarize predicted information; for instance, algorithms to build pathloss maps are proposed in~\cite{kasparick2015kernel}. In this paper, the authors propose two kernel-based adaptive algorithms, namely  the adaptive projected subgradient method and the multikernel approach with adaptive model selection. Numerical evaluation is done for both a urban scenario and a campus network scenario, using real measurements. 
The performance of the algorithms is evaluated assuming perfect knowledge of the users' trajectories.

\subsubsection{Combined channel and mobility context}
Channel quality and mobility information are jointly predicted in~\cite{nicholson2008breadcrumbs}. The authors combine information on visited locations and corresponding achieved link quality to provide \textit{connectivity forecast}. A Markov model is implemented in order to forecast future channel conditions.
Location prediction accuracy is approximately $70\%$ for a prediction window of $20$ seconds. However, the location information has quite a  coarse granularity (of about $100$ m). In terms of bandwidth, the proposed model, evaluated on a real dataset, shows an accuracy within $10$ KB/s for over $50\%$  of the evaluation period, and within $50$ KB/s for over $80\%$  of the time. 
In~\cite{naimi2014anticipation}, prediction is employed to adjust the routing metrics in ad hoc wireless networks. In particular, the metrics considered in the paper are the average number of retransmissions needed and the time expected to transmit a data packet. The solution anticipates the future signal strength using linear regression on the history of the link quality measurements. Simulations show that the packet delivery ratio is close to $100\%$, even though it drops to $20\%$ using classical methods.

When the information used to drive the prediction is affected by errors, it is important to account for the magnitude of the error. This has been considered, for instance, in~\cite{muppirisetty2015spatial} \rev{and~\cite{muppirisetty2016channel}}, where the impact of location uncertainties is taken into account. Namely, the authors of~\cite{muppirisetty2015spatial} show that classical \ac{GP} wrongly predicts the channel gain in presence of errors, while uncertain \ac{GP}, which explicitly accounts for location uncertainty, outperforms the former in both learning and predicting the received power. Gains are shown also for a simple proactive resource allocation scenario. 
\rev{Similarly, the same authors propose in~\cite{muppirisetty2015proactive} a proactive scheduling mechanism that exploits the statistical properties of user demand and channel conditions. Furthermore, the model captures the impact of prediction uncertainties and assesses the optimal gain obtained by the proactive resource scheduler. The authors also propose an asymptotically optimal policy that attains the optimal gain rapidly as the prediction window size increases.}
Uncertainties are also dealt with in~\cite{bui2015mobile}, where a resource allocation algorithm for mobile networks that leverages link quality prediction is proposed. Time series filtering techniques (\ac{ARMA}) are used to predict near term link quality, whereas medium to long term prediction is based on statistical models. The authors propose a resource allocation optimization framework under imperfect prediction of future available capacity. Simulations are done using a real dataset and show that the proposed solution outperforms the limited horizon optimizer  (i.e., when the prediction is done only for the upcoming few seconds) by $10-15\%$. Resource allocation is also addressed in~\cite{margolies2014exploiting}, which extends the standard \ac{PF} scheduler of 4G networks to account for data rate prediction obtained through adaptive radio maps.

\subsubsection{Channel-assisted video optimization}
In~\cite{wang2013ames}, the authors propose an adaptive mobile video streaming framework, which stores video in the cloud and offers to each user a continuous video streaming adapted to the fluctuations of the link quality. The paper proposes a mechanism to predict the potential available bandwidth in the next time window (of a duration of a few seconds) based on the measurements of the link quality done in the previous time window. A prototype implementation of the proposed framework is used to evaluate the performance. This shows that the prediction has a relative error of about $10\%$ for very short time windows (a couple of seconds) but becomes relatively poor for larger time windows. The video performance is evaluated in terms of ``click-to-play'' delay, which is halved with the proposed approach.
A Markov model is used in~\cite{bao2015bitrate}, where information on both channel and buffer states is combined to optimize mobile video streaming. Both an optimal policy as well as a fast heuristic are proposed. A drive test was conducted to evaluate the performance of the proposed solution. In particular, the authors show the proportional dependency between utility and buffer size, as well as the complexity of the two algorithms. Furthermore, a Markov model is adopted to represent different user's achievable rates~\cite{seetharam2015managing} and channel states~\cite{hosseini2015not}. The transition matrix is derived empirically to minimize the number of video stalls and their duration over a $10$-second horizon. 

\rev{Video calls are considered in~\cite{kurdoglu2016realtime}. Namely, a cross-layer design for proactive congestion control, named Rebera, is proposed. The system measures the real-time available bandwidth and uses a linear adaptive filter to estimate the future capacity. Furthermore, it ensures that the video sending rate never exceeds the predicted values, thereby preventing self-congestion and reducing delays. Performance results with respect to today's solutions are given for both a testbed and a real cellular network. 
In~\cite{liu2016hop}, the authors propose a hop-by-hop video quality adaptation scheme at the router level  to improve the performance of adaptive video streaming in~\acp{CCN}. In this context, the routers monitor network conditions by estimating the end-to-end bandwidth and proactively decrease the video quality when network congestion occurs. Performance is evaluated considering a realistic large-scale network topology and it is shown that the proposed solution outperforms state of the art schemes in terms of both  playback quality and average delay.}

\rev{\subsubsection{Video optimization under uncertainty}
For the video optimization use case, some works also assess the impact of uncertain predictions. 
In~\cite{blobel2015anticipatory}, the authors propose a stochastic model of prediction errors, based on~\cite{bui2014model}, and introduce an online scheduler that is aware of prediction errors. Namely, based on the expected prediction accuracy, the algorithm determines whether to consider or discard the predicted data rate. A similar model for prediction errors is introduced in~\cite{tsilimantos2016anticipatory}. In this case, a \ac{LP}  formulation is proposed to trade off spectral efficiency and stalling time. The proposed solution  shows good gains with respect to the case without prediction, even when errors occur. \ac{LP} is used also in~\cite{atawia2014robust} to minimize the base station airtime with the constraint of no video interruption. In this case, uncertainties are modeled  by using a fuzzy approach. Furthermore, in order to keep track of the previous values of the error, a Kalman filter is used. Simulations are run using synthetic data and show the effect of channel variability on video degradation and average airtime. 
In~\cite{mangla2016video},  bandwidth prediction is exploited to increase the quality of video streaming. Both perfect and uncertain prediction are considered and a robust heuristic is proposed to mitigate the effect of prediction errors when adapting the video bitrate. 
In~\cite{atawia2015chance, atawia2016joint}, a predictive resource allocation robust to rate uncertainties is proposed. The authors propose a framework that provides quality guarantees with the objective of minimizing energy consumption. Both optimal gradient-based and real-time guided heuristic solutions are presented. In~\cite{atawia2015chance} both Gaussian and Bernstein approximation are used to model rate uncertainties, whereas~\cite{atawia2016joint} considers only the former one. \revv{Similarly, \cite{hossain2004link}~provides predictive \ac{QoS} over wireless \ac{ATM} networks: given the TDMA nature of these networks, these schemes optimize the number of allocated time slots depending on the characteristics of the traffic stream and the wireless link.}} 

\subsubsection{Efficiency bounds and approximations for multimedia streaming applications}
A few papers~(\cite{abouzeid2014energy, abouzeid2014efficient, bui2015anticipatory, bui2015anticipatoryb, draxler2013cross, draxler2015smarterphones, valentin2014anticipatory, zou2015can}) investigate resource allocation optimization assuming that the future channel state is perfectly known. While addressing different objectives, these papers share similar methods: they first devise a problem formulation from which an optimal solution can be obtained (using standard optimization techniques), then they propose sub-optimal approaches and on-line algorithms to obtain an approximation of the optimal solution. Furthermore, all these papers leverage a buffer to counteract the randomness of the channel. For instance, in case a given amount of information has to be gathered within a deadline, the buffer allows the system to optimize (for a given objective function) the resource allocation while meeting the deadline.

In this regard, energy-efficiency is the primary objective in~\cite{abouzeid2014energy, abouzeid2014efficient}, which is optimized by allowing the network base stations to be switched off once the users' streaming requirements have been satisfied. Simulations show that an energy saving up to $80 \%$ with respect to the baseline approach can be achieved and that the performance of the heuristic solution is quite close to the optimal (but impractical) \ac{MILP} approach. 
Buffer size is investigated in~\cite{valentin2014anticipatory}, where the author introduces a linear formulation that minimizes the amount for resources assigned to non-real time video streaming with constraints on the user's playout buffer. Results are shown for a scenario with both video and best effort users and highlight the gain in terms of required resources to serve the video users as well as data rate for the best effort users. 

The trade-off between streaming interruption time and average quality is investigated in~\cite{draxler2013cross, draxler2015smarterphones} by devising a mixed-integer quadratically constrained problem which computes the optimal download time and quality for video segments. Then, the authors propose a set of heuristics tailored to greedily optimize segment scheduling according to a specific objective function, e.g., maximum quality, minimum streaming interruption, or fairness.
Similar objectives are tackled in~\cite{bui2015anticipatory, bui2015anticipatoryb} in a lexicographic approach, so that streaming continuity is always prioritized over quality. They first propose a heuristic for the lateness-quality problem that performs almost as good as the \ac{MILP} formulation. Then, they extend the \ac{MILP} formulation to include \ac{QoS} guarantees and they introduce an iterative approximation based on a simpler \ac{LP} formulation. 
A further heuristic approach is devised in~\cite{zou2015can} and accounts for the buffer and channel state prediction. The proposed approach maximizes the streaming quality while guaranteeing that there are no interruptions. 
\rev{
\subsubsection{Cognitive radio maps}
\acp{CR} are context-aware wireless devices that adapt their functionalities to changes in the environment. They have been recently used~\cite{xing2013spectrum, wei2013construction, yilmaz2013radio} to obtained the so-called \ac{REM}: a multi-dimensional database containing a wide set of information ranging from regulations to spectrum usage. \newline\indent
For instance, \ac{REM} are used to predict spectrum availability in \ac{CR}~\cite{xing2013spectrum}: the paper exploits cognitive maps to provide contextual information for predictive machine learning approaches such as \ac{HMM}, \ac{ANN} and regression techniques. The construction of these maps is discussed in~\cite{wei2013construction} and the references therein, while their use as enabler for \ac{CR} networks is analyzed in~\cite{yilmaz2013radio}. \newline\indent
In the context of anticipatory networking, \acp{REM} are often used as a source of contextual information for the actual prediction technique adopted, rather than as prediction tools themselves.
\revv{
\cite{hoyhtya2016spectrum, chen2016survey}~present two surveys of methodologies and measurement campaigns of spectrum occupancy. In particular, \cite{hoyhtya2016spectrum} proposes a conservative approach to account for measurement uncertainty, while~\cite{chen2016survey} exploits predictors to provide the future channel status. In addition, prediction through machine learning approaches is addressed in~\cite{thilina2013machine}, where different techniques are compared to assess future channel availability. \newline\indent
Imperfect measurements are dealt with in~\cite{khan2016opportunistic}, which models the problem as a repeated game and maximizes the total network payoff. 
However, in cognitive networks, the channel status depends on the activity of primary users. \cite{saleem2014primary}~surveys the models proposed so far to describe primary users activity and that can be used to drive prediction in this area. Once the activity of primary users is available or predicted, it is possible to control the activity of secondary users in order to guarantee the agreed \ac{QoS} to the former~\cite{monemi2015characterizing, monemi2016characterization}. These papers compute the feasible cognitive interference region in order to allow secondary users' communication respecting primary users' rights. The utilization of spectrum opportunity describes the probability of a secondary user to exploit a free communication slot~\cite{ozger2016utilization}. \newline\indent
A similar form of opportunistic spectrum usage goes under the name of white space~\cite{akhtar2016white}: i.e., channels that are unused at specific location and time. \ac{CR}s can take advantage of these frequencies thanks to dynamic spectrum access. Finally, \cite{khan2016cognitive} describes how to exploit \ac{CR} to realize a complete smart grid scenario; \cite{bukhari2016survey} describes how to exploit channel bonding to increase the bandwidth and decrease the delay of \ac{CR}.
}
} 



\subsection{Traffic Context}
This section overviews some of the approaches that focus on traffic and throughput prediction. Although related to the previous context, the papers discussed in this section leverage information collected from higher layers of the protocol stack. For instance, solutions falling in this category try to predict, among other parameters, the number of active users in the network and the amount of traffic they are going to produce. Similarly, but from the perspective of a single user, the prediction can target the data rate that a streaming application is going to achieve in the near term.

We grouped these papers in three main classes: pure analysis of mobile traffic; traffic prediction for networking optimization; and direct throughput prediction. 

\subsubsection{Traffic analysis and characterization}
\label{subsub:traffic-analysis}
The analysis of mobile traffic is fundamental for long-term network optimization and re-configuration. To this end, several pieces of work have addressed such research topics in the recent past. 

The work in~\cite{paul2011understanding} targets the creation of regressors for different performance indicators at different spatio-temporal granularity for mobile cellular networks. Namely, the authors focus on the characterization of  per-device throughput,  base station throughput and device mobility. A one-week nation-wide cellular network dataset is collected through proprietary traffic inspection tools placed in the operator network and are used to characterize the per-user traffic, cell-aggregate traffic and to perform further spatio-temporal correlation analysis. 

A similar scope is addressed by~\cite{shafiq2011characterizing} which, on the other hand, focuses more on core network measurements. Flow level mobile device traffic data are collected from a cellular operator's core network and are used to characterize the IP traffic patterns of mobile cellular devices. 

More recently, the authors of~\cite{sayeed2015cloud} studied traffic prediction in cloud analytics and prove that optimizing the choice of metrics and parameters can lead to accurate prediction even under high latency. This prediction is exploited at the application/\ac{TCP} layer to improve the performance of the application avoiding buffer overflows and/or congestion.

\subsubsection{Traffic prediction}
\label{subsub:traffic-prediction}

Several applications can benefit from the prediction of traffic performance features. For instance, a predictive framework that anticipates the arrival of upcoming requests is used in~\cite{tadrous2013proactive} to prefetch the needed content at the mobile terminal. 
%
The authors propose a theoretical framework to assess how the outage probability scales with the prediction horizon. 
The theoretical framework accounts for prediction errors and multicast delivery. 
Along the same line, queue modeling~\cite{huang2014backpressure} and analysis~\cite{abedini2014content} is used to predict the upcoming workloads in a lookahead time window. 
Leveraging the workload prediction, a multi-slot joint power control and scheduling problem is formulated to find the optimal assignment that minimizes the total cost~\cite{huang2014backpressure} or maximizes the \ac{QoS}~\cite{abedini2014content}.

Multimedia optimization is the focus in~\cite{xu2013proteus}. By predicting throughput, packet loss and transmission delay half a second in advance, the authors propose to dynamically adjust application-level parameters of the reference video streaming or video conferencing services including the compression ratio of the video codec, the forward error correction code rate and the size of the de-jittering buffer. 
Traffic prediction is also addressed in~\cite{samulevicius2015most}, where the authors propose to use a database of events (concerts, gatherings, etc.) to improve the quality of the traffic prediction in case of unexpected traffic patterns and in~\cite{lee2013generalized}, where a general predictive control framework along with Kalman filter is proposed to counteract the impact of network delay and packet loss. 
The objective of~\cite{sekar2013developing} is to build a model for user engagement as a function of performance metrics in the context of video streaming services. 
The authors use a supervised learning approach based on average bitrate, join time, buffering ratio and buffering to estimate the user engagement.
Finally, inter-download time can be modeled~\cite{beister2014predicting} and subsequently predicted for quality optimization.

\rev{The work in~\cite{pollakis2016anticipatory} targets energy-efficient resource scheduling in mobile radio networks. The authors introduce a \ac{MNLP} which returns on a slot basis the  optimal allocation of resources to users and the optimal users-cell association pattern. The proposed model leverages optimal traffic predictors to obtain the expected traffic conditions in the following slots.  
Radio resource allocation in mobile radio networks is addressed also in~\cite{yu2014predictive} and later by the same authors in~\cite{yu2016power}; the target is to design a predictive framework to optimally orchestrate the resource allocation and network selection in case one operator owns multiple access networks. The predictive framework  aims at minimizing the expected time average power consumption while keeping the network (user queues) stable. 
The core contribution of~\cite{du2016traffic, du2016resource} is the use of deep learning techniques to predict the upcoming video traffic sessions; the prediction outcome is then used to proactively allocate the resources of video servers to these future traffic demands.
}

\subsubsection{Throughput prediction}
\label{subsub:throughput-prediction}
Rather than predicting the expected traffic or optimizing the network based on traffic prediction, the work in this section targets the prediction/optimization based on the expected throughput. A common characteristic of the work described here is that the spatio-temporal correlation is exploited in the prediction phase of the expected throughput. 

Quite a few early works studied how to effectively predict the obtainable data rate. In particular, long term prediction~\cite{papagiannaki2003long} with 12-hour granularity allows to estimate aggregate demands up to 6 months in advance. Shorter and variable time scales are studied in~\cite{sadek2004multi, zhou2005network} adopting \ac{ARIMA} and \ac{GARCH} techniques.

In~\cite{abouzeid2013predictive}, the authors propose a dynamic framework to allocate downlink radio resources across multiple cells of 4G systems. The proposed framework leverages context information of three types: radio maps, user's location and mobility, as well as application-related information. The authors assume that a forecast of this information is available and can be used to optimize the resource allocation in the network. The performance of the proposed solution is evaluated through simulation for the specific use case of video streaming. 
Geo-localized radio maps are also exploited in~\cite{yao2012improving}. Here the optimization is performed at the application layer by letting adaptive video streaming clients and servers dynamically change the streaming rate on the basis of the current bandwidth prediction from the bandwidth maps. 
The empirical collection of geo-localized data rate measures is also addressed in~\cite{riiser2013commute} which introduces a dataset of adaptive \ac{HTTP} sessions performed by mobile users. 

\rev{The work in~\cite{millan2015tracking} considers the problem of predicting end-to-end quality of multi-hop paths in community WiFi networks. The end-to-end quality is measured by a linear combination of the expected transmission count across all the links composing the multi-hop path. The authors resort to a real data set of a WiFi community network and test several predictors for the end-to-end quality. }

\rev{The anticipation of the upcoming throughput values is often applied to the optimization of adaptive video streaming services. In this context, Yin \emph{et al.}~\cite{yin2015control} leverage throughput prediction to optimally adapt the bit rate of video encoders; here, prediction is based on the harmonic mean of the last $k$ throughput samples.}

\rev{In~\cite{yi2016cs2p, jiang2016cfa} the authors build on the conjecture that video sessions sharing the same critical features have similar \ac{QoE} (e.g., re-buffering, startup latency, etc.). Consequently, first clustering techniques are applied to group similar video sessions, and then throughput predictors based on \acp{HMM} are applied to each cluster to dynamically adapt the bit rate of the video encoder to the predicted throughput samples.}

\rev{The work in~\cite{zahran2016oscar} resorts to a model-based throughput predictor in which the throughput of a \ac{DASH}-based video streaming service is assumed to be a random variable with Beta-like distribution whose parameters are empirically estimated within an observation time window. Building on this estimate, the authors propose a \ac{MNLP} with a concave objective
function and linear constraints. The program is implemented as a multiple choice knapsack problem and  solved using commercial solvers.  Along the same lines, the optimization of a \ac{DASH}-based video streaming service is addressed in~\cite{wang2016squad}, where the authors propose an adaptive video streaming framework based on a smoothed rate estimate for the video sessions.}

\rev{The work in~\cite{miller2015control} considers the scenario where a small cell is used to deliver video content to a highly dense set of users. The video delivery can also be supported in a distributed way by end-user devices storing content locally. A control-theoretic framework is proposed to dynamically set the video quality of the downloaded content while enforcing stability of the system.}

\subsection{Social Context}

The work on anticipatory networking leveraging social context exploits \emph{ex ante} or \emph{ex post} information on social-type relationships between agents in the networking environment. Such information may include: the network of social ties and connections, the user's preference on contents, measures on user's centrality in a social network, and measures on users' mobility habits. The aforementioned context information is leveraged in three main application scenarios: caching at the edge of mobile networks, mobility prediction, and downlink resource allocation in mobile networks. 

\subsubsection{Social-assisted caching}
Motivated by the need of limiting the load in the backhaul of 5G networks, references~\cite{bastug2013proactive, bastug2014living, bastug2014anticipatory} propose two schemes to proactively move contents closer to the end users. In~\cite{bastug2013proactive}, caching happens at the small cells, whereas in~\cite{bastug2014living, bastug2014anticipatory} contents can be proactively downloaded by a subset of end users which then re-distribute them via \ac{D2D} communication. The authors first define two optimization problems which target the load reduction in the backhaul (caching at small cells) and in the small cell (caching at end users), respectively, then heuristic algorithms based on machine learning tools are proposed to obtain sub-optimal solutions in reasonable processing time. The heuristic first collects users' content rating/preferences to predict the popularity matrix $\mat{P}_m$. Then, content is placed at each small cell in a greedy way starting from the most popular ones until a storage budget is hit. 
The first algorithmic step of caching at the end users is to identify the $K$ most connected users and to cluster the remaining ones in communities. Then it is possible to characterize the content preference distributions within each community and greedily place contents at the cluster heads.  In~\cite{bastug2014anticipatory}, the prediction leverages additional information on the underlying structure of content popularity within the communities of users. 
\rev{Joint mobility and popularity prediction for content caching at small cell base stations is studied in~\cite{siris2016exploiting}. Here, the authors propose a heuristic caching scheme that determines whether a particular content item should be cached at a particular base station by jointly predicting the mobility pattern of users that request that item as well as its popularity, where popularity prediction is performed using the inter-arrival times of consecutive requests for that object. They conclude that the joint scheme outperforms caching with only mobility and only popularity models.}

A similar problem is addressed in~\cite{golrezaei2012femtocaching}: the authors consider a distributed network of femto base stations, which can be leveraged to cache videos. The authors study where to cache videos such that the average sum delay across all the end users is minimized for a given video content popularity distribution, a given storage capacity and an arbitrary model for the wireless link.  A greedy heuristic is then proposed to reduce the computational complexity. 

\rev{In~\cite{tadrous2015optimal,tadrous2015joint}, it is argued that proactive caching of delay intolerant content based on user preferences is subject to prediction uncertainties that affect the performance of any caching scheme. In~\cite{tadrous2015optimal}, these uncertainties are modeled as probability distributions of content requests over a given time period. The authors provide lower bounds on the content delivery cost given that the probability distribution for the requests is available. They also derive caching policies that achieve this lower bound asymptotically. It is shown that under uniform uncertainty, the proposed policy breaks down to equally spreading the amount of predicted content data over the horizon of the prediction window. Another approach to solve the same problem is used in~\cite{tadrous2015joint}, where personalized content pricing schemes are deployed by the service provider based on user preferences in order to enhance the certainty about future demand. The authors model the pricing problem as an optimization problem. Due to the non-convex nature of their model, they use an iterative sub-optimal solution that separates price allocation and proactive download decisions.}
\rev{
\subsubsection{Social-assisted matching game theory} 
Matching game theory~\cite{gu2015matching} can be used to allocate networks resources between users and base stations, when social attributes are used to profile users. For instance, by letting users and base stations rank one another to capture users' similarities in terms of interests, activities and interactions, it is possible to create social utility functions controlling a distributed matching game. In~\cite{semiari2015context}, a self-organizing, context-aware framework for \ac{D2D} resource allocation is proposed that exploits the likelihood of strongly connected users to request similar contents. The solution is shown to be computationally feasible and to offer substantial benefits when users' social similarities are present. A similar approach is used in~\cite{semiari2016context} to deal with joint millimeter and micro wave dual base station resource allocation, in~\cite{namvar2014context} for user base station association in small cell networks, and in~\cite{zhang2015social} to optimize \ac{D2D} offloading techniques.
Caching in small cell networks can also be addressed as a many-to-many matching game~\cite{hamidouche2014many}: by matching video popularity among users most frequently served by a given server it is possible to devise caching policies that minimize end-users' delays. Simulations show the approach is effective in small cell networks.
}

\subsubsection{Social-assisted mobility prediction}

Motivated by the need to reduce the active scanning overhead in IEEE 802.11 networks, the authors of~\cite{wanalertlak2011behavior} propose a mobility prediction tool to anticipate the next access point a WiFi user is moving to. The proposed solution is based on context information on the handoffs which were performed in the past; specifically, the system stores centrally a time varying handoff table which is then fed into an ARIMA predictor which returns the likelihood of a given user to handoff to a specific access point. The quality of the predictor is measured in terms of signaling reduction due to active scanning.  

The prediction of user mobility is also addressed in~\cite{noulas2012mining}. The authors leverage information coming from the social platform Foursquare to predict user mobility on coarse granularity. The \emph{next check-in problem} is formulated to determine the next place in an urban environment which will be most likely visited by a user. The authors build a time-stamped dataset of  ``check-ins'' performed by Foursquare users over a period of one month across several venues worldwide. A set of features is then defined to represent user mobility including user mobility features (e.g., number of historical visits to specific venues or categories of venues, number of historical visits that friends have done to specific venues), global mobility features (e.g., popularity of venues, distance between venues, transition frequency between couples of venues), and temporal features which measures the historical check-ins over specific time periods.  Such a feature set is then used to train a supervised classification problem to predict the next check-in venue. Linear regression and M5 decision trees are used in this regard. The work is mostly speculative and does not address directly any specific application/use of the proposed mobility prediction tool.  

Along the same lines, the mobility of users in urban environments is characterized in~\cite{calabrese2010human}. Different from the previous work which only exploits social information, the authors also leverage physical information about the current position of moving users. A probabilistic model of the mobile users' behavior is built and trained on a real life dataset of user mobility traces. A social-assisted mobility prediction model is proposed in~\cite{bapierre2011variable}, where a variable-order Markov model is developed and trained on both temporal features (i.e., when users were at specific locations) and social ones (i.e., when friends of specific users were at a given location). The accuracy of the proposed model is cross-validated on two user-mobility datasets. 

\classtable

\subsubsection{Social-assisted radio resource allocation}
The optimization of elastic traffic in the downlink of mobile radio networks is addressed in~\cite{proebster2012context, proebster2011context}. The key tenet is to provide to the downlink scheduler ``richer'' context to make better decisions in the allocation of the radio resources. Besides classical network-side context including the cell load and the current channel quality indicator which are widely used in the literature to steer the scheduling, the authors propose to include user-side features which generically capture the satisfaction degree of the user for the reference application. Namely, the authors introduce the concept of a \emph{transaction}, which represents the atomic data download requested by the end user (e.g., a web page download via \ac{HTTP}, an object download via \ac{HTTP} or a file download via \ac{FTP}). For each transaction and for each application, a utility function is defined capturing the user's sensitivity with respect to the transmission delay and the expected completion time. The functional form of this utility function depends on the type of application which ``generated'' the transaction; as an example, the authors make the distinction between transactions from applications which are running in the foreground and the background on the user's terminal.   For the sake of presentation, a parametric logistic function is  used to represent the aforementioned utility. The authors then formulate an optimization problem to maximize the sum utility across all the users and transactions in a given mobile radio cell and design a greedy heuristic to obtain a sub-optimal solution in reasonable computing time. The proposed algorithm is validated  against state-of-the-art scheduling solutions (\ac{PF} / weighted \ac{PF} scheduling) through simulation on synthetic data mimicking realistic user distributions, mobility patterns and traffic patterns.  

\rev{In order to predict the spatial traffic of base stations in a cellular network, \cite{yi2016spatial} applies the idea of social networks to base stations. Here, the base stations themselves create a social network and a social graph is created between them based on the spatial correlation of the traffic of each of them. The correlation is calculated using the Pearson coefficient. Based on the topology of the social graph, the most important base stations are identified and used for traffic prediction of the entire network, which is done using \ac{SVM}. The authors conclude that with the traffic data of less than 10\% of the base stations, effective prediction with less than 20\% mean error can be achieved.}

Social-oriented techniques related to the popularity of the end users are leveraged also in~\cite{tsiropoulos2011impact} where the authors target the performance optimization of downlink resource allocation in future generation networks. The utility maximization problem is formulated with the utility being a combination (product) of a network-oriented term (available bandwidth) and a social-oriented term (social distance). The social-oriented term is defined to be the degree centrality measure~\cite{jackson2008social} for a specific user. The proposed problem is sub-optimally solved through a heuristic which is finally validated using synthetic data.

\subsection{Summary}
\label{sec:chal:cont}

Hereafter, we summarize the main takeaways of the section in terms of application and objective for which different context types can be used. Table~\ref{tab:prediction_class} provides a synthesis of the main considerations: each context is associated with its typical applications, prediction methodologies (ordered by decreasing popularity), optimization approaches and general remarks.

\subsubsection{Mobility prediction} 
It has been shown that predictability of user mobility can be potentially very high \rev{(93\% potential predictability in user mobility as stated in~\cite{song2010limits}), despite the significant differences in the travel patterns}. As a matter of fact, many papers study how to forecast users' mobility by means of a variety of techniques. \rev{For predicting trajectories, characterized by sequences of discretized locations indicated by cell \acp{ID} or road segments, fixed-order Markov models or variable-order Markov models are the most promising tools, while for continuous trajectories, regression techniques are widely used. To enhance the prediction accuracy,} the most popular ones leverage geographic information: \ac{GPS} data, cell records and received signal strength are used to obtain precise and frequent data sampling to locate users on a map. However, the movements of an individual are largely influenced by those of other individuals via social relations. Several papers analyze social information and location check-ins to find recurrent patterns. For this second case usually a sparser dataset is available and may limit the accuracy of the prediction. 

\subsubsection{Network efficiency} Predicting and optimizing network efficiency (i.e., increasing the performance of the network while using the same amount of resources) is the most frequent objective in anticipatory networking. We found papers exploiting all four types of context to achieve this. As such, objectives and constraints cover the whole attribute space. Improving network efficiency is likely to become the main driver for including anticipatory networking solutions in next generation networks.

\subsubsection{Multimedia streaming} The main source of data traffic in 4G networks has been multimedia streaming and, in particular, video on demand. 5G networks are expected to continue and even increase this trend. As a consequence, several anticipatory networking solutions focus on the optimization of this service. All the context types have been used to this extent and each has a different merit: social information is needed to predict when a given user is going to request a given content, combined geographic and social information allows the network to cache that content closer to where it will be required and physical channel information can be used to optimize the resource assignment.

\subsubsection{Network offloading} Mobility prediction can be used to handover communications between different technologies to decrease network congestion, improve user experience, reduce users' costs and increase energy efficiency.

\subsubsection{Cognitive networking} Physical channel prediction can be exploited for cognitive networking and for network mapping. The former application allows secondary users to access a shared medium when primary subscribers left resource unused, thus, predicting when this is going to happen will highly improve the effectiveness of the solution. The latter, instead, exploits link information to build networking maps that can provide other applications with an estimate of communication quality at a given time and place.

\subsubsection{Throughput- and traffic-based applications} Traffic information is usually studied to be, first, modeled and, subsequently, predicted. Traffic models and predictors are then used to improve networking efficiency by means of resource allocation, traffic shaping and network planning.

\section{Prediction Methodologies for Anticipatory Networking}
\label{sec:prediction}

In this section, we present some selected prediction methods for the types of context introduced in Section~\ref{sec:guidelines}. 
The selected methods are classified into four main categories: {\it time series methods}, {\it similarity-based classification}, {\it regression analysis}, and {\it statistical methods for probabilistic modeling}. Their mathematical principles and the application to inferring and predicting the aforementioned contextual information are introduced in Sections~\ref{subsec:Prediction_TimeSeries}, \ref{subsec:Classification}, \ref{subsec:Regression}, and \ref{subsec:Probabilistic}, respectively.

The goal of the prediction handbook is to show \textit{which methods work in which situation}. 
In fact, selecting the appropriate prediction method requires to analyze the prediction variables and the model constraints with respect to the application scenario (see Section \ref{sec:guidelines}). 
This section concludes with a series of takeaways that summarize some general principles for selection of prediction methods based on the scenario analysis.

\subsection{Time Series Predictive Modeling}\label{subsec:Prediction_TimeSeries}
A time series is a set of time-stamped data entries which allows a natural association of data collected on a regular or irregular time basis. 
In wireless networks, large volumes of data are stored as time series and frequently show temporal correlation.
For example, the trajectory of the mobile device can be characterized by successive time-stamped locations obtained from geographical measurements;
 individual social behavior can be expressed through time-evolving events;
 traffic loads modeled in time series can be leveraged for network planning and controlling. 
Fig.~\ref{fig:TS_CellLoad} and \ref{fig:TS_AggrLoad} illustrate two time series of per-cell and per-city aggregated uplink and downlink data traffic, where temporal correlation is clearly recognizable.

In the following, we introduce the two most widely used time series models based on linear dynamic systems: 1) \acf{ARMA}, and 2) Kalman filters. Examples of context prediction in wireless networks are given and their extensions to nonlinear systems are briefly discussed.

\figtimeseries

\subsubsection{Autoregressive and moving average models}\label{subsubsec:TS_Stationary_Linear}

Consider a univariate time series $\{X_t: t\in\set{T}\}$, where $\set{T}$ denotes the set of time indices. The general \ac{ARMA} model, denoted by $\arma(p,q)$, has $p$ \ac{AR} terms and $q$ \ac{MA} terms, given by
\begin{equation}
X_t = Z_t + \sum_{i = 1}^p\phi_i X_{t-i} + \sum_{j=1}^q \theta_j Z_{t-j}
\label{eqn:TS_ARMA}
\end{equation}
where $Z_t$ is the process of the white noise errors, and $\{\phi_i\}_{i=1}^p$ and $\{\theta_j\}_{j=1}^q$ are the parameters. The \ac{ARMA} model is a generalization of the simpler \ac{AR} and \ac{MA} models that can be obtained for $q = 0$ and $p=0$ respectively.
Using the {\it lag operator} $L^i X_t := X_{t-i}$ the model becomes
\begin{equation}
\phi(L)X_t = \theta(L) Z_t
\label{eqn:TS_ARMA_Lag}
\end{equation}
where $\phi(L):=1-\sum_{i=1}^p\phi_i L^i$ and $\theta(L):=1 + \sum_{j=1}^q \theta_jL^j$.

The fitting procedure of such processes assumes {\it stationarity}.
However, this property is seldom verified in practice and {\it non-stationary} time series need to be stationarized through differencing and logging. The \ac{ARIMA} model generalizes \ac{ARMA} models for the case of non-stationary time series: a non seasonal ARIMA model $\arima(p,d,q)$ after $d$ differentiations reduces to an $\arma(p,q)$ of the form
\begin{equation}
\phi(L)\Delta^d X_t = \theta(L) Z_t,
\label{eqn:ARIMA}
\end{equation}
where $\Delta^d = (1-L)^d$ denotes the $d$th difference operator.

Numerous studies have been done on prediction of traffic load in wireless or IP backbone networks using autoregressive models. The stationarity analysis often provides important clues for selecting the appropriate model.  For instance, in~\cite{papagiannaki2003long} a low-order ARIMA model is applied to capture the non-stationary short memory process of traffic load, while in \cite{sadek2004multi}  a Gegenbauer ARMA model is used to specify long memory processes under the assumption of stationarity. 
Similar models are applied to mobility- or channel-related contexts. In~\cite{wanalertlak2011behavior}, an exponential weighted moving average, equivalent to $\arima(0,1,1)$, is used to forecast handoffs. In \cite{tie2011anticipatory, jiang2013tracking}, \ac{AR} models are applied to predict future signal-to-noise ratio values and user positions, respectively.
If the variance of the data varies with time, as in~\cite{zhou2005network} for data traffic, and can be expressed using an \ac{ARMA}, then the whole model is referred to as \ac{GARCH}.

\subsubsection{Kalman filter}\label{subsubsec:Kalman}
Kalman filters are widely applied in time series analysis for linear dynamic systems, which track the estimated system state and its uncertainty variance. 
In the anticipatory networking literature, Kalman filters have been mainly adopted to model the linear dependence of the system states based on historical data.

Consider a multivariate time series $\{\ve{x}_t\in\field{R}^n: t\in\set{T}\}$, the Kalman filter addresses the problem of estimating state $\ve{x}_t$ that is governed by the linear stochastic difference equation
\begin{equation}
\ve{x}_t=\mat{A}_t\ve{x}_{t-1} + \mat{B}_t\ve{u}_{t}+\ve{w}_{t}, \ t = 0,1,\ldots,
\label{eqn:Kalman_state}
\end{equation}
where $\mat{A}_t\in\field{R}^{n\times n}$ expresses the state transition, and $\mat{B}_t \in\field{R}^{n\times l}$ relates the optional control input $\ve{u}_t\in\field{R}^l$ to the state $\ve{x}_t\in\field{R}^n$. The random variable $\ve{w}_t\sim \set{N}(\ve{0}, \mat{Q}_t)$ represents a multivariate normal noise process with covariance matrix $\mat{Q}_t\in\field{R}^{n\times n}$. 
The observation $\ve{z}_t\in\field{R}^m$ of the true state $\ve{x}_t$ is given by
\begin{equation}
\ve{z}_{t}=\mat{H}_t\ve{x}_t + \ve{v}_t,
\label{eqn:Kalman_measure}
\end{equation}
where $\mat{H}_t\in\field{R}^{m \times n}$ maps the true state
space into the observed space.
The random variable $\ve{v}_t$ is the observation noise process following
$\ve{v}_t\sim \set{N}(\ve{0}, \mat{R}_t)$ with covariance $\mat{R}_t\in\field{R}^{n\times n}$.
Kalman filters iterate between 1) predicting the system state with Eq. (\ref{eqn:Kalman_state}) and 2) updating the model according to Eq. (\ref{eqn:Kalman_measure}) to refine the previous prediction. The interested reader is referred to \cite{harvey1990forecasting} for more details. 

In~\cite{zaidi2005real, yang2013broadcasting}, Kalman filters are used to study users' mobility.
Wireless channel gains are studied in~\cite{dallanese2011channel} with \ac{KKF}, while the authors of~\cite{okutani1984dynamic} adopt the technique to predict short-term traffic volume.
The extended Kalman filter adapts the standard model to nonlinear systems via online Taylor expansion. According to~\cite{pappas2014extended}, this improves shadow/fading estimation.

\subsection{Similarity-based Classification}\label{subsec:Classification}

Similarity-based classification aims to find inherent structures within a dataset. 
The core rationale is that similarity patterns in a dataset can be used to predict unknown data or missing features. Recommendation systems are a typical application where users give a score to items and the system tries to infer similarities among users and scores to predict the missing entries.

These techniques are unsupervised learning methods, since categories are not predetermined, but are inferred from the data. They are applied to datasets exhibiting one or more of the following properties: 1) entries of the dataset have many attributes, 2) no law is known to link the different features, and 3) no classification is available to manually label the dataset.

In what follows, we briefly review the similarity-based classification tools that have been used in the anticipatory networking literature accounted for in this survey.

\subsubsection{Collaborative filtering}
\label{sec:predotherCF}

Recommendation systems usually adopt \acf{CF} to predict unknown opinions according to user's and/or content's similarities. While a thorough survey is available in~\cite{lee2012comparative}, here, we just introduce the main concepts related to anticipatory networking.

\ac{CF} predicts the missing entries of a $n_c \times n_u$ matrix $\mat{Y} \in \mathcal{A}^{n_c \times n_u}$, mapping $n_c$ users to $n_u$ contents through their opinions which are taken from an alphabet $\mathcal{A}$ of possible ratings. Thus, the entry $y_{ik}, i\in\{1,\dots,n_c\}, k\in\{1,\dots,n_u\}$ expresses how much user $k$ likes content $i$. An auxiliary matrix $\mat{R} \in [0,1]^{n_c \times n_u}$ expresses whether user $k$ evaluated content $i$ ($r_{ik}=1$) or not ($r_{ik}=0$).

To predict the missing entries of $\mat{Y}$ the feature learning approach exploits a set of $n_f$ features to represent contents' and users' similarities and defines two matrices $\mat{X} \in [0,1]^{n_c \times n_f}$ and $\mat{\Theta}\in\mathcal{A}^{n_u \times n_f}$, whose entries $x_{ij}$ and $\theta_{kj}$ represent how much content $i$ is represented by feature $j$ and how high user $k$ would rate a content completely defined by feature $j$, respectively. The new matrices aim to map $\mat{Y}$ in the feature space and they can be computed by:
\begin{eqnarray}
\label{eq:cfopt}
& \underset{\mat{X},\mat{\Theta}}{\textrm{argmin}} & \sum_{i,k:r_{ik}=1} (  \ve{x}_{i \ast} \ve{\theta}_{k \ast}^T  - y_{ik})^2,
\end{eqnarray}
where $\ve{x}_{i \ast}:= (\col_i \mat{X}^T)^T$ denotes the $i$-th row of matrix $\mat{X}$. Note that in \eqref{eq:cfopt} the regularization terms are omitted. Solving \eqref{eq:cfopt} amounts to obtain a matrix $\tilde{\mat{Y}} = \mat{X}\mat{\Theta}^T$ which best approximates $\mat{Y}$ according to the available information ($i,k:r_{ik}=1$). Finally, $\tilde{y}_{ik} = \ve{x}_{i\ast}\ve{\theta}_{k\ast}^T$ predicts how user $k$ with parameters $\ve{\theta}_{k\ast}$ rates content $i$ having feature vector $\ve{x}_{i\ast}$.

Other applications of \ac{CF} are, for instance, network caching optimization~\cite{bastug2014think,dutta2015predictive}, where communication efficiency is optimized by storing contents where and when they are predicted to be consumed. Similarly, location-based services~\cite{noulas2012mining} predict where and what to serve to a given user. 

\subsubsection{Clustering}
\label{sec:predotherClus}

Clustering techniques are meant to group elements that share similar characteristics. The following provides an introduction to $K$-means, which is among the most commonly-used clustering techniques in anticipatory networking. The interested reader is referred to~\cite{xu2005survey} for a complete review.

$K$-means splits a given dataset into $K$ groups without any prior information about the group structure. The basic idea is to associate each observation point from a dataset $\mathcal{X} := \{\ve{x}_i\in \mathbb{R}^n: i = 1, \ldots, M\}$, to one of the centroids in set $\set{M} := \{\ve{\mu}_j \in \mathbb{R}^n: j=1,\dots,K\}$. The centroids are optimized by minimizing the intra-cluster sum of squares (sum of distance of each point in the cluster to the $K$ centroids), given by
\begin{eqnarray}
\label{eq:clusteropt}
 & \underset{\set{C},\set{M}}{\textrm{minimize}} & \sum_{j=1}^K \sum_{i=1}^M  c_{ij}\|\ve{x}_i - \ve{\mu}_j \|^2,
\end{eqnarray}
where $\set{C} := \{c_{ij} \in \{0,1\}: i = 1,\dots,M, j=1,\dots,K\}$ associates entry $\ve{x}_i$ to centroid $\ve{\mu}_j$. No entry can be associated to multiple centroids ($\sum_{j=1}^K c_{ij} = 1, \forall i \in \set{M}$). 

Clustering is applied in anticipatory networking to build a data-driven link model~\cite{tarsa2015taming}, to find similarities within vehicular paths~\cite{froehlich2008route}, to identify social events~\cite{samulevicius2015most} that might impact network performance, and to identify device types~\cite{shafiq2011characterizing}.

\subsubsection{Decision Trees}
\label{sec:predotherDT}

A supervised version of clustering is {\it decision tree learning} (the interested reader is referred to~\cite{murthy1998automatic} for a survey on the topic). 
Assuming that each input observation is mapped to a consequence on its target value (such as reward, utility, cost, etc.), 
the goal of decision tree learning is to build a set of rules to map the observations to their target values. Each decision branches the tree into different paths that lead to leaves representing the class labels.
With prior knowledge, decision trees can be exploited for location-based services~\cite{noulas2012mining}, for identifying trajectory similarities~\cite{monreale2009wherenext}, and for predicting the \ac{QoE} for multimedia streams~\cite{sekar2013developing}. For continuous target variables,  regression trees can be used to learn trends in network performance~\cite{xu2013proteus}.

\figfunctional

\subsection{Regression Analysis}\label{subsec:Regression}
When the interest lies in understanding the relationship between different variables, regression analysis is used to predict dependent variables from a number of independent variables by means of so-called regression functions. In the following, we introduce three regression techniques, which are able to capture complex nonlinear relationships, namely {\it functional regression}, {\it support vector machines} and {\it artificial neural networks}.

\subsubsection{Functional regression}\label{subsubsec:functional}
Functional data often arise from measurements, where each point is expressed as a function over a physical continuum (e.g., Fig.~\ref{fig:FDA_wifiload} illustrates the example of aggregated WiFi traffic as a function of the hour of the day). Functional regression has two interesting properties: smoothness allows to study derivatives, which may reveal important aspects of the processes generating the data, and the mapping between original data and the functional space may reduce the dimensionality of the problem and, as a consequence, the computational complexity \cite{ramsay2006functional}.
The commonly encountered form of function prediction regression model (scalar-on-function) is given by \cite{ramsay1991some}:
\begin{equation}
Y_i = B_0 + \int X_i(z)B(z)dz + E_i
\label{eqn:FLM_2}
\end{equation}
where $Y_i, i = 1, \ldots, M$ is a continuous response, $X_i(z)$ is a functional predictor over the variable $z$, $B(z)$ is the functional coefficient, $B_0$ is the intercept, and $E_i$ is the residual error.

Functional regression methods are applied in~\cite{sayeed2015cloud} to predict traffic-related \ac{LTE} metrics (e.g., throughput, modulation and coding scheme, and used resources) showing that cloud analytics of short-term \ac{LTE} metrics is feasible. In~\cite{mozer2000predicting}, functional regression is used to study churn rate of mobile subscribers to maximize the carrier profitability.
\figsvm

\subsubsection{Support vector machines}\label{sec:svm}
\ac{SVM} is a supervised learning technique that constructs a hyperplane or set of hyperplanes (linear or nonlinear) in a high- or infinite-dimensional space, which can be used for classification, regression, or other tasks. 
In this survey we introduce the SVM for classification, and the same principle is used by \ac{SVM} for regression. 
Consider a training dataset $\{(\ve{x}_i,y_i): \ve{x}_i\in \mathbb{R}^n,y_i\in\{-1,1\}, i = 1, \ldots, M\}$, where $\mathbf{x}_i$ is the $i$-th training vector and $y_i$ the label of its class. 
First, let us assume that the data is linearly separable and define the linear separating hyperplane as $\ve{w}\cdot\ve{x} - b = 0$, where $\ve{w}\cdot\ve{x}$ is the Euclidean inner product.
The optimal hyperplane is the one that maximizes the {\it margin} (i.e., distance from the hyperplane to the instances closest to it on either side), which can be found by solving the following optimization problem:
\begin{eqnarray} \label{eq:svm}
\nonumber
&\text{minimize}& \frac{1}{2}||\mathbf{w}||^2 \\
&\text{subject to}&y_i(\mathbf{x}_i\cdot\mathbf{w} + b) - 1\geq0~\forall i \in \{1,\dots,M\}.
\end{eqnarray}  
Fig.~\ref{fig:svm:lin} shows an example of linear \ac{SVM} classifier separating two classes in $\mathbb{R}^2$.

If the data is not linearly separable, the training points are projected to a high-dimensional space $\mathcal{H}$ through a nonlinear transformation $\ve{\phi}:R^n\rightarrow\mathcal{H}$. Then, a linear model in the new space is built, which corresponds to a nonlinear model in the original space. 
Since the solution of (\ref{eq:svm}) consists of inner products of training data $\ve{x}_i\cdot\ve{x}_j$, for all $i,j$, in the new space the solution is in the form of $\ve{\phi}(\ve{x}_i)\cdot \ve{\phi}(\ve{x}_j)$.   
The {\it kernel trick} is applied to replace the inner product of basis functions by a {\it kernel function} $K(\ve{x}_i,\ve{x}_j) = \ve{\phi}(\ve{x}_i)\cdot \ve{\phi}(\ve{x}_j)$ between instances in the original input space, without explicitly building the transformation $\ve{\phi}$.

The Gaussian kernel $K(\ve{x},\ve{y}) := \exp(\gamma||\ve{x}-\ve{y}||^2)$ is one of the most widely used kernels in the literature. For example, it is used in~\cite{chen2013predicting} to predict user mobility. 
In~\cite{kasparick2015kernel}, the authors propose an algorithm for reconstructing coverage maps from path-loss measurements using a kernel method.
Nevertheless, choosing an appropriate kernel for a given prediction task remains one of the main challenges.

\subsubsection{Artificial neural networks}\label{subsubsec:neuralNetworks}
\ac{ANN} is a supervised machine learning solution for both regression and classification. An \ac{ANN} is a network of nodes, or \emph{neurons}, grouped into three layers (input, hidden and output), which allows for nonlinear classification. Ideally, it can achieve zero training error. 

Consider a training dataset $\{(\ve{x}_i,y_i): \ve{x}_i\in \mathbb{R}^n, i = 1, \ldots, M\}$.
Each hidden node $h_l$ approximates a so-called logistic function in the form $h_l = 1/(1+\exp(-\ve{\omega}_l\cdot\mathbf{x}))$, where $\ve{\omega}_l$ is a weight vector.
The outputs of the hidden nodes are processed by the output nodes to approximate $\ve{y}$. These nodes use linear and logistic functions for regression and classification, respectively.
In the linear case, the approximated output is represented as:
\begin{equation}
\ve{\hat{y}}=\sum_{l=1}^Lh_lv_l =\sum_{l=1}^L\frac{1}{1+\exp(-\ve{\omega}_l\cdot\mathbf{x})}v_l,
\end{equation}
where $L$ is the number of hidden nodes and $v_l$ is the weight vector of the output layer. The training of an \ac{ANN} can be performed by means of the \emph{backpropagation} method that finds weights for both layers to minimize the mean squared error between the training labels $y$ and their approximations $\hat{y}$. In the anticipatory networking literature, \acp{ANN} have been used for example to predict mobility in mobile ad-hoc networks~\cite{ghouti2013mobility, kaaniche2010mobility}.

For both \acp{SVM} and \acp{ANN}, as for other supervised learning approaches, no prior knowledge about the system is required but a large training set has to be acquired for parameter setting in the predictive model. A careful analysis needs to be performed while processing the training data in order to avoid both overfitting and underlearning.
\subsection{Statistical Methods for Probabilistic Forecasting}
\label{subsec:Probabilistic}

Probabilistic forecasting involves the use of information at hand to make statements about the likely course of future events. 
In the following subsections, we introduce two probabilistic forecasting techniques: {\it Markovian models} and {\it Bayesian inference}.

\subsubsection{Markovian models}\label{subsubsec:Markovian}
These models can be applied to any system for which state transitions only depend on the current state. 
In the following we briefly discuss the basic concepts of discrete, and continuous time \acp{MC} and their respective applications to anticipatory networking.

A \ac{DTMC} is a discrete time stochastic process $X_n (n\in\mathbb{N})$, where a state $X_n$ takes a finite number of values from a set $\mathcal{X}$ in each time slot. 
The Markovian property for a \ac{DTMC} transitioning from any time slot $k$ to $k+1$ is expressed as follows: 
\begin{equation}
 P(X_{k+1} = j|X_{k}=i) = p_{ij}(k).
\label{eq:markov}
\end{equation}

For a stationary \ac{DTMC}, the subscript $k$ is omitted and the transition matrix $\mathbf{P}$, where $p_{ij}$ represents the transition probability from state $i$ to state $j$, completely describes the model.
Empirical measurements on mobility and traffic evolution can be accurately predicted using a \ac{DTMC} with low computational complexity~\cite{chon2011mobility, barth2011mobility, bapierre2011variable, chon2012evaluating, shafiq2011characterizing}.  
However, obtaining the transition probabilities of the system requires a variable training period, which depends on the prediction goal. In practice, the data collection period can be in the order of one~\cite{shafiq2011characterizing} or even multiple weeks~\cite{nicholson2008breadcrumbs,barth2012combining}.

A \ac{DTMC} assumes the time the system spends in each state is equal for all states. This time depends on the prediction application and can range from a few hundred milliseconds to predict wireless channel quality~\cite{hosseini2015not}, to tens of seconds for user mobility prediction~\cite{nicholson2008breadcrumbs,barth2011mobility}, to hours for Internet traffic~\cite{shafiq2011characterizing}. 
For tractability reason, the state space is often compressed by means of simple heuristics~\cite{barth2012combining, beister2014predicting, nicholson2008breadcrumbs}, $K$-means clustering~\cite{hosseini2015not, bapierre2011variable}, equal probability classification~\cite{beister2014predicting}, and density-based clustering~\cite{bapierre2011variable}.  

Eq.~(\ref{eq:markov}) defines a first order \ac{DTMC} and can be extended to the $l$-th order (i.e., transition probabilities depend on the $l$ previous states).
By Using higher order, DTMCs can increase the accuracy of the prediction at the expense of a longer training time and an increased computational complexity \cite{chon2012evaluating,bapierre2011variable,barth2011mobility}. 

If the sojourn time of each state is relevant to the prediction, the system can be modeled as a \ac{CTMC}. The Markovian property is preserved in \ac{CTMC} when the sojourn time is exponentially distributed, as in~\cite{gidofalvi2012and}. When the sojourn time has an arbitrary distribution, it becomes a Markov renewal process as described in~\cite{lee2006modeling,abu2010application}.

If the transition probabilities cannot be directly measured, but only the output of the system is quantifiable (dependent on the state), hidden Markov models allow to map the output state space to the unobservable model that governs the system. As an example, the inter-download times of video segments are predicted in~\cite{beister2014predicting}, where the output sequences are the inter-download times of the already downloaded segments and the states are the instants of the next download request.

\tableprediction

\subsubsection{Bayesian inference}\label{subsubsec:Bayesian}
This approach allows to make statements about what is unknown, by conditioning on what is known. Bayesian prediction can be summarized in the following steps: 1) define a {\it model} that expresses qualitative aspects of our knowledge but has unknown parameters, 2) specify a {\it prior} probability distribution for the unknown parameters, 3) compute the {\it posterior} probability distribution for the parameters, given the observed data, and 4) make predictions by averaging over the posterior distribution.

Given a set of observed data $\set{D}:=\{(\ve{x}_i, \ve{y}_i): i = 1, \ldots, M\}$ consisting of a set of input samples $\set{X}: = \{\ve{x}_i\in\field{R}^p: i = 1, \ldots, M\}$ and a set of output samples $\set{Y} := \{\ve{y}_i\in\field{R}^q: i = 1, \ldots, M\}$, inference in Bayesian models is based on the {\it posterior distribution} over the parameters, given by the {\it Bayes' rule}:
\begin{align}
p(\ve{\theta}|\set{D}) & = \frac{p(\set{Y}|\set{X}, \ve{\theta})p(\ve{\theta})}{p(\set{Y}|\set{X})}\propto p(\set{Y}|\set{X}, \ve{\theta})p(\ve{\theta}), \label{eqn:Bayes}
\end{align}
where $\ve{\theta}$ is the unknown parameter vector. 

Two recent works adopting the Bayesian framework are~\cite{muppirisetty2015spatial} and~\cite{liao2015channel}. The former focuses on spatial prediction of the wireless channel, building a $2$D non-stationary random field accounting for pathloss, shadowing and multipath. The latter exploits spatial and temporal correlation to develop a general prediction model for the channel gain of mobile users. 

\subsection{Summary}\label{subsec:GeneralPrinciple}
Hereafter, we provide some guidelines for selecting the appropriate prediction methods depending on the application scenario or context of interest.

\subsubsection{Applications and data}
The predicted context is the most important information that drives decision making in anticipatory optimization problems (see Section~\ref{sec:optimization}). Thus, the selection of the prediction method shall take into consideration the objectives of the application and the constraints imposed by the available data.

\paragraph{Choosing the outputs} Applications define the properties of the predicted variables, such as dimension, granularity, accuracy, and range. For example, large granularity or high data aggregation (such as frequently visited location, social behavior pattern) is best dealt with similarity-based classification methods which provide sufficiently accurate prediction without the complexity of other model-based regression techniques.

\paragraph{System model and data} The application environment is equally important as its outputs, which determines the constraints of modeling. Often, an accurate analysis of the scenario might highlight linearity, deterministic and/or causal laws among the variables that can further improve the prediction accuracy. Moreover, the quality of dataset heavily affects the prediction accuracy. Different methods exhibit different level of robustness to noisy data. 

\subsubsection{Guidelines for selecting methods} To choose the correct tool among the aforementioned set, we study the rationale for adopting each of them in the literature and derive the following practical guidelines.

\paragraph{Model-based methods} When a physical model exists, model-based regression techniques based on closed-form expressions can be used to obtain an accurate prediction. They are usually preferable for long-term forecast and exhibit good resilience to poor data quality.

\paragraph{Time series-based methods} These are the most convenient tools when the information is abundant and shows strong temporal correlation. Under these conditions, time series methods provide simple means to obtain multiple scale prediction of moderate to high precision.

\paragraph{Causal methods} If the data exhibits large and fast variations, causality laws can be key to obtain robust predictions. In particular, if a causal relationship can be observed between the variables of interest and the other observable data, causal models usually outperform pure data-driven models. 

\paragraph{Probabilistic models} If the physical model of the prediction variable is either unavailable or too complex to be used, probabilistic models offer robust prediction based on the observation of a sufficient amount of data. In addition, probabilistic methods are capable of quantifying the uncertainty of the prediction, based on the probability density function of the predicted state.

\subsubsection{Prediction summary} Table~\ref{tab:ObjectiveAndConstaints_predict} characterizes each prediction method with respect to {\it properties of the context} and {\it constraints} presented in Section~\ref{sec:guidelines}. 
Note that the methods for predicting a multivariate process can be applied to univariate processes without loss of generality. The granularity of variables and the prediction range are described using
qualitative attributes such as {\bf S}hort, {\bf M}edium, {\bf L}arge, and {\bf any} instead of explicit values. For example, for the time series of traffic load per cell, S, M and L time scales are generally defined by minutes, tens of minutes and hours, respectively, while for the time series of channel gain, they can be seen as milliseconds, hundreds of milliseconds and seconds, respectively. The sixth column reports the prediction type, that can be driven by {\bf data}, {\bf models} or {\bf both}. Linearity indicates whether it is required ({\bf Y}) or not ({\bf N}) or applicable in {\bf both} cases. The side information column states whether out-of-band information can ({\bf both}), cannot ({\bf N}) or must ({\bf Y}) be used to build the model. Finally, the quality column reports whether the predictor is {\bf weak} or {\bf robust} against insufficient or unreliable dataset.  

\section{Optimization Techniques for Anticipatory Networking}
\label{sec:optimization}

This section identifies the main optimization techniques adopted by anticipatory networking solutions to achieve their objectives. Disregarding the particular domain of each work, the common denominator is to leverage some future knowledge obtained by means of prediction to drive the system optimization. How this optimization is performed depends both on the ultimate objectives and how data are predicted and stored.

In general, we found two main strategies for optimization: (1) adopting a well-known optimization framework to model the problem and (2) designing a novel solution (most often) based on heuristic considerations about the problem. The two strategies are not mutually exclusive and often, when known approaches lead to too complex or impractical solutions, they are mixed in order to provide feasible approximation of the original problem. 

Heuristic approaches usually consist of (1) algorithms that allow for fast computation of an approximation of the solution of a more complex problem (e.g., convex optimization) and (2) greedy approaches that can be proven optimal under some set of assumptions. Both approaches trade optimality for complexity and most often are able to obtain performance quite close to the optimal one. However, heuristic approaches are tailored to the specific application and are usually difficult to be generalized or to be adapted for different scenarios, thus they cannot be directly applied to new applications if the new requirements do not match those of the original scenario. 

In what follows, we focus on optimization methods only and we will provide some introductory descriptions of the most relevant ones used for anticipatory networking. The objective is to provide the reader with a minimum set of tools to understand the methodologies and to highlight the main properties and applications. 

\subsection{Convex Optimization}

Convex optimization is a field that studies the problem of minimizing a convex function over convex sets. The interested reader can refer to~\cite{boyd2004convex} for convex optimization theory and algorithms. Hereafter, we will adopt Boyd's notation~\cite{boyd2004convex} to introduce definitions and formulations that frequently appear in anticipatory networking papers. 

The inputs are often referred to as the optimization variables of the problem and defined as the vector $\ve{x} = (x_1,\dots,x_n)$. In order to compute the best configuration or, more precisely, to optimize the variables, an objective is defined: this usually corresponds to minimizing a function of the optimization variables, $f_0:\mathbb{R}^n \rightarrow \mathbb{R}$. The feasible set of input configurations is usually defined through a set of $m$ constraints $f_i(x) \leq b_i$, $i = 1,\dots,m$, with $f_i:\mathbb{R}^n \rightarrow \mathbb{R}$. 
The general formulation of the problem is
\begin{eqnarray}
\label{eq:genopt}
 & \textrm{minimize} & f_0(\ve{x}) \nonumber \\
 & \textrm{subject to} & f_i \leq b_i, \;\; i = 1,\dots,m.
\end{eqnarray}

The solution to the optimization problem is an optimal vector $\ve{x}^*$ that provides the smallest value of the objective function, while satisfying all the constraints.


The convexity property (i.e., objective and constraint functions satisfy $f_i(a\ve{x} + (1-a)\ve{y}) \leq af_i(\ve{x}) + (1-a)f_i(\ve{y})$ for all $\ve{x},\ve{y} \in \mathbb{R}^n$ and $a \in [0,1]$) can be exploited in order to derive efficient algorithms that allows for fast computation of the optimal solution.
Furthermore, if the optimization function and the constraints are linear, i.e., $f_i(a\ve{x} + b\ve{y}) = af_i(\ve{x}) + bf_i(\ve{y})$ for all $\ve{x},\ve{y} \in \mathbb{R}^n$ and $a,b \in \mathbb{R}$, the problem belongs to the class of \emph{linear optimization}. 
For this class, highly efficient solvers exist, thanks to their inherently simple structure. 
Within the linear optimization class, three subclasses are of particular interest for anticipatory networking: least-squares problems, linear programs and mixed-integer linear programs. 

\emph{Least-squares} problems can be thought of as distance minimization problems. 
They have no constraints ($m=0$) and their general formulation is:
\begin{eqnarray}
\label{eq:ls}
 \textrm{minimize} & f_0(\ve{x}) = ||\mat{A}\ve{x} - \ve{b}||_2^2,
\end{eqnarray}
where $A \in \mathbb{R}^{k \times n}$, with $k \geq n$ and $||x||_2$ is the Euclidean norm. Notably, problems of this class have an analytical solution $\ve{x}=(\mat{A}^T\mat{A})^{-1}\mat{A}^T\ve{b}$ (where superscript $^T$ denotes the transpose) derived from reducing the problem to the set of linear equations $\mat{A}^T\mat{A}\ve{x} = \mat{A}^T\ve{b}$. 


\emph{Linear programming} (LP) problems are characterized by linear objective function and constraints and are written as
\begin{eqnarray}
\label{eq:lpgen}
 & \textrm{minimize} & \ve{c}^T\ve{x} \nonumber \\
 & \textrm{subject to} & \mat{A}^T\ve{x} \leq b,
\end{eqnarray}
where $\ve{c} \in \mathbb{R}^n$, $\mat{A} \in \mathbb{R}^{n \times m}$ and $\ve{b} \in \mathbb{R}^n$ are the parameters of the problem. Although, there is no analytical closed-form solution to LP problems, a variety of efficient algorithms are available to compute the optimal vector $\ve{x}^*$.
When the optimization variable is a vector of integers $x \in \mathbb{Z}^n$, the class of problems is called \emph{integer linear programming} (ILP), while the class of \emph{mixed-integers linear programming} (MILP) allows for both integer and real variables to co-exist. These last two classes of problems can be shown to be NP-hard (while \ac{LP} is P complete) and their solution often implies combinatorial aspects. See~\cite{schrijver1998theory}  for more details on integer optimization.

In anticipatory networking, we find that resource allocation problems are often modeled as \ac{LP}, \ac{ILP} or \ac{MILP}, by setting the amount of resources to be allocated as the optimization variable and accounting for prediction in the constraints of the problem. In~\cite{abouzeid2014energy}, prediction of the channel gain is exploited to optimize the energy efficiency of the network. Time is modeled as a finite number of slots corresponding to the look-ahead time of the prediction. 
When dealing with multimedia streaming, the data buffer is usually modeled in the constraints of the problem by linking the state at a given time slot to the previous slot. The solver will then choose whether to use resources in the current slot or use what has been accumulated in the buffer, as in, e.g.,~\cite{draxler2015smarterphones}. 
Admission control is often used to enforce quality-of-service, e.g.,~\cite{bui2015anticipatory,chen2015rate}, with the drawback of introducing integer variables in the optimization function. In these cases, 
 the optimal ILP/MILP formulation is followed by a fast heuristic that enables the implementation of real-time algorithms.


\subsection{Model Predictive Control}\label{sec:mpc}
Model Predictive Control (MPC) is a control theoretic approach that optimizes the sequence of actions in a dynamic system by using the process model of that system within a finite time horizon. Therefore, the process model, i.e., the process that turns the system from one state to the next, should be known. In each time slot $t$, the system state, $\ve{x}(t)$, is defined as a vector of attributes that define the relevant properties of the system. At each state, the control action, $\ve{u}(t)$, turns the system to the next state $\ve{x}(t+1)$ and results in the output $\ve{y}(t+1)$. In case the system is linear, both the next state and the output can be determined as follows:
\begin{eqnarray}
 \ve{x}(t+1) & = & \mat{A}\ve{x}(t) + \mat{B}\ve{u}(t) + \ve{\psi}(t) \\
 \ve{y}(t) & = & \mat{C}\ve{x}(t) + \ve{\epsilon}(t),
\end{eqnarray}
where $\ve{\psi}(t)$ and $\ve{\epsilon}(t)$ are usually zero mean random variables used to model the effect of disturbances on the input and output, respectively, and $\mat{A}$, $\mat{B}$, and $\mat{C}$ are matrices determined by the system model.

At each time slot, 
the next $N$ states and their respective outputs are predicted and a cost function $J(\cdot)$ is minimized to determine the optimal control action $\ve{u}^*(t)$ at $t = t_0$:
\begin{equation}\label{eq:mpc_opt}
\ve{u}^*(t_0)=\arg\underset{\ve{u}(t_0)}\min J(\ve{\hat x}(t_0),\ve{u}(t_0)),
\end{equation}
where $\ve{\hat x}(t_0)$ is the set of all the predicted states from $t = t_0 + 1$ to $t = t_0 + N$, including the observed state at $t = t_0$. 
The expression in \eqref{eq:mpc_opt} essentially states that the optimal action of the current time slot is computed based on the predicted states of a finite time horizon in the future.
In other words, in each time slot the MPC sequentially performs a $N$ step lookahead open loop optimization of which only the first step is implemented \cite{qin2003survey}. 

This approach has been adopted for on-line prediction and optimization of wireless networks \cite{bianchi2013networked,lee2013generalized}. 
Since the process model (for the prediction of future states and outputs) is available in this kind of systems, autoregressive methods can be used along with Kalman filtering \cite{lee2013generalized}, or max-min MPC formulation \cite{witheephanich2014min}. 
In~\cite{bianchi2013networked}, Kalman filtering is compared to other methods such as mean and median value estimation, Markov chains, and exponential averaging filters.  

Optimization based on MPC relies on a finite horizon. The length of the horizon determines the trade-off between complexity and accuracy. Longer horizons need further look ahead and more complex prediction but in turn result in a more foresighted control action \cite{witheephanich2014min}. Reducing the horizon reduces the complexity while resulting in a more myopic action. This trade-off is examined in \cite{bianchi2013networked} by proposing an algorithm that adaptively adjusts the horizon length. In general, the prediction horizon is kept to a fairly low number (1 step in \cite{witheephanich2014min} and 6 steps in \cite{lee2013generalized}) to avoid high computation overhead.

It is worth noting that MPC methods can be extended to the nonlinear case. In this case, the prediction accuracy and control optimality increase at the cost of more complex algorithms to find the  solution~\cite{qin2003survey}. Another benefit of these approaches is their applicability to non-stationary problems. 
 
\subsection{Markov Decision Process}\label{sec:mdp}

Markov Decision Process (MDP) is an efficient tool for optimizing sequential decision making in stochastic environments. Unlike MPCs, MDPs can only be applied to stationary systems where a priori information about the dynamics of the system as well as the state-action space is available. 

A MDP consists of a four tuple $(\set{X},\set{U},\mat{P},r)$, where $\set{X}$ and $\set{U}$ represent the set of all achievable states in the system and the set of all actions that can be performed in each of the states, respectively. Time is assumed to be slotted and in any time slot $t$, the system is in state $x_t \in \set{X}$ from which it can take an action $u_t$ from the set $U_{x_t} \in \set{U}$. Due to the assumption of stationarity, we can omit the time subscript for states and actions. Upon taking action $u$ in state $x$, the system moves to the next state $x'\in\set{X}$ with  transition probability $\mathbf{P}(x'|x,u)$ and receives a reward equal to $r(x,u,x')$. The transition probabilities are predicted and modeled as a Markov Chain prior to solving the MDP and preserve the Markovian behavior of the system.

The goal is to find the optimal policy $\pi^*: \set{X}\rightarrow\set{U}$  (i.e., optimal sequence of actions that must be taken from any initial state) in order to maximize the long term discounted average reward $\mathbb{E}\left(\sum_{t = 0}^\infty \gamma^tr(x_t,u_t,x_{t+1})\right)$, where $0 \leq \gamma < 1$ is called \textit{discount factor} and determines how myopic (if closer to zero) or foresighted (if closer to 1) the decision process should be. 
In order to derive the optimal policy, each state is assigned to a value function $V^\pi(x)$, which is defined as the long term discounted sum of rewards obtained by following policy $\pi$ from state $x$ onwards. The goal of MDP algorithms is to find $V^{\pi^*}(x) (\forall x\in\set{X})$.
Given that the Markovian property holds, it has been proved that the optimal value functions follow the Bellman optimality criterion described below \cite{puterman2014markov} :
\begin{multline}
V^{\pi^*}(x) = \\
= \max_{u\in\set{U}}\sum_{x'\in\set{X'}}\left(r(x,u,x') + \gamma \mathbf{P}(x'|x,u)V^{\pi^*}(x')\right)   \\
\forall x\in\set{X},
\end{multline}
where $\set{X'}\subset\set{X}$ is the set of states for which $\mathbf{P}(x'|x,u)>0$. In order to solve the above equation set, linear programming or dynamic programming techniques can be used, in which the optimal policy is derived by simple iterative algorithms such as policy iteration and value iteration  \cite{puterman2014markov}. 

MDPs are very efficient for several problems, especially in the framework of anticipatory networking, due to their wide applicability and ease of implementation. MDP-based optimized download policies for adaptive video transmission under varying channel and network conditions are presented in~\cite{hosseini2015not,bao2015bitrate,chen2013markov}. 

\taboptimization


In order to avoid large state spaces (which limit the applicability of MDPs), there are cases where the accuracy of the model must be compromised for simplicity. In~\cite{chen2013markov}, a large video receiver buffer is modeled for storing video on demand but only a small portion of the buffer is used in the optimization, while the rest of the buffer follows a heuristic download policy. \cite{hosseini2015not,bao2015bitrate} solve this problem by increasing the duration of the time slot such that more video can be downloaded in each slot and, therefore, the buffer is filled entirely based on the optimal policy. This, in turn, comes at the cost of lower accuracy, since the assumption is that the system is static within the duration of a time slot. 
Heuristic approaches are also adopted for on-line applications. For instance, creating decision trees with low depth from the MDP outputs is proposed in~\cite{hosseini2015not}. Simpler heuristics are also applied to the MDP outputs in \cite{bao2015bitrate,chen2013markov,dutta2015predictive}.

If any of the assumptions discussed above does not hold, or if the state space of the system is too large, MDPs and their respective dynamic programming solution algorithms fail. However, there are alternative techniques to solve this kind of problems. For instance, if the system dynamics follow a Markov Renewal Process instead of a \ac{MC}, a semi \ac{MDP} is solved instead of the regular one~\cite{puterman2014markov}. In non-stationary systems, for which the dynamics cannot be predicted a priori or the reward function is not known beforehand, reinforcement learning~\cite{sutton1998reinforcement} can be applied and the optimization turns into an on-line unsupervised learning problem. Large state spaces can be dealt with using value function approximation, where the value function of the \ac{MDP} is approximated as a linear function, a neural network, or a decision tree \cite{sutton1998reinforcement}. If different subsets of state attributes have independent effects on the overall reward, i.e., multi user resource allocation, the problem can be modeled as a weakly coupled \ac{MDP}~\cite{fu2010systematic} and can be decomposed into smaller and more tractable \acp{MDP}.

\rev{
\subsection{Game theoretic approaches}
Although small in number, the papers adopting a game theoretic framework offer an alternative approach to optimization. In fact, while the approaches described in the previous subsections strive to compute the optimal solution of an often complex problem formulation, game theory defines policies that allow the system to converge towards a so-called equilibrium, where no player can modify her action to improve her utility. In mobile networks, game theory is applied in the form of matching games~\cite{gu2015matching}, where system players (e.g. users) have to be matched with network resources (e.g. base stations or resource blocks). \newline\indent
Three types of matching games can be used depending on the application scenario: 1) one-to-one matching, where each user can be matched with at most one resource (as in~\cite{semiari2015context}, which optimizes \ac{D2D} communication in small cell scenarios); 2) many-to-one matching, where either multiple resources can be assigned to a single user (as in~\cite{semiari2016context} for small cell resource allocation), or multiple users can be matched to a single resource (as in~\cite{namvar2014context} for user-cell association); 3) many-to-many matching, where multiple users can be matched with multiple resource (as in~\cite{hamidouche2014many} where videos are associated to caching servers).
}

\subsection{Summary}
\label{sec:chal:opt}

This section (and Table~\ref{tab:class_opt_summary}) summarizes the main takeaways of this optimization handbook. 

\subsubsection{Convex Optimization methods} These methods are often combined with time series analysis or ideal prediction. The main reason is that they are used to determine performance bounds when the solving time is not a system constraint. Thus, convex optimization is suggested as a benchmark for large scale prediction. This may have to be replaced by fast heuristics in case the optimization tool needs to work in real-time. An exception to this is \ac{LP} for which very efficient algorithms exist that can compute a solution in polynomial time. 
In contrast, convex optimization methods should be preferred when dealing with high precision and continuous output. They require the complete dataset and show a reliability comparable to that of the used predictor.

\subsubsection{Model Predictive Control} \ac{MPC} combines prediction and optimization to minimize the control error by tuning both the prediction and the control parameters. Therefore, it can be coupled with any predictor. The main drawback of this approach is that, by definition, prediction and optimization cannot be decoupled and must be evaluated at each iteration. This makes the solution computationally very heavy and it is generally difficult to obtain real-time algorithms based on \ac{MPC}.
The close coupling between prediction and optimization makes it possible to adopt the method for any application for which a predictor can be designed with the only additional constraint being the execution time. Objectives and constraints are usually those imposed by the used predictor.

\subsubsection{Markov Decision Processes} \acp{MDP} are characterized by a statistical description of the system state and they usually model the system evolution through probabilistic predictors. As such, they best fit to scenarios that show similar objective functions and constraints as those of probabilistic predictors.
Thus, \acp{MDP} are the ideal choice when the optimization objective aims at obtaining stationary policies (i.e., policies that can be applied independently of the system time). This translates to low precision and high reliability. Moreover, even though they require a computationally heavy phase to optimize the policies, once the policies are obtained, fast algorithms can easily be applied.
\rev{
\subsubsection{Game theory} Matching games prove to be effective solutions that, without struggling to compute an overly complex optimal configuration, let the system converge towards a stable equilibrium which satisfies all the players (i.e., no action can be taken to improve the utility of any player). These are the preferable solutions for those applications where the computational capability is a stringent constraint and where fairness is important for the system quality.
}

\revv{
\section{Applicability of Anticipatory Networking to other Wireless Networks}
\label{sec:network}}

\tabnetwork

\revv{So far this survey mainly focused on current cellular networks. In this section we analyze how different types of mobile wireless networks can take advantage of anticipatory networking solutions. Although each type would deserve a dedicated survey, in what follows we provide brief summaries of the distinctive features, the application scenarios, the expected benefits and the challenges related to the implementation of anticipatory networking for each of them. Table~\ref{tab:network} summarizes the discussion of this section.}

\revv{\subsection{5G Cellular Networks}
LTE and LTE-advanced represent the fourth generation of mobile cellular networks and, as it emerged from the analyses of the previous sections, they can already benefit from predictive optimization. Since the fifth generation is expected to improve on its predecessors in every aspect~\cite{hossain20155g}, not only is anticipatory networking applicable, but also it will provide even greater benefits. 
\subsubsection{Characteristics}
The next generation of mobile cellular networks will provide faster communications, improved users \ac{QoE}, shorter communication delays, higher reliability and improved energy savings. Among the solutions envisioned to realize these improvements, cell densification, mm-wave bands, massive MIMO, unified multi-technology frame structure and architecture and network function virtualization are the ones that are going to have a substantial impact on existing and future use case scenarios. In fact, a denser infrastructure is going to decrease the average time mobile users spend in a specific cell; the directionality of communications in higher portion of the spectrum will increase the importance of localization and tracking functionalities; while the increase of communicating elements and the de-localization of radio access functionalities are going to impact on channel models and network resource management.
\subsubsection{Advantages}
The performance of 5G cellular networks will strongly depend on their knowledge of the exact user positions (e.g., localization for mm-wave, resource management for network function virtualization). As a consequence, predictive solutions that provide the system with accurate information about users' current and future positions, trajectories, traffic profiles and content request probabilities are likely to be the most desirable aspects of anticipatory solutions. \newline\indent
For what concerns 5G applications, we believe network caching and cloud \ac{RAN} will also greatly benefit from this. In fact, the former can exploit prediction to decide which content to store in which specific part of the network to serve a given user profile, while the latter can, for instance, forecast when to instantiate a number of virtual machines to face an increase of the network traffic. 
\subsubsection{Challenges}
The upcoming 5G technologies will also bring new challenges to the basic mechanisms of anticipatory networking. In particular, we see mm-wave, massive MIMO and cell densification as disruptive technologies for the current methods used for predictive optimization. In this regard, mm-waves channel model is going to impact how to forecast future signal quality and achievable data rates while network densification and massive MIMO will challenge the scalability of prediction techniques due to the sheer size of the information needed to describe and exchange them.
}

\revv{\subsection{Mobile ad hoc networks}
\ac{MANET} consist of mobile wireless devices connected to one another without a fixed infrastructure~\cite{giordano2002mobile}. As a consequence, they share some characteristics with cellular networks but have some unique features due to the variable topology. These networks are the most practical form of communication when an infrastructure is absent or it has been compromised by a disruptive event.
\subsubsection{Characteristics}
The dynamic nature of \acp{MANET} causes the path between any two nodes to vary over time and require adaptive routing mechanisms that allow, on one hand, to maintain the connectivity among all the network nodes and, on the other hand, to balance the load in the different areas of the network. In addition, adaptive discovery and management functionalities are needed  to allow new devices and services to be added to an existing network and to report problems and missing links/nodes. When a \ac{MANET} extends over an area larger than the communication range of the devices, transmissions must be relayed from one node to another in order to allow messages to reach their destinations.
\subsubsection{Advantages}
Knowing nodes' positions in advance and being able to track their trajectories enable advanced routing functionalities: in fact, additional paths can be created before a missing link interrupts a route without waiting for a new discovery procedure to be performed. Also, routing tables can be readily adapted when shorter routes appear. In a similar way, management procedure can be enhanced by knowing in advance the traffic being produced by a given node or area of the network or by forecasting which service is going to be needed in a given part of the network.
\subsubsection{Challenges}
The absence of a fixed infrastructure is the main source of challenges that are distinctive of \acp{MANET}. For instance, it is not possible to have known databases collecting users' and devices' information to build prediction models nor centralized optimization services can be provided or they may suffer from delays in delivering solutions and/or information to the whole network. Moreover, the topology variability makes map-based prediction techniques difficult or impossible to apply.
}

\revv{\subsection{Cognitive Radio Networks}
\ac{CR} networks consist of devices that exploit channels that are unused at specific locations and times~\cite{chen2016survey}, but that are usually allocated to primary users (i.e. users that can legitimately communicate using a given channel). \ac{CR} devices are usually referred to as secondary users as their operations must not interfere with those performed by the primary users.
\subsubsection{Characteristics}
The main distinctive feature of \ac{CR} devices is that they need to scan for primary users' activity before attempting any communication in order not to disrupt legitimate transmissions. This scanning/sensing activity decreases the amount of time secondary users' can spend on actual communications and, thus, it reduces their throughput. On the other hand, a \ac{CR} network is usually able to build accurate spectrum occupancy models fusing the information coming from different devices.
\subsubsection{Advantages}
Prediction capabilities are already envisioned for \ac{CR} networks, in fact, it is easily understandable that being able to predict when primary users are going to occupy their channel will decrease the amount of sensing needed to decide when a secondary user is allowed to transmit. Not only can spectrum occupancy maps be used to predict the upcoming channel state, but also, content information and predictive models available to primary users can be exploited by secondary users to reduce their interference probability. Therefore, allowing secondary users to access primary user information is profitable for both: if \ac{CR} are able to improve their throughput by more precisely picking spectrum holes, primary users will be more protected from secondary interference.
\subsubsection{Challenges}
Although anticipatory \ac{CR} can be seen as symbiotic to primary users, their operations introduce a non trivial feedback in the resulting system. In fact, those models that are valid when primary users operate only may be no longer valid when secondary users contribute. However, given that those models are usually built using information about primary users only, it will be impossible with the current techniques to create or modify prediction and optimization solutions that take into consideration secondary users. As such, the whole anticipatory infrastructure needs to account for \ac{CR} in order to allow prediction-based schemes to work for primary and secondary users.
}

\revv{\subsection{Device-to-Device}
\ac{D2D} communication refers to the use of direct communication between mobile phones to support the operations of a cellular network~\cite{asadi2014survey}. 
In addition, since \ac{D2D} must not interfere with the regular cellular network operations it can be seen as secondary users to the main communications. Therefore, they share characteristics that are specific to \acp{MANET} and \ac{CR} networks.
\subsubsection{Characteristics}
\ac{D2D} communications are characterized by a complex topology where the usual star network overlies a mesh network. Also, the devices may use different \ac{RAN}s in the mesh network: for instance they can exploit the same cellular technology (inband) or other wireless solutions such as direct-WiFi.
\subsubsection{Advantages}
Given the similarities to \acp{MANET} and \acp{CR}, \ac{D2D} communications can take advantage from anticipatory networking mostly to mitigate interference related problems and to improve the resource and power allocation.
\subsubsection{Challenges}
While we do not expect \ac{D2D} communications to pose distinctive challenges to the implementation of anticipatory networking that are not listed in the previous sections, that will make the adoption of current prediction models less straightforward. In fact, prediction-based optimization and other anticipatory schemes will be made more complex due to the possible coexistence of multiple technologies and the primary/secondary interference and interactions, which will require to also predict \ac{D2D} channels, in addition to primary.
}


\revv{\subsection{Internet of Things}
Nowadays, thanks to the miniaturization and the progressive decrease of computational and communicating chipsets, more and more ordinary objects are being equipped with micro-CPUs and are connected to the Internet~\cite{alfuqaha2015internet, zanella2014internet, xu2014internet}: in such a way smart cities and smart industries, among a variety of other enhanced scenarios, can be realized. The typical device in the \ac{IoT} is capable of performing one or a set of measurements and/or actuations on the real world. They are usually constrained in their capabilities: for instance, they can be battery powered or equipped with low data rate radios or their computational power may be limited.
\subsubsection{Characteristics}
Due to the wide definition of the entities that populate the \ac{IoT}, many of its features have been already described in the preceding subsections. For instance, \ac{IoT} communications often involve \ac{D2D} aspects, they can be \ac{CR} if they are able to sense spectrum and they can be considered part of a \ac{MANET} if they are mobile. However, the most unique features that are only present in \ac{IoT} devices are
that they involve \ac{M2M} type communication and that devices are typically constrained. Moreover, although the number of smart things is expected to grow exponentially in the next decade, their traffic is not going to grow as fast as that, e.g., the one generated by mobile cellular networks. In fact, \ac{IoT} traffic is expected to be mainly due to monitoring, control and detection activities, which are characterized by limited throughput and almost deterministic transmission frequency.
\subsubsection{Advantages}
Anticipatory networking and prediction-based optimization can be applied to many aspects of the \ac{IoT}. For instance, devices that harvest their energy from renewable sources may predict the source availability and optimize their operations according to that. Furthermore, data prediction models can be used to compress the data produced by devices by sending only the difference from the forecast or the same models can be used to identify anomalies or prevent disruptive events before they can cause serious problems. Finally, due to the almost deterministic periodicity of data production, their communication can be easily modeled and accounted for to mitigate their impact on the overall system.
\subsubsection{Challenges}
Scalability is one of the main challenges in \ac{IoT}. In fact, due to the variety of device types, the difference in their capabilities, requirements and applications, the amount of information needed to represent and model the \ac{IoT} is huge and the obtained benefits must more than compensate for the cost related to its realization. Moreover, the \ac{IoT} is impacted by most of the challenges and problems discussed above for the other network types.
}

\revv{\section{On the impact of Anticipatory Networking on the Protocol Stack}
\label{sec:protocol}}

\revv{In this section, we address another important aspect of anticipatory networking solutions: where to implement them in the ISO/OSI protocol stack~\cite{zimmermann1980osi} and which layers contribute to their realizations.}

\revv{\subsection{Physical}
We do not expect anticipatory networking solutions to modify how the physical layer is designed and managed. In fact, in order to apply prediction-based schemes, some form of interaction is required between two or more entities of the system. As a consequence, the physical layer, which defines how information is transferred to bits and wave-form~\cite{zimmermann1980osi}, might provide different profiles to allow for predictive techniques to be applied in the higher layers, but will not directly implement any of them.
}

\revv{\subsection{Data Link}
The data link layer is the first entry point for predictive solutions. In particular, this layer implements \ac{MAC} functionalities. Therefore, resource management~\cite{lu2013optimizing} and admission control~\cite{bui2015anticipatoryb} procedures are likely to greatly benefit from anticipatory optimization. Also, we envision that anticipatory networking to be even more important in next generation networks: in particular, channel estimation and beam steering solutions are going to be key for the success of mm-wave a massive MIMO communications~\cite{hossain20155g}.
}

\revv{\subsection{Network}
The network layer contains two of the functionalities that can benefit the most from prediction: routing and caching~\cite{bastug2014living, naimi2014anticipation}. In fact, by knowing users' mobility and traffic in advance it is possible to optimize routes and caching location to maximize network performance and save resources. For instance, it is possible to build alternative paths before the existing ones deteriorate and break and popular contents may be moved across the network according to where they will be requested with higher probability.}

\revv{\subsection{Transport}
This layer is mainly concerned with end-to-end message delivery and the two most popular protocols are \ac{TCP} and \ac{UDP}: the former guarantees reliable communications, while the latter is a lightweight best-effort solution. Anticipatory networking solutions are easily implemented here~\cite{calabrese2010human, abouzeid2015evaluating}, in particular, when error correction and retransmissions are driven by network metrics such as, among others, \ac{RTT} and \ac{BER}. Prediction models can be used to react to changes in the network conditions before they reach a disruptive state and recovery actions have to be taken. In addition, modern transport solutions, such as multipath-\ac{TCP}, can exploit predictive optimization to manage the traffic flows along the different routes and improve the \ac{QoS}.}

\revv{\subsection{Session, Presentation and Application}
Since these layers are concerned with connection management between end-points (session), syntax mapping between different protocols (presentation) and interaction with users and software (application), they are the least preferable to implement anticipatory networking solutions. However, in order to allow applications to exploit predictive mechanisms, these three layers will act as a connection point to provide application with the needed context information and to allow them to configure the needed services and parameters for the application requirements. For instance, in Section~\ref{sec:classification}.A.6 we described geographically-assisted video optimization~\cite{draxler2015smarterphones, hosseini2015not} where mobile phone applications modulated the request video bit rate to optimize the playback of the video itself, or geo-assisted applications~\cite{noulas2012mining} that exploits social and contextual information to enhance their services.}

\rev{\section{Issues, Challenges, and Research Directions}
  \label{sec:challenges}}

We conclude the paper by providing some insights on how anticipatory optimization will enable new 5G use cases and by detailing the open challenges of anticipatory networking in order to be successfully applied in 5G.

\rev{\subsection{Context related analyses}
\subsubsection{Geographic context}
Geographic context is essential to achieve seamless service. Depending on the optimization objective, a mobility state can be defined with different granularity in multiple dimensions (location, time, speed, etc.). For example, for handover optimization it is sufficient to predict the staying time in the current serving cell and the next serving cell of the user. Medium to large spatial granularity such as cell \ac{ID} or cell coverage area can be considered as a state, and a trajectory can be characterized by a discrete sequence of cell \acp{ID} over time. State-space models such as Markov chains, \ac{HMM} and Kalman filters fit the system modeling,
while requiring large training samples and considerable insight to make the model compact and tractable. An alternative is the variable-order Markov models, including a variety of lossless compression algorithms (some of the most used belong to \acl{LZ} family), where Shannon's entropy measure is identified as a basis for comparing user mobility models. 
Such an information-theoretic approach enables adaptive online learning of the model, to reduce update paging cost. Moving from discrete to continuous models, which are applied to assist the prediction of other system metrics with high granularity, e.g., link gain or capacity, regression techniques are widely used.
To enhance the prediction accuracy, a priori knowledge can be exploited to provide additional constraints on the content and form of the model,  based on street layouts, traffic density, user profiles, etc. However, finding the right trade-off between the model accuracy and complexity is challenging. An effective solution is to decompose the state space and to introduce localized models,
e.g., to use distinct models for weekdays and weekends, or urban and rural areas.}

\rev{
Although mobility prediction has been shown to be viable, it has not been widely adopted in practical systems. This is because, unlike location-aware applications with users' permission to use their location information, mobile service providers must not violate the privacy and security of mobile users. To facilitate the next generation of user-centric networks, new interaction protocols and platforms need to be developed for enabling more user-friendly agreements on the data usage between the service providers and the mobile users.}

\rev{Furthermore, next generation wireless networks introduce ultra-dense small cells and high frequencies such as mmWaves. The transmission range gets shorter and transmission often occurs in line-of-sight conditions. Thus, 2D geographic context with a coarse level of accuracy is not sufficient to fully utilize the future radio techniques and resources. This trend opens the door for new research directions in inference and prediction of 3D geographic context, by utilizing advanced feedback from sensors in user equipments such as accelerometers, magnetometers, and gyroscopes.}
\rev{
\subsubsection{Link context}
When predicting link context, i.e., channel quality and its parameters, linear time series models have the potential to provide the best tradeoff between performance and complexity. When the channel changes slowly, e.g., because users are static or pedestrian, it is convenient to exploit the temporal correlation of historic measurements of the users' channel and implement linear auto-regressive prediction. This can be quite accurate for very short prediction horizons and at the same time simple enough to be implemented in real time systems. Kalman filters can also be used to track errors and their variance, based on previous measurements, thus handling uncertainties. 
However, time series and linear models are not robust to fast changes. Therefore, in high mobility scenarios, more complex models are needed. One possible approach is to exploit the spatio-temporal correlation between location and channel quality. By combining the prediction of the channel qualities with the prediction of the user's trajectory, regression analysis, e.g., \acp{SVM}, can be employed to build accurate radio maps to estimate the long term average channel quality, which accounts for pathloss and slow fading, but neglects fast fading variations. 
Ideally, one should have two predictions available: a very accurate short term prediction and an approximate long term prediction.}

\rev{Usually, such prediction is exploited to optimize the scheduling, i.e., resource allocation over time or frequency. Convex and linear optimization are often used when prediction is assumed to be perfect. In contrast, Markov models are applied when a probabilistic forecasting is available. 
Despite the great benefits that link context can potentially bring to resource (and more generally network) optimization, today's networks do not yet have the proper infrastructure to collect, share, process and distribute link context. Furthermore, proper methods are needed not only to gather data from users, but also, to discard irrelevant or redundant measurements as well as to handle sparsity or gaps in the collected data.  
}
\rev{
\subsubsection{Traffic context}
Traffic and throughput prediction has a concrete impact on the optimization of different services of different networks at different time scales.}

\rev{Network-wide and for long time scales, linear time series models are already used to predict the macroscopic traffic patterns of mobile radio cells for medium/long-term management and optimization of the radio resources. At faster time scales and for specific radio cells or groups of radio cells, the probabilistic forecasting of the upcoming traffic, e.g., by using Markovian models, can be exploited to solve short-term problems including the radio resource allocation among users and the cell assignment problem. }

\rev{Throughput prediction tools are then naturally coupled with video streaming services in mobile radio networks which have embedded rate adaptation capabilities. In this context, a good practice is to use simple yet effective look-ahead video throughput predictors based on time windows which are often coupled with clustering approaches to group similar video sessions. Deep learning techniques are also proposed to predict the throughput of video sessions, which offer improved performance at the price of a much higher complexity. }     

\rev{The data coming from traffic/throughput prediction can be effectively coupled with application/scenario-specific optimization frameworks. When targeting network-wide efficiency, centralized optimization approaches seem to be superior and more widely used. As an example, the problem of radio resource allocation in mobile radio networks is effectively representable and solvable though convex optimization techniques in semi-real-time scenario. In contrast, when the optimization has to be performed with the granularity of the technology-specific time slot, sub-optimal heuristics are preferable. Besides resorting to optimization approaches, control theoretic modeling is extremely powerful in all those cases where the optimization objective includes traffic (and queue) stability.  
}
\rev{
\subsubsection{Social context}
We can conclude that leveraging the social context of data transmission results in gains for proactive caching of multimedia content and can improve resource allocation by predicting the social behavior of users. For the former, determining the popularity of content plays a crucial role. Collaborative filtering is a well-known approach for this purpose. However, due to the heavy tail nature of content popularity, trying to use this kind of models for a broad class of content will usually not lead to good results. However, for more specific and limited classes of content, i.e., localized advertisement, where a particular item is likely to be requested by a large number of users, popularity prediction is an appealing solution. In general, proactive caching requires that content is stored on caches close to the edge network in order not to put excessive load on the core network. For optimizing resource allocation using social behavior, the social interaction of different users can be used to create social graphs that determine the level of activity of each user and thereby make it possible to predict the amount of resources each user will need. Network utility maximization and heuristic methods are the most popular techniques for this context. Due to the complexity of modeling the social behavior of users, they are useful for wireless networks that either expose a great deal of measurable social interaction (device-to-device communication, dense cellular networks with small cells, local wireless networks in a sports stadium), or when resources are very scarce.}

\subsection{Anticipation-enabled use cases}
\label{sec:chal:opp}

Future networks are envisioned to cater to a large variety of new services and applications. Broadband access in dense areas, massive sensor networks, tactile Internet and ultra-reliable communications are only a few of the use cases detailed in~\cite{NGMN}. The network capabilities of today's systems (i.e., 4G systems) are not able to support such requirements. Therefore, 5G systems will be designed to guarantee an efficient and flexible use (and sharing) of wireless resources, supported by a native software defined network and/or network function virtualization architecture~\cite{NGMN}. 
Big data analysis and context awareness are not only enablers for new value added services but, combined with the power of anticipatory optimization, can play a role in the 5G technology. 

\subsubsection{Mobility management} Network densification will be used in 5G systems in order to cope with the tremendous growth of traffic volume. As a drawback, mobility management will become more difficult. Additionally, it is foreseen that mobility in 5G will be on-demand~\cite{NGMN}, i.e., provided for and customized to the specific service that needs it. In this sense, being able to predict the user's context (e.g., requested service) and his mobility behavior can be extremely useful in order to speed up handover procedures and to enable seamless connectivity. Furthermore, since individual mobility is highly social, social context and mobility information will be jointly used to perform predictions for a group of socially related individuals. 


\subsubsection{Network sharing} 5G systems will support resource and network sharing among different stakeholders, e.g., operators, infrastructure providers, service providers. The effectiveness of such sharing mechanisms relies on the ability of each player to predict the evolution of his own network, e.g., expected network load, anticipated user's link quality and prediction of the requested services. Wireless sharing mechanisms can strongly benefit from the added value provided by anticipation, especially when prediction is available at fine granularity, e.g., in a multi-operator scheduler~\cite{malanchini2016wireless}.

\subsubsection{Extreme real-time communications} Tactile Internet is only one of the applications that will require a very low latency (i.e., in the order of some milliseconds). Allocating resources and guaranteeing such low end-to-end delay will be very challenging. 5G systems will support such requirements by means of a new physical layer (e.g., a new air interface). However, this will not be enough if not combined with context information used to prioritize control information (e.g., used to move virtual or real objects in real time) over content~\cite{fettweis2014tactile}. Knowledge about the information that is transmitted and its specific requirements will be crucial in order to assign priorities and meet the expected quality-of-experience in a combined effort of physical and higher layers. 

\subsubsection{Ultra-reliable communications} Reliability is mentioned in several 5G white papers, e.g. in~\cite{NGMN}, as necessary prerequisite for  lifeline communications and e-health services, e.g., remote surgery. A recent work~\cite{suryaprakash2016reliability} proposed a quantified definition of reliability in  wireless access networks. As outlined here, a posteriori evaluation of the achieved reliability is not enough in order to meet the expected target, which in some cases is as high as $99.999\%$. To this end, it is mandatory to design resource allocation mechanisms that account for (and are able to anticipate the impact on) reliability in advance.  

\subsection{Open challenges}
\label{sec:chal:supp}
While the literature surveyed so far clearly points out how anticipatory networking can enhance current networks, this section discusses several problems that need to be solved for its wider adoption. \rev{In particular, we identified four functionalities that are going to play an important role in the adoption of anticipatory networking in 5G networks:
\begin{itemize}
  \item {\bf Measurements and information collection:} in order to provide means to obtain and share context information, future networks need to provide trusted mechanisms to manage the information exchange. 
  \item {\bf Data analysis and prediction:} information databases need interoperable procedures to make sure that processing and forecasting tools are usable with many possible information sources	.
  \item {\bf Optimization and decision making:} data and procedures are then exploited to derive system management policies. 
  \item {\bf Execution:} finally, in contrast to current procedures, anticipatory execution engines need to take into account the impact of the decisions made in the past and re-evaluate their costs and rewards in hindsight of the actual evolution of the system.
\end{itemize}
For instance, scheduling and load balancing are two processes that greatly profit from anticipatory networking and cannot be realized without a comprehensive integration of the four aforementioned functionalities in future generation networks. The realization of these functionalities poses the following important challenges.
}

\subsubsection{Privacy and security} In our opinion, one of the main hindrances for anticipatory networking to become part of next generation networks is related to how users feel about sharing data and being profiled. While voluntarily sharing personal information has become a daily habit, many disapprove that companies create profiles using their data~\cite{singer2015sharing}. In a similar way, there might be a strong resistance against a new technology that, even though in an anonymous way, collects and analyzes users' behavior to anticipate users' decisions. 
Standards and procedures need to be studied to enforce users' privacy, data anonymity and an adequate security level for information storage.
\rev{In addition, data ownership and control need to be defined and regulated in order to allow users and providers to interact in a trusted environment, where the former can decide the level of information disclosure and the latter can operate within shared agreements.}

\subsubsection{Network functions and interfaces} Many of the applications that are likely to benefit from anticipatory networking capabilities (i.e. decision making and execution) require unprecedented interactions among information producers, analyzers and consumers. A simple example is provided by predictive media streaming optimizers, which need to obtain content information from the related database and user streaming information from the user and/or the network operator. This information is then analyzed and fed to a streaming provider that optimizes its service accordingly. While ad hoc services can be realized exploiting the current networking functionalities, next generation applications, such as the extreme real-time communications mentioned above, will greatly benefit from a tighter coupling between context information and communication interfaces. \rev{We believe that the potential of anticipatory functionalities can be used in communication system and they could be applied to other domains, such as public transportation and smart city management.}

\subsubsection{Next generation architecture} 5G networks are currently being discussed and, while much attention is paid to increasing the network capacity and virtualizing the network functions, we believe that the current infrastructure should be enhanced with repositories for context information and application profiles~\cite{wan2014context} to assist the realization of novel predictive applications. As per the previous concerns above, sharing sensible information, even in an anonymized way, will require particular care in terms of users' privacy and database accessibility. \rev{We believe that anticipatory networking can potentially improve every kind of mobile networks: cellular networks will likely be the first to exploit this paradigm, because they already own the information needed to enable the predictive frameworks and it is only a matter of time and regulations to make it a reality. Once it will be integrated in cellular networks, other systems, such as public WiFi deployments, device-to-device solutions and the Internet of Things, will be able to participate in the infrastructure to exploit forecasting functionalities; in particular, we believe this will be applied to smart cities and multi-modal transportation.}
\rev{\subsubsection{Impact of prediction errors}
When making and using predictions, one should carefully estimate its accuracy, which is itself a challenge. It might be potentially more harmful to use a wrong prediction than not using prediction at all. Usually, a good accuracy can be obtained for a short prediction horizon, which, however, should not be too short, otherwise the optimization algorithms cannot benefit from it. Therefore, a good balance between prediction horizon and accuracy must be found in order to provide gains. In contrast, over medium/long term periods, metrics can usually be predicted in terms of statistical behavior only. 
Furthermore, to build robust algorithms that are able to deal with uncertainties, proper prediction error models should be derived. In the existing literature, uncertainties are mainly modeled as Gaussian random variables. Despite the practicability of such an assumption, more complex error models should be derived to take into account the source (e.g., location and/or channel quality) as well as the cause (e.g., GPS accuracy and/or fast fading effect) of errors.}

\rev{\section{Conclusions}}
\label{sec:conclusions}

\rev{This survey analyzed the literature on anticipatory networking for mobile networks. We provided a thorough analysis of application scenarios categorized by the contextual information used to build the predictive framework. The most relevant prediction and optimization techniques adopted in the literature have been described and commented in two handbooks that have the twofold objective of supporting researchers to advance in the field and providing standardization and regulation bodies with a common ground on anticipatory networking solutions. While the core of this survey is devoted to mobile cellular networks, we also analyzed applicability and advantages of anticipatory networking solution to other types of wireless networks and at the different layers of the protocol stack.
Finally, we analyzed benefits and disadvantages of the proposed solutions, the most promising application scenarios for 5G networks, and the challenges that are yet to be faced to adopt anticipatory networking paradigms.}

\rev{To conclude, while the literature reviewed in this works suggests that anticipatory networking is a quite mature approach to improve the performance of mobile networks, we believe that issues (mainly at the system level) still need to be solved to realize its potential.
In particular, most of the work which has been evaluated in this survey tends to focus on the benefit of anticipation, while overlooking possible problems and disadvantages in the anticipatory networking framework. }

\rev{All the main components of anticipatory networking, the context database and the prediction/anticipation intelligence,  must be effectively integrated into the mobile network architecture which poses challenges at different levels. First, new interfaces and communication paradigms must be defined for data collection from both end users and sources external to the mobile network itself; second, the management of the context databases brings an additional burden in terms of required bandwidth and processing power for several network elements which may lead to scalability issues as well as security and privacy concerns. }
\rev{To this extent, a thorough and comprehensive cost-benefit analysis for specific anticipatory networking scenarios is, in our opinion, a required next step for the research in the field. }


\ifCLASSOPTIONcaptionsoff
  \newpage
\fi

\rev{\section{List of Acronyms}
\begin{acronym}[ConvOpt]
    \acro{16-QAM}{16 Quadrature Amplitude Modulation}
    \acro{64-QAM}{64 Quadrature Amplitude Modulation}
    \acro{256-QAM}{256 Quadrature Amplitude Modulation}
    \acro{ACF}{Autocorrelation Function}
    \acro{AGRS}{Anticipatory Generalized Rate Scheduling}
    \acro{AMOS}{Anticipatory Multi-Operator Scheduling}
    \acro{ANN}{Artificial Neural Network}
    \acro{AR}{AutoRegressive}
    \acro{ARIMA}{AutoRegressive Integrated and Moving Average}
    \acro{ARMA}{AutoRegressive and Moving Average}
    \acro{ARQ}{Automatic Repeat Request}
    \acro{aDA}{advanced Dynamic Algorithm}
    \acro{ADC}{Analog-to-Digital Converter}
    \acro{APP}{Application layer}
    \acro{ASIC}{Application Specific Integrated Circuits}
    \acro{ATM}{Asynchronous Transfer Mode}
    \acro{AWGN}{Additive White Gaussian Noise}
    \acro{BER}{Bit Error Rate}
    \acro{BPSK}{Binary Phase Shift Keying}
    \acro{BS}{Base Station}
    \acro{BTS}{Base Transceiver Station}
    \acro{CBR}{Constant Bit Rate}
    \acro{CC}{Chase Combining}
    \acro{CCN}{Content Centric Network}
    \acro{CDF}{Cumulative Distribution Function}
    \acro{CDMA}{Code-Division-Multiple-Access}
    \acro{CF}{Collaborative Filtering}
    \acro{CIC}{Cascaded Integrator-Comb}
    \acro{CIF}{Common Intermediate Format}
    \acro{CNR}{Channel Gain-to-Noise Ratio}
    \acro{ConvOpt}{Convex Optimization}
    \acro{CR}{Cognitive Radio}
    \acro{CRC}{Cyclic Redundancy Check}
    \acro{CSI}{Channel State Information}
    \acro{CQI}{Channel Quality Indicator}
    \acro{CS}{Carrier Sensing}
    \acro{CTM}{Continuous Time Markov}
    \acro{CTMC}{Continuous Time Markov Chain}
    \acro{D2D}{device-to-device}
    \acro{DAC}{Digital-to-Analog Converter}
    \acro{DASH}{Dynamic Adaptive Streaming over HTTP}
    \acro{DoF}{Degree-of-Freedom}
    \acro{DCT}{Discrete Cosine Transform}
    \acro{DIV}{Distortion In Interval}
    \acro{DLC}{Data Link Control layer}
    \acro{DSP}{Digital Signal Processor}
    \acro{DTM}{Discrete Time Markov}
    \acro{DTMC}{Discrete Time Markov Chain}
    \acro{EGC}{Equal Gain Combining}
    \acro{EKF}{Extended Kalman Filter}
    \acro{ELM}{Extreme Learning Machine}
    \acro{ETX}{Expected Transmission Count}
    \acro{EWMA}{Exponential Weighted Moving Average}
    \acro{FDMA}{Frequency Division Multiple Access}
    \acro{FEC}{Forward Error Correction}
    \acro{FER}{Frame Error Rate}
    \acro{FS}{Frame Selection}
    \acro{FIFO}{First-In-First-Out}
    \acro{FPGA}{Field Programmable Gate Array}
    \acro{FSC}{Frame Check Sequences}
    \acro{FTP}{File Transfer Protocol}
    \acro{GARCH}{Generalized AutoRegressive Conditionally Heteroskedastic}
    \acro{GMSK}{Gaussian Minimum Shift Keying}
    \acro{GRS}{Generalized Rate Scheduling}
    \acro{GP}{Gaussian Process}
    \acro{GPS}{Global Positioning System}
    \acro{GoP}{Group of Pictures}
    \acro{GSR}{GNU Software Radio}
    \acro{HMM}{Hidden Markov Models}
    \acro{HTTP}{Hypertext Transfer Protocol}
    \acro{HTML}{Hypertext Mark-up Language}
    \acro{ICI}{Inter-carrier Interference}
    \acro{ID}{identity}
    \acro{IEEE}{Institute of Electrical and Electronics Engineers, Inc.}
    \acro{ILP}{Integer Linear Programming}
    \acro{IoT}{Internet-of-Things}
    \acro{IP}{Internet Protocol}
    \acro{IPC}{Inter-Process Communication}
    \acro{ISI}{Inter-symbol Interference}
    \acro{ISM}{industrial, scientific and medical}
    \acro{KKF}{Kriged Kalman Filter}
    \acro{KPI}{Key Performance Indicator}
    \acro{LTE}{Long Term Evolution}
    \acro{LLC}{Logical Link Control layer}
    \acro{LOS}{Line Of Sight}
    \acro{LP}{Linear Programming}
    \acro{LZ}{Lempel-Ziv}
    \acro{M2M}{Machine-to-Machine}
    \acro{MA}{Moving Average}
    \acro{MAC}{Medium Access Control}
    \acro{MANET}{Mobile Ad-hoc Networks}
    \acro{MC}{Markov Chain}
    \acro{MCM}{Multi Carrier Modulation}
    \acro{MIMO}{Multiple-Input Multiple-Output}
    \acro{MISO}{Multiple-Input Single-Output}
    \acro{MILP}{Mixed-Integer Linear Programming}
    \acro{MNLP}{Mixed Non-Linear Program}
    \acro{MOS}{Multi-Operator Scheduling}
    \acro{MPC}{Model Predictive Control}
    \acro{MDP}{Markov Decision Process}
    \acro{MPEG}{Moving Pictures Expert Group}
    \acro{MRC}{Maximum Ratio Combining}
    \acro{MRP}{Markov Renewal Process}
    \acro{SC}{Selection Combining}
    \acro{MSB}{Most Significant Bit}
    \acro{MSS}{Maximum Segment Size}
    \acro{MTU}{Maximum Transmission Unit}
    \acro{NAV}{Network Allocation Vector}
    \acro{NCR}{Non-Cooperative Relaying}
    \acro{NFV}{Network Function Virtualization}
    \acro{NLOS}{Non-Line Of Sight}
    \acro{NPS}{Network Path Selection}
    \acro{OFDM}{Orthogonal Frequency Division Multiplexing}
    \acro{OS}{Operating System}
    \acro{OR}{Opportunistic Relaying\slash{}Routing}
    \acro{PCA}{Principal Component Analysis}
    \acro{PDF}{Probability Density Function}
    \acro{PDU}{Protocol Data Unit}
    \acro{PER}{Packet Error Rate}
    \acro{PF}{Proportionally Fair}
    \acro{PHY}{Physical layer}
    \acro{PPT}{point-to-point}
    \acro{PRB}{Physical Resource Block}
    \acro{PSC}{Packet Selection Combining}
    \acro{PSNR}{Peak Signal-to-Noise Ratio}
    \acro{QCIF}{Quarter CIF}
    \acro{QoE}{Quality-of-Experience}
    \acro{QoS}{Quality-of-Service}
    \acro{QPSK}{Quadrature Phase Shift Keying}
    \acro{RAN}{Radio Access Network}
    \acro{REM}{Radio Environment Map}
    \acro{RCPC}{Rate-Compatible Punctured Convolutional}
    \acro{RF}{Radio Frequency}
    \acro{RMS}{root mean square}
    \acro{RRM}{Radio Resource Management}
    \acro{RTT}{Round Trip Time}
    \acro{SDF}{Selection Decode-and-Forward}
    \acro{SDR}{Software Defined Radio}
    \acro{SR}{Software Radio}
    \acro{SEP}{Symbol Error Probability}
    \acro{SCM}{Single Carrier Modulation}
    \acro{SDC}{Selection Diversity Combining}
    \acro{SDN}{Software Defined Network}
    \acro{SIFS}{Short Inter-Frame Space}
    \acro{SMDP}{Semi-Markov Decision Process}
    \acro{SNR}{Signal-to-Noise Ratio}
    \acro{SINR}{Signal-to-Interference-plus-Noise Ratio}
    \acro{SVM}{Support Vector Machine}
    \acro{TCP}{Transmission Control Protocol}
    \acro{TDMA}{Time Division Multiple Access}
    \acro{TC}{Topological Coordinate}
    \acro{TCP}{Transport Control Protocol}
    \acro{TPM}{Topology Preserving Map}
    \acro{UDP}{User Datagram Protocol}
    \acro{USRP}{Universal Software Radio Peripheral}
    \acro{VBR}{Variable Bit Rate}
    \acro{VFA}{Value Function Approximation}
    \acro{VQM}{Video Queue Management}
    \acro{WCMDP}{Weakly Coupled \ac{MDP}}
    \acro{WLAN}{Wireless Local Area Network}
    \acro{WMAN}{Wireless Metropolitan Area Network}
    \acro{WSN}{Wireless Sensor Network}
    \acro{WT}{Wireless Terminal}
    \acro{WWW}{World-Wide-Web}
\end{acronym}}

\bibliographystyle{IEEEtran}
\bibliography{main}

\begin{thebibliography}{100}
\providecommand{\url}[1]{#1}
\csname url@samestyle\endcsname
\providecommand{\newblock}{\relax}
\providecommand{\bibinfo}[2]{#2}
\providecommand{\BIBentrySTDinterwordspacing}{\spaceskip=0pt\relax}
\providecommand{\BIBentryALTinterwordstretchfactor}{4}
\providecommand{\BIBentryALTinterwordspacing}{\spaceskip=\fontdimen2\font plus
\BIBentryALTinterwordstretchfactor\fontdimen3\font minus
  \fontdimen4\font\relax}
\providecommand{\BIBforeignlanguage}[2]{{%
\expandafter\ifx\csname l@#1\endcsname\relax
\typeout{** WARNING: IEEEtran.bst: No hyphenation pattern has been}%
\typeout{** loaded for the language `#1'. Using the pattern for}%
\typeout{** the default language instead.}%
\else
\language=\csname l@#1\endcsname
\fi
#2}}
\providecommand{\BIBdecl}{\relax}
\BIBdecl

\bibitem{zheng2016big}
K.~Zheng, Z.~Yang, K.~Zhang, P.~Chatzimisios, K.~Yang, and W.~Xiang, ``Big
  data-driven optimization for mobile networks toward {5G},'' \emph{IEEE
  Network}, vol.~30, no.~1, pp. 44--51, 2016.

\bibitem{makris2013survey}
P.~Makris, D.~N. Skoutas, and C.~Skianis, ``{A survey on context-aware mobile
  and wireless networking: On networking and computing environments'
  integration},'' \emph{IEEE Communications Surveys \& Tutorials}, vol.~15,
  no.~1, pp. 362--386, 2013.

\bibitem{pejovic2015anticipatory}
V.~Pejovic and M.~Musolesi, ``{Anticipatory mobile computing: A survey of the
  state of the art and research challenges},'' \emph{ACM Computing Surveys
  (CSUR)}, vol.~47, no.~3, p.~47, 2015.

\bibitem{boucheron2005theory}
S.~Boucheron, O.~Bousquet, and G.~Lugosi, ``Theory of classification: A survey
  of some recent advances,'' \emph{ESAIM: probability and statistics}, vol.~9,
  pp. 323--375, 2005.

\bibitem{liu2015empirical}
Y.~Liu and J.~Y. Lee, ``An empirical study of throughput prediction in mobile
  data networks,'' in \emph{IEEE Global Communications Conference (GLOBECOM)},
  2015, pp. 1--6.

\bibitem{nguyen2008survey}
T.~T. Nguyen and G.~Armitage, ``A survey of techniques for internet traffic
  classification using machine learning,'' \emph{IEEE Communications Surveys \&
  Tutorials}, vol.~10, no.~4, pp. 56--76, 2008.

\bibitem{jin2013understanding}
L.~Jin, Y.~Chen, T.~Wang, P.~Hui, and A.~V. Vasilakos, ``Understanding user
  behavior in online social networks: A survey,'' \emph{IEEE Communications
  Magazine}, vol.~51, no.~9, pp. 144--150, 2013.

\bibitem{barakovic2013survey}
S.~Barakovi{\'c} and L.~Skorin-Kapov, ``{Survey and challenges of QoE
  management issues in wireless networks},'' \emph{Hindawi Journal of Computer
  Networks and Communications}, 2013.

\bibitem{hoyhtya2016spectrum}
M.~H{\"o}yhty{\"a}, A.~M{\"a}mmel{\"a}, M.~Eskola, M.~Matinmikko,
  J.~Kalliovaara, J.~Ojaniemi, J.~Suutala, R.~Ekman, R.~Bacchus, and
  D.~Roberson, ``Spectrum occupancy measurements: A survey and use of
  interference maps,'' \emph{IEEE Communications Surveys \& Tutorials},
  vol.~18, no.~4, pp. 2386--2414, 2016.

\bibitem{chen2016survey}
Y.~Chen and H.-S. Oh, ``A survey of measurement-based spectrum occupancy
  modeling for cognitive radios,'' \emph{IEEE Communications Surveys \&
  Tutorials}, vol.~18, no.~1, pp. 848--859, 2016.

\bibitem{song2010limits}
C.~Song, Z.~Qu, N.~Blumm, and A.-L. Barab{\'a}si, ``{Limits of predictability
  in human mobility},'' \emph{Science}, vol. 327, no. 5968, pp. 1018--1021,
  2010.

\bibitem{lu2013approaching}
X.~Lu, E.~Wetter, N.~Bharti, A.~J. Tatem, and L.~Bengtsson, ``{Approaching the
  limit of predictability in human mobility},'' \emph{Nature Scientific
  reports}, vol.~3, 2013.

\bibitem{jiang2013tracking}
Y.~Jiang, D.~C. Dhanapala, and A.~P. Jayasumana, ``{Tracking and prediction of
  mobility without physical distance measurements in sensor networks},'' in
  \emph{{IEEE International Conference on Communications (ICC)}}, 2013, pp.
  1845--1850.

\bibitem{ghouti2013mobility}
L.~Ghouti, T.~R. Sheltami, and K.~S. Alutaibi, ``{Mobility prediction in mobile
  ad hoc networks using extreme learning machines},'' \emph{Procedia Computer
  Science}, vol.~19, pp. 305--312, 2013.

\bibitem{chen2013predicting}
X.~Chen, F.~M{\'e}riaux, and S.~Valentin, ``{Predicting a user's next cell with
  supervised learning based on channel states},'' in \emph{{IEEE Signal
  Processing Advances in Wireless Communications (SPAWC)}}, 2013, pp. 36--40.

\bibitem{xiong2014mpaas}
H.~Xiong, D.~Zhang, D.~Zhang, V.~Gauthier, K.~Yang, and M.~Becker, ``{{MPaaS}:
  Mobility prediction as a service in telecom cloud},'' \emph{Springer
  Information Systems Frontiers}, vol.~16, no.~1, pp. 59--75, 2014.

\bibitem{lee2006modeling}
J.-K. Lee and J.~C. Hou, ``{Modeling steady-state and transient behaviors of
  user mobility: formulation, analysis, and application},'' in \emph{{ACM
  international symposium on Mobile ad hoc networking and computing
  (MobiHoc)}}, 2006, pp. 85--96.

\bibitem{abu2010application}
H.~Abu-Ghazaleh and A.~S. Alfa, ``{Application of mobility prediction in
  wireless networks using Markov renewal theory},'' \emph{IEEE Transactions on
  Vehicular Technology}, vol.~59, no.~2, pp. 788--802, 2010.

\bibitem{barth2011mobility}
D.~Barth, S.~Bellahsene, and L.~Kloul, ``{Mobility prediction using mobile user
  profiles},'' in \emph{{IEEE Modeling, Analysis \& Simulation of Computer and
  Telecommunication Systems (MASCOTS)}}, 2011, pp. 286--294.

\bibitem{barth2012combining}
------, ``{Combining local and global profiles for mobility prediction in LTE
  femtocells},'' in \emph{{ACM Modeling, analysis and simulation of wireless
  and mobile systems (MSWiM)}}, 2012, pp. 333--342.

\bibitem{gidofalvi2012and}
G.~Gid{\'o}falvi and F.~Dong, ``{When and where next: Individual mobility
  prediction},'' in \emph{{ACM SIGSPATIAL International Workshop on Mobile
  Geographic Information Systems}}, 2012, pp. 57--64.

\bibitem{chon2013understanding}
Y.~Chon, N.~D. Lane, Y.~Kim, F.~Zhao, and H.~Cha, ``{Understanding the coverage
  and scalability of place-centric crowdsensing},'' in \emph{{ACM international
  joint conference on Pervasive and ubiquitous computing (Ubicomp)}}, 2013, pp.
  3--12.

\bibitem{chon2012evaluating}
Y.~Chon, H.~Shin, E.~Talipov, and H.~Cha, ``{Evaluating mobility models for
  temporal prediction with high-granularity mobility data},'' in \emph{{IEEE
  Pervasive Computing and Communications (PerCom)}}, 2012, pp. 206--212.

\bibitem{chon2014smartdc}
Y.~Chon, E.~Talipov, H.~Shin, and H.~Cha, ``{{SmartDC}: Mobility
  prediction-based adaptive duty cycling for everyday location monitoring},''
  \emph{IEEE Transactions on Mobile Computing}, vol.~13, no.~3, pp. 512--525,
  2014.

\bibitem{chon2014adaptive}
Y.~Chon, Y.~Kim, H.~Shin, and H.~Cha, ``Adaptive duty cycling for place-centric
  mobility monitoring using zero-cost information in smartphone,'' \emph{IEEE
  Transactions on Mobile Computing}, vol.~13, no.~8, pp. 1694--1706, 2014.

\bibitem{chon2011mobility}
Y.~Chon, E.~Talipov, H.~Shin, and H.~Cha, ``{Mobility prediction-based
  smartphone energy optimization for everyday location monitoring},'' in
  \emph{{ACM conference on embedded networked sensor systems (SenSys)}}, 2011,
  pp. 82--95.

\bibitem{akyildiz2004predictive}
I.~F. Akyildiz and W.~Wang, ``The predictive user mobility profile framework
  for wireless multimedia networks,'' \emph{IEEE/ACM Transactions on Networking
  (TON)}, vol.~12, no.~6, pp. 1021--1035, 2004.

\bibitem{scellato2011nextplace}
S.~Scellato, M.~Musolesi, C.~Mascolo, V.~Latora, and A.~T. Campbell,
  ``{Nextplace: a spatio-temporal prediction framework for pervasive
  systems},'' in \emph{{Springer Pervasive Computing}}, 2011, vol. 6696, pp.
  152--169.

\bibitem{de2013interdependence}
M.~De~Domenico, A.~Lima, and M.~Musolesi, ``{Interdependence and predictability
  of human mobility and social interactions},'' \emph{Elsevier Pervasive and
  Mobile Computing}, vol.~9, no.~6, pp. 798--807, 2013.

\bibitem{fazio2016pattern}
P.~Fazio, M.~Tropea, F.~De~Rango, and M.~Voznak, ``Pattern prediction and
  passive bandwidth management for hand-over optimization in {QoS} cellular
  networks with vehicular mobility,'' \emph{IEEE Transactions on Mobile
  Computing}.

\bibitem{abouzeid2015evaluating}
H.~Abou-zeid, H.~S. Hassanein, Z.~Tanveer, and N.~AbuAli, ``Evaluating mobile
  signal and location predictability along public transportation routes,'' in
  \emph{IEEE Wireless Communications and Networking Conference (WCNC)}, 2015,
  pp. 1195--1200.

\bibitem{yang2013broadcasting}
J.~Yang and Z.~Fei, ``{Broadcasting with prediction and selective forwarding in
  vehicular networks},'' \emph{Hindawi International journal of distributed
  sensor networks}, vol. 2013.

\bibitem{sridaran2013location}
A.~Sridharan and J.~Bolot, ``{Location patterns of mobile users: A large-scale
  study},'' in \emph{{IEEE INFOCOM}}, 2013, pp. 1007--1015.

\bibitem{froehlich2008route}
J.~Froehlich and J.~Krumm, ``{Route prediction from trip observations},'' SAE
  Technical Paper, Tech. Rep., 2008.

\bibitem{monreale2009wherenext}
A.~Monreale, F.~Pinelli, R.~Trasarti, and F.~Giannotti, ``{WhereNext: a
  location predictor on trajectory pattern mining},'' in \emph{{ACM
  international conference on Knowledge discovery and data mining (SIGKDD)}},
  2009, pp. 637--646.

\bibitem{geopkdd}
\BIBentryALTinterwordspacing
``{{GeoPKDD: Geographic Privacy-aware Knowledge Discovery and Delivery}},''
  2005-2008. [Online]. Available: \url{http://www.geopkdd.eu}
\BIBentrySTDinterwordspacing

\bibitem{bui2014model}
N.~Bui, F.~Michelinakis, and J.~Widmer, ``{A Model for Throughput Prediction
  for Mobile Users},'' in \emph{{European Wireless 2014}}, 2014, pp. 1--6.

\bibitem{liao2015channel}
Q.~Liao, S.~Valentin, and S.~Stanczak, ``{Channel gain prediction in wireless
  networks based on spatial-temporal correlation},'' in \emph{{IEEE Signal
  Processing Advances in Wireless Communications (SPAWC)}}, 2015, pp. 400--404.

\bibitem{momentum}
\BIBentryALTinterwordspacing
``{MOMENTUM, “MOdels and siMulations for nEtwork plaNning and conTrol of
  UMts},'' 2004. [Online]. Available: \url{http://www.zib.de/momentum}
\BIBentrySTDinterwordspacing

\bibitem{wanalertlak2011behavior}
W.~Wanalertlak, B.~Lee, C.~Yu, M.~Kim, S.-M. Park, and W.-T. Kim,
  ``{Behavior-based mobility prediction for seamless handoffs in mobile
  wireless networks},'' \emph{Springer Wireless Networks}, vol.~17, no.~3, pp.
  645--658, 2011.

\bibitem{riiser2012video}
H.~Riiser, T.~Endestad, P.~Vigmostad, C.~Griwodz, and P.~Halvorsen, ``{Video
  streaming using a location-based bandwidth-lookup service for bitrate
  planning},'' \emph{ACM Transactions on Multimedia Computing, Communications,
  and Applications (TOMM)}, vol.~8, no.~3, pp. 24:1--24:19, 2012.

\bibitem{lu2013optimizing}
Z.~Lu and G.~De~Veciana, ``{Optimizing stored video delivery for mobile
  networks: The value of knowing the future},'' in \emph{{IEEE INFOCOM}}, 2013,
  pp. 2706--2714.

\bibitem{abouzeid2013optimal}
H.~Abou-zeid, H.~S. Hassanein, and S.~Valentin, ``{Optimal predictive resource
  allocation: Exploiting mobility patterns and radio maps},'' in \emph{IEEE
  Global Communications Conference (GLOBECOM)}, 2013, pp. 4877--4882.

\bibitem{margolies2014exploiting}
R.~Margolies, A.~Sridharan, V.~Aggarwal, R.~Jana, N.~Shankaranarayanan,
  V.~Vaishampayan, G.~Zussman \emph{et~al.}, ``{Exploiting mobility in
  proportional fair cellular scheduling: Measurements and algorithms},'' in
  \emph{{IEEE INFOCOM}}, 2014, pp. 1339--1347.

\bibitem{siris2013enhancing}
V.~A. Siris and D.~Kalyvas, ``{Enhancing mobile data offloading with mobility
  prediction and prefetching},'' \emph{ACM SIGMOBILE Mobile Computing and
  Communications Review}, vol.~17, no.~1, pp. 22--29, 2013.

\bibitem{hao2014gtube}
J.~Hao, R.~Zimmermann, and H.~Ma, ``{Gtube: Geo-predictive video streaming over
  http in mobile environments},'' in \emph{{ACM Multimedia Systems Conference
  (MMSys)}}, 2014, pp. 259--270.

\bibitem{tie2011anticipatory}
X.~Tie, A.~Seetharam, A.~Venkataramani, D.~Ganesan, and D.~L. Goeckel,
  ``{Anticipatory wireless bitrate control for blocks},'' in \emph{{ACM
  COnference on emerging Networking EXperiments and Technologies (CoNEXT)}},
  2011, p.~9.

\bibitem{piacentini2010path}
M.~Piacentini and F.~Rinaldi, ``{Path loss prediction in urban environment
  using learning machines and dimensionality reduction techniques},''
  \emph{Springer Computational Management Science}, vol.~8, no.~4, pp.
  371--385, 2011.

\bibitem{dallanese2011channel}
E.~Dall'Anese, S.-J. Kim, and G.~B. Giannakis, ``{Channel gain map tracking via
  distributed Kriging},'' \emph{IEEE Transactions on Vehicular Technology},
  vol.~60, no.~3, pp. 1205--1211, 2011.

\bibitem{yin2011prediction}
S.~Yin, D.~Chen, Q.~Zhang, and S.~Li, ``{Prediction-based throughput
  optimization for dynamic spectrum access},'' \emph{IEEE Transactions on
  Vehicular Technology}, vol.~60, no.~3, pp. 1284--1289, 2011.

\bibitem{tarsa2015taming}
S.~J. Tarsa, M.~Comiter, M.~B. Crouse, B.~McDanel, and H.~Kung, ``{Taming
  Wireless Fluctuations by Predictive Queuing Using a Sparse-Coding Link-State
  Model},'' in \emph{{ACM international symposium on Mobile ad hoc networking
  and computing (MobiHoc)}}, 2015, pp. 287--296.

\bibitem{kasparick2015kernel}
M.~Kasparick, R.~L. Cavalcante, S.~Valentin, S.~Stanczak, and M.~Yukawa,
  ``{Kernel-based adaptive online reconstruction of coverage maps with side
  information},'' \emph{IEEE Transactions on Vehicular Technology}, vol.~65,
  no.~7, pp. 5461--5473, 2015.

\bibitem{nicholson2008breadcrumbs}
A.~J. Nicholson and B.~D. Noble, ``{Breadcrumbs: forecasting mobile
  connectivity},'' in \emph{{ACM international conference on Mobile computing
  and networking (MobiCom)}}, 2008, pp. 46--57.

\bibitem{naimi2014anticipation}
S.~Naimi, A.~Busson, V.~V{\`e}que, L.~B.~H. Slama, and R.~Bouallegue,
  ``{Anticipation of ETX metric to manage mobility in ad hoc wireless
  networks},'' in \emph{{Springer Ad-hoc, Mobile, and Wireless Networks}},
  2014, pp. 29--42.

\bibitem{muppirisetty2015spatial}
L.~S. Muppirisetty, T.~Svensson, and H.~Wymeersch, ``{Spatial wireless channel
  prediction under location uncertainty},'' \emph{IEEE Transactions on Wireless
  Communications}, vol.~15, no.~2, pp. 1031--1044, 2016.

\bibitem{muppirisetty2016channel}
M.~Fr, L.~S. Muppirisetty, H.~Wymeersch \emph{et~al.}, ``{Channel gain
  prediction for multi-agent networks in the presence of location
  uncertainty},'' in \emph{IEEE International Conference on Acoustics, Speech
  and Signal Processing (ICASSP)}, 2016, pp. 3911--3915.

\bibitem{muppirisetty2015proactive}
L.~S. Muppirisetty, J.~Tadrous, A.~Eryilmaz, and H.~Wymeersch, ``On proactive
  caching with demand and channel uncertainties,'' in \emph{IEEE Conference on
  Communication, Control, and Computing (Allerton)}, 2015, pp. 1174--1181.

\bibitem{bui2015mobile}
N.~Bui and J.~Widmer, ``{Mobile network resource optimization under imperfect
  prediction},'' in \emph{{IEEE World of Wireless, Mobile and Multimedia
  Networks (WoWMoM)}}, 2015, pp. 1--9.

\bibitem{wang2013ames}
X.~Wang, M.~Chen, T.~T. Kwon, L.~Yang, and V.~Leung, ``{AMES-Cloud: a framework
  of adaptive mobile video streaming and efficient social video sharing in the
  clouds},'' \emph{IEEE Transactions on Multimedia}, vol.~15, no.~4, pp.
  811--820, 2013.

\bibitem{bao2015bitrate}
W.~Bao and S.~Valentin, ``{Bitrate adaptation for mobile video streaming based
  on buffer and channel state},'' in \emph{{IEEE International Conference on
  Communications (ICC)}}, 2015, pp. 3076--3081.

\bibitem{seetharam2015managing}
A.~Seetharam, P.~Dutta, V.~Arya, J.~Kurose, M.~Chetlur, and S.~Kalyanaraman,
  ``{On managing quality of experience of multiple video streams in wireless
  networks},'' \emph{IEEE Transactions on Mobile Computing}, vol.~14, no.~3,
  pp. 619--631, 2015.

\bibitem{hosseini2015not}
S.~A. Hosseini, F.~Fund, and S.~S. Panwar, ``{{(Not)} yet another policy for
  scalable video delivery to mobile users},'' in \emph{{ACM International
  Workshop on Mobile Video (MoVid)}}, 2015, pp. 17--22.

\bibitem{kurdoglu2016realtime}
E.~Kurdoglu, Y.~Liu, Y.~Wang, Y.~Shi, C.~Gu, and J.~Lyu, ``{Real-time bandwidth
  prediction and rate adaptation for video calls over cellular networks},'' in
  \emph{ACM International Conference on Multimedia Systems (MMSys)}, 2016,
  p.~12.

\bibitem{liu2016hop}
Z.~Liu and Y.~Wei, ``Hop-by-hop adaptive video streaming in content centric
  network,'' in \emph{IEEE International Conference on Communications (ICC)},
  2016, pp. 1--7.

\bibitem{blobel2015anticipatory}
M.~Dr{\"a}xler, J.~Blobel, and H.~Karl, ``Anticipatory download scheduling in
  wireless video streaming with uncertain data rate prediction,'' in \emph{IFIP
  Wireless and Mobile Networking Conference (WMNC)}, 2015, pp. 136--143.

\bibitem{tsilimantos2016anticipatory}
D.~Tsilimantos, A.~Nogales-G{\'o}mez, and S.~Valentin, ``{Anticipatory Radio
  Resource Management for Mobile Video Streaming with Linear Programming},'' in
  \emph{{IEEE International Conference on Communications (ICC)}}, 2016.

\bibitem{atawia2014robust}
R.~Atawia, H.~Abou-zeid, H.~S. Hassanein, and A.~Noureldin, ``Robust resource
  allocation for predictive video streaming under channel uncertainty,'' in
  \emph{IEEE Global Communications Conference (GLOBECOM)}, 2014, pp.
  4683--4688.

\bibitem{mangla2016video}
T.~Mangla, N.~Theera-Ampornpunt, M.~Ammar, E.~Zegura, and S.~Bagchi, ``Video
  through a crystal ball: effect of bandwidth prediction quality on adaptive
  streaming in mobile environments,'' in \emph{ACM International Workshop on
  Mobile Video (MoVid)}, 2016, p.~1.

\bibitem{atawia2015chance}
R.~Atawia, H.~Abou-zeid, H.~S. Hassanein, and A.~Noureldin,
  ``Chance-constrained qos satisfaction for predictive video streaming,'' in
  \emph{IEEE Local Computer Networks (LCN)}, 2015, pp. 253--260.

\bibitem{atawia2016joint}
------, ``Joint chance-constrained predictive resource allocation for
  energy-efficient video streaming,'' \emph{IEEE Journal on Selected Areas in
  Communications (JSAC)}, vol.~34, no.~5, pp. 1389--1404, 2016.

\bibitem{hossain2004link}
E.~Hossain and V.~K. Bhargava, ``Link-level traffic scheduling for providing
  predictive qos in wireless multimedia networks,'' \emph{IEEE Transactions on
  Multimedia}, vol.~6, no.~1, pp. 199--217, 2004.

\bibitem{abouzeid2014energy}
H.~Abou-zeid, H.~S. Hassanein, and S.~Valentin, ``{Energy-efficient adaptive
  video transmission: Exploiting rate predictions in wireless networks},''
  \emph{IEEE Transactions on Vehicular Technology}, vol.~63, no.~5, pp.
  2013--2026, 2014.

\bibitem{abouzeid2014efficient}
H.~Abou-zeid and H.~S. Hassanein, ``{Efficient lookahead resource allocation
  for stored video delivery in multi-cell networks},'' in \emph{{IEEE Wireless
  Communications and Networking Conference (WCNC)}}, 2014, pp. 1909--1914.

\bibitem{bui2015anticipatory}
N.~Bui, I.~Malanchini, and J.~Widmer, ``{Anticipatory admission control and
  resource allocation for media streaming in mobile networks},'' in \emph{{ACM
  Modeling, analysis and simulation of wireless and mobile systems (MSWiM)}},
  2015.

\bibitem{bui2015anticipatoryb}
N.~Bui, S.~Valentin, and J.~Widmer, ``{Anticipatory quality-resource allocation
  for multi-user mobile video streaming},'' in \emph{{IEEE Workshop on
  Communication and Networking Techniques for Contemporary Video (CNCTV)}},
  2015.

\bibitem{draxler2013cross}
M.~Dr{\"a}xler and H.~Karl, ``{Cross-layer scheduling for multi-quality video
  streaming in cellular wireless networks},'' in \emph{{IEEE International
  Wireless Communications and Mobile Computing Conference (IWCMC)}}, 2013, pp.
  1181--1186.

\bibitem{draxler2015smarterphones}
M.~Dr{\"a}xler, J.~Blobel, P.~Dreimann, S.~Valentin, and H.~Karl,
  ``{SmarterPhones: Anticipatory download scheduling for wireless video
  streaming},'' in \emph{{IEEE International Conference and Workshops on
  Networked Systems (NetSys)}}, 2015, pp. 1--8.

\bibitem{valentin2014anticipatory}
S.~Valentin, ``{Anticipatory resource allocation for wireless video
  streaming},'' in \emph{{IEEE International Conference on Communication
  Systems (ICCS)}}, 2014, pp. 107--111.

\bibitem{zou2015can}
X.~K. Zou, J.~Erman, V.~Gopalakrishnan, E.~Halepovic, R.~Jana, X.~Jin,
  J.~Rexford, and R.~K. Sinha, ``{Can accurate predictions improve video
  streaming in cellular networks?}'' in \emph{{ACM International Workshop on
  Mobile Computing Systems and Applications (HotMobile)}}, 2015, pp. 57--62.

\bibitem{xing2013spectrum}
X.~Xing, T.~Jing, W.~Cheng, Y.~Huo, and X.~Cheng, ``Spectrum prediction in
  cognitive radio networks,'' \emph{IEEE Wireless Communications}, vol.~20,
  no.~2, pp. 90--96, 2013.

\bibitem{wei2013construction}
Z.~Wei, Q.~Zhang, Z.~Feng, W.~Li, and T.~A. Gulliver, ``On the construction of
  radio environment maps for cognitive radio networks,'' in \emph{IEEE Wireless
  Communications and Networking Conference (WCNC)}, 2013, pp. 4504--4509.

\bibitem{yilmaz2013radio}
H.~B. Yilmaz, T.~Tugcu, F.~Alag{\"o}z, and S.~Bayhan, ``Radio environment map
  as enabler for practical cognitive radio networks,'' \emph{IEEE
  Communications Magazine}, vol.~51, no.~12, pp. 162--169, 2013.

\bibitem{thilina2013machine}
K.~M. Thilina, K.~W. Choi, N.~Saquib, and E.~Hossain, ``Machine learning
  techniques for cooperative spectrum sensing in cognitive radio networks,''
  \emph{IEEE Journal on selected areas in communications}, vol.~31, no.~11, pp.
  2209--2221, 2013.

\bibitem{khan2016opportunistic}
Z.~Khan, J.~J. Lehtom{\"a}ki, L.~A. DaSilva, E.~Hossain, and M.~Latva-Aho,
  ``Opportunistic channel selection by cognitive wireless nodes under imperfect
  observations and limited memory: a repeated game model,'' \emph{IEEE
  Transactions on Mobile Computing}, vol.~15, no.~1, pp. 173--187, 2016.

\bibitem{saleem2014primary}
Y.~Saleem and M.~H. Rehmani, ``Primary radio user activity models for cognitive
  radio networks: A survey,'' \emph{Journal of Network and Computer
  Applications}, vol.~43, pp. 1--16, 2014.

\bibitem{monemi2015characterizing}
M.~Monemi, M.~Rasti, and E.~Hossain, ``Characterizing feasible interference
  region for underlay cognitive radio networks,'' in \emph{{IEEE International
  Conference on Communications (ICC)}}.\hskip 1em plus 0.5em minus 0.4em\relax
  IEEE, 2015, pp. 7603--7608.

\bibitem{monemi2016characterization}
------, ``On characterization of feasible interference regions in cognitive
  radio networks,'' \emph{IEEE Transactions on Communications}, vol.~64, no.~2,
  pp. 511--524, 2016.

\bibitem{ozger2016utilization}
M.~Ozger and O.~B. Akan, ``On the utilization of spectrum opportunity in
  cognitive radio networks,'' \emph{IEEE Communications Letters}, vol.~20,
  no.~1, pp. 157--160, 2016.

\bibitem{akhtar2016white}
F.~Akhtar, M.~H. Rehmani, and M.~Reisslein, ``White space: Definitional
  perspectives and their role in exploiting spectrum opportunities,''
  \emph{Telecommunications Policy}, vol.~40, no.~4, pp. 319--331, 2016.

\bibitem{khan2016cognitive}
A.~A. Khan, M.~H. Rehmani, and M.~Reisslein, ``Cognitive radio for smart grids:
  Survey of architectures, spectrum sensing mechanisms, and networking
  protocols,'' \emph{IEEE Communications Surveys \& Tutorials}, vol.~18, no.~1,
  pp. 860--898, 2016.

\bibitem{bukhari2016survey}
S.~H.~R. Bukhari, M.~H. Rehmani, and S.~Siraj, ``A survey of channel bonding
  for wireless networks and guidelines of channel bonding for futuristic
  cognitive radio sensor networks,'' \emph{IEEE Communications Surveys \&
  Tutorials}, vol.~18, no.~2, pp. 924--948, 2016.

\bibitem{paul2011understanding}
U.~Paul, A.~P. Subramanian, M.~M. Buddhikot, and S.~R. Das, ``{Understanding
  traffic dynamics in cellular data networks},'' in \emph{{IEEE INFOCOM}},
  2011, pp. 882--890.

\bibitem{shafiq2011characterizing}
M.~Z. Shafiq, L.~Ji, A.~X. Liu, and J.~Wang, ``{Characterizing and modeling
  internet traffic dynamics of cellular devices},'' in \emph{{ACM joint
  international conference on Measurement and modeling of computer systems
  (SIGMETRICS)}}, 2011, pp. 305--316.

\bibitem{sayeed2015cloud}
Z.~Sayeed, Q.~Liao, D.~Faucher, E.~Grinshpun, and S.~Sharma, ``Cloud analytics
  for wireless metric prediction-framework and performance,'' in \emph{IEEE
  International Conference on Cloud Computing (CLOUD)}, 2015, pp. 995--998.

\bibitem{tadrous2013proactive}
J.~Tadrous, A.~Eryilmaz, and H.~El~Gamal, ``{Proactive resource allocation:
  Harnessing the diversity and multicast gains},'' \emph{IEEE Transactions on
  Information Theory}, vol.~59, no.~8, pp. 4833--4854, 2013.

\bibitem{huang2014backpressure}
L.~Huang, S.~Zhang, M.~Chen, and X.~Liu, ``{When backpressure meets predictive
  scheduling},'' in \emph{{ACM international symposium on Mobile ad hoc
  networking and computing (MobiHoc)}}, 2014, pp. 33--42.

\bibitem{abedini2014content}
N.~Abedini and S.~Shakkottai, ``{Content caching and scheduling in wireless
  networks with elastic and inelastic traffic},'' \emph{IEEE/ACM Transactions
  on Networking}, vol.~22, no.~3, pp. 864--874, 2014.

\bibitem{xu2013proteus}
Q.~Xu, S.~Mehrotra, Z.~Mao, and J.~Li, ``{{PROTEUS}: Network Performance
  Forecast for Real-time, Interactive Mobile Applications},'' in \emph{{ACM
  international conference on Mobile systems, applications, and services
  (MobiSys)}}, 2013, pp. 347--360.

\bibitem{samulevicius2015most}
S.~Samulevicius, T.~B. Pedersen, and T.~B. Sorensen, ``{MOST: mobile broadband
  network optimization using planned spatio-temporal events},'' in \emph{{IEEE
  Vehicular Technology Conference (VTC Spring)}}, 2015, pp. 1--5.

\bibitem{lee2013generalized}
M.-F.~R. Lee, F.-H.~S. Chiu, H.-C. Huang, and C.~Ivancsits, ``{Generalized
  predictive control in a wireless networked control system},'' \emph{Hindawi
  International Journal of Distributed Sensor Networks}, 2013.

\bibitem{sekar2013developing}
A.~B.~V. Sekar, A.~Akella, S.~S.~I. Stoica, and H.~Zhang, ``Developing a
  predictive model of quality of experience for internet video,'' in \emph{{ACM
  SIGCOMM}}, 2013, pp. 339--350.

\bibitem{beister2014predicting}
F.~Beister and H.~Karl, ``{Predicting mobile video inter-download times with
  hidden Markov models},'' in \emph{{IEEE International Conference on Wireless
  and Mobile Computing, Networking and Communications (WiMob)}}, 2014, pp.
  359--364.

\bibitem{pollakis2016anticipatory}
E.~Pollakis and S.~Stanczak, ``Anticipatory networking for energy savings in
  {5G} systems,'' \emph{VDE ITG-Fachbericht-WSA}, 2016.

\bibitem{yu2014predictive}
H.~Yu, M.~H. Cheung, L.~Huang, and J.~Huang, ``Predictive delay-aware network
  selection in data offloading,'' in \emph{IEEE Global Communications
  Conference (GLOBECOM)}, 2014, pp. 1376--1381.

\bibitem{yu2016power}
------, ``Power-delay tradeoff with predictive scheduling in integrated
  cellular and wi-fi networks,'' \emph{IEEE Journal on Selected Areas in
  Communications (JSAC)}, vol.~34, no.~4, pp. 735--742, 2016.

\bibitem{du2016traffic}
J.~Du, C.~Jiang, Y.~Qian, Z.~Han, and Y.~Ren, ``Traffic prediction based
  resource configuration in space-based systems,'' in \emph{IEEE International
  Conference on Communications (ICC)}, 2016, pp. 1--6.

\bibitem{du2016resource}
------, ``Resource allocation with video traffic prediction in cloud-based
  space systems,'' \emph{IEEE Transactions on Multimedia}, vol.~18, no.~5, pp.
  820--830, 2016.

\bibitem{papagiannaki2003long}
K.~Papagiannaki, N.~Taft, Z.-L. Zhang, and C.~Diot, ``{Long-term forecasting of
  internet backbone traffic: Observations and initial models},'' in \emph{{IEEE
  INFOCOM}}, 2003, pp. 1178--1188.

\bibitem{sadek2004multi}
N.~Sadek and A.~Khotanzad, ``{Multi-scale high-speed network traffic prediction
  using k-factor Gegenbauer ARMA model},'' in \emph{{IEEE International
  Conference on Communications (ICC)}}, vol.~4, 2004, pp. 2148--2152.

\bibitem{zhou2005network}
B.~Zhou, D.~He, Z.~Sun, and W.~H. Ng, ``{Network traffic modeling and
  prediction with ARIMA/GARCH},'' in \emph{{HET-NETs Conference}}, 2005, pp.
  1--10.

\bibitem{abouzeid2013predictive}
H.~Abou-Zeid and H.~S. Hassanein, ``{Predictive green wireless access:
  Exploiting mobility and application information},'' \emph{IEEE Wireless
  Communications}, vol.~20, no.~5, pp. 92--99, 2013.

\bibitem{yao2012improving}
J.~Yao, S.~S. Kanhere, and M.~Hassan, ``{Improving {QoS} in high-speed mobility
  using bandwidth maps},'' \emph{IEEE Transactions on Mobile Computing},
  vol.~11, no.~4, pp. 603--617, 2012.

\bibitem{riiser2013commute}
H.~Riiser, P.~Vigmostad, C.~Griwodz, and P.~Halvorsen, ``Commute path bandwidth
  traces from {3G} networks: Analysis and applications,'' in \emph{{ACM
  Multimedia Systems Conference (MMSys)}}, 2013, pp. 114--118.

\bibitem{millan2015tracking}
P.~Millan, C.~Molina, E.~Dimogerontakis, L.~Navarro, R.~Meseguer, B.~Braem, and
  C.~Blondia, ``Tracking and predicting end-to-end quality in wireless
  community networks,'' in \emph{IEEE International Conference on Future
  Internet of Things and Cloud (FiCloud)}, 2015, pp. 794--799.

\bibitem{yin2015control}
X.~Yin, A.~Jindal, V.~Sekar, and B.~Sinopoli, ``{A control-theoretic approach
  for dynamic adaptive video streaming over HTTP},'' \emph{ACM SIGCOMM Computer
  Communication Review}, vol.~45, no.~4, pp. 325--338, 2015.

\bibitem{yi2016cs2p}
Y.~Sun, X.~Yin, J.~Jiang, V.~Sekar, F.~Lin, N.~Wang, T.~Liu, and B.~Sinopoli,
  ``{CS2P}: Improving video bitrate selection and adaptation with data-driven
  throughput prediction,'' in \emph{ACM SIGCOMM}, 2016, pp. 272--285.

\bibitem{jiang2016cfa}
J.~Jiang, V.~Sekar, H.~Milner, D.~Shepherd, I.~Stoica, and H.~Zhang, ``Cfa: a
  practical prediction system for video qoe optimization,'' in \emph{USENIX
  Symposium on Networked Systems Design and Implementation (NSDI 16)}, 2016,
  pp. 137--150.

\bibitem{zahran2016oscar}
A.~H. Zahran, J.~Quinlan, D.~Raca, C.~J. Sreenan, E.~Halepovic, R.~K. Sinha,
  R.~Jana, and V.~Gopalakrishnan, ``{OSCAR: an optimized stall-cautious
  adaptive bitrate streaming algorithm for mobile networks},'' in \emph{ACM
  International Workshop on Mobile Video (MoVid)}, 2016, p.~2.

\bibitem{wang2016squad}
C.~Wang, A.~Rizk, and M.~Zink, ``Squad: a spectrum-based quality adaptation for
  dynamic adaptive streaming over http,'' in \emph{ACM International Conference
  on Multimedia Systems (MMSys)}, 2016, p.~1.

\bibitem{miller2015control}
K.~Miller, D.~Bethanabhotla, G.~Caire, and A.~Wolisz, ``A control-theoretic
  approach to adaptive video streaming in dense wireless networks,'' \emph{IEEE
  Transactions on Multimedia}, vol.~17, no.~8, pp. 1309--1322, 2015.

\bibitem{bastug2013proactive}
E.~Ba{\c{s}}tu{\u{g}}, J.-L. Gu{\'e}n{\'e}go, and M.~Debbah, ``{Proactive small
  cell networks},'' in \emph{{IEEE International Conference on
  Telecommunications (ICT)}}, 2013.

\bibitem{bastug2014living}
E.~Ba{\c{s}}tu{\u{g}}, M.~Bennis, and M.~Debbah, ``{Living on the edge: The
  role of proactive caching in {5G} wireless networks},'' \emph{IEEE
  Communications Magazine}, vol.~52, no.~8, pp. 82--89, 2014.

\bibitem{bastug2014anticipatory}
------, ``{Anticipatory caching in small cell networks: A transfer learning
  approach},'' in \emph{{1st KuVS Workshop on Anticipatory Networks}}, 2014.

\bibitem{siris2016exploiting}
V.~A. Siris, X.~Vasilakos, and D.~Dimopoulos, ``Exploiting mobility prediction
  for mobility \& popularity caching and dash adaptation,'' in \emph{IEEE World
  of Wireless, Mobile and Multimedia Networks (WoWMoM)}, 2016, pp. 1--8.

\bibitem{golrezaei2012femtocaching}
N.~Golrezaei, K.~Shanmugam, A.~G. Dimakis, A.~F. Molisch, and G.~Caire,
  ``{Femtocaching: Wireless video content delivery through distributed caching
  helpers},'' in \emph{{IEEE INFOCOM}}, 2012, pp. 1107--1115.

\bibitem{tadrous2015optimal}
J.~Tadrous and A.~Eryilmaz, ``On optimal proactive caching for mobile networks
  with demand uncertainties,'' \emph{IEEE/ACM Transactions on Networking},
  vol.~24, no.~5, pp. 2715--2727, 2015.

\bibitem{tadrous2015joint}
J.~Tadrous, A.~Eryilmaz, and H.~El~Gamal, ``Joint smart pricing and proactive
  content caching for mobile services,'' \emph{IEEE/ACM Transactions on
  Networking}, vol.~24, no.~4, pp. 2357--2371, 2015.

\bibitem{gu2015matching}
Y.~Gu, W.~Saad, M.~Bennis, M.~Debbah, and Z.~Han, ``Matching theory for future
  wireless networks: fundamentals and applications,'' \emph{IEEE Communications
  Magazine}, vol.~53, no.~5, pp. 52--59, 2015.

\bibitem{semiari2015context}
O.~Semiari, W.~Saad, S.~Valentin, M.~Bennis, and H.~V. Poor, ``Context-aware
  small cell networks: How social metrics improve wireless resource
  allocation,'' \emph{IEEE Transactions on Wireless Communications}, vol.~14,
  no.~11, pp. 5927--5940, 2015.

\bibitem{semiari2016context}
O.~Semiari, W.~Saad, and M.~Bennis, ``Context-aware scheduling of joint
  millimeter wave and microwave resources for dual-mode base stations,'' in
  \emph{{IEEE International Conference on Communications (ICC)}}, 2016.

\bibitem{namvar2014context}
N.~Namvar, W.~Saad, B.~Maham, and S.~Valentin, ``A context-aware matching game
  for user association in wireless small cell networks,'' in \emph{2014 IEEE
  International Conference on Acoustics, Speech and Signal Processing
  (ICASSP)}.\hskip 1em plus 0.5em minus 0.4em\relax IEEE, 2014, pp. 439--443.

\bibitem{zhang2015social}
Y.~Zhang, E.~Pan, L.~Song, W.~Saad, Z.~Dawy, and Z.~Han, ``Social network aware
  device-to-device communication in wireless networks,'' \emph{IEEE
  Transactions on Wireless Communications}, vol.~14, no.~1, pp. 177--190, 2015.

\bibitem{hamidouche2014many}
K.~Hamidouche, W.~Saad, and M.~Debbah, ``Many-to-many matching games for
  proactive social-caching in wireless small cell networks,'' in \emph{IEEE
  International Symposium on Modeling and Optimization in Mobile, Ad Hoc, and
  Wireless Networks (WiOpt)}, 2014, pp. 569--574.

\bibitem{noulas2012mining}
A.~Noulas, S.~Scellato, N.~Lathia, and C.~Mascolo, ``{Mining user mobility
  features for next place prediction in location-based services},'' in
  \emph{{IEEE International Conference on Data Mining (ICDM)}}, 2012, pp.
  1038--1043.

\bibitem{calabrese2010human}
F.~Calabrese, G.~D. Lorenzo, and C.~Ratti, ``{Human mobility prediction based
  on individual and collective geographical preferences},'' in \emph{{IEEE
  International Conference on Intelligent Transportation Systems (ITSC)}},
  2010, pp. 312--317.

\bibitem{bapierre2011variable}
H.~Bapierre, G.~Groh, and S.~Theiner, ``{A variable order Markov model approach
  for mobility prediction},'' \emph{Pervasive Computing}, pp. 8--16, 2011.

\bibitem{proebster2012context}
M.~Proebster, M.~Kaschub, T.~Werthmann, and S.~Valentin, ``{Context-aware
  resource allocation for cellular wireless networks},'' \emph{EURASIP Journal
  on Wireless Communications and Networking}, vol. 2012, p. 2012:216.

\bibitem{proebster2011context}
M.~Proebster, M.~Kaschub, and S.~Valentin, ``{Context-aware resource allocation
  to improve the quality of service of heterogeneous traffic},'' in \emph{{IEEE
  International Conference on Communications (ICC)}}, 2011, pp. 1--6.

\bibitem{yi2016spatial}
Z.~Yi, X.~Dong, X.~Zhang, and W.~Wang, ``Spatial traffic prediction for
  wireless cellular system based on base stations social network,'' in
  \emph{IEEE Systems Conference (SysCon)}, 2016, pp. 1--5.

\bibitem{tsiropoulos2011impact}
G.~Tsiropoulos, D.~G. Stratogiannis, N.~Mantas, and M.~Louta, ``{The impact of
  social distance on utility based resource allocation in next generation
  networks},'' in \emph{{IEEE International Congress on Ultra Modern
  Telecommunications and Control Systems and Workshops (ICUMT)}}, 2011.

\bibitem{jackson2008social}
M.~O. Jackson, \emph{{Social and economic networks}}.\hskip 1em plus 0.5em
  minus 0.4em\relax Princeton, NJ, USA: Princeton University Press, 2008.

\bibitem{TelecomItalia}
\BIBentryALTinterwordspacing
{Telecom Italia}, ``Big data challenge 2015.'' [Online]. Available:
  \url{http://aris.me/contents/teaching/data-mining-2015/project/BigDataChallengeData.html}
\BIBentrySTDinterwordspacing

\bibitem{harvey1990forecasting}
A.~C. Harvey, \emph{{Forecasting, structural time series models and the Kalman
  filter}}.\hskip 1em plus 0.5em minus 0.4em\relax Cambridge university press,
  1990.

\bibitem{zaidi2005real}
Z.~R. Zaidi and B.~L. Mark, ``{Real-time mobility tracking algorithms for
  cellular networks based on Kalman filtering},'' \emph{{IEEE Transactions on
  Mobile Computing}}, vol.~4, no.~2, pp. 195--208, 2005.

\bibitem{okutani1984dynamic}
I.~Okutani and Y.~J. Stephanedes, ``{Dynamic prediction of traffic volume
  through Kalman filtering theory},'' \emph{Elsevier Transportation Research
  Part B: Methodological}, vol.~18, no.~1, pp. 1--11, 1984.

\bibitem{pappas2014extended}
G.~Pappas and M.~Zohdy, ``Extended {Kalman} filtering and pathloss modeling for
  shadow power parameter estimation in mobile wireless communications,''
  \emph{International Journal on Smart Sensing and Intelligent Systems},
  vol.~7, no.~2, pp. 898--924, 2014.

\bibitem{lee2012comparative}
J.~Lee, M.~Sun, and G.~Lebanon, ``{A comparative study of collaborative
  filtering algorithms},'' \emph{arXiv preprint arXiv:1205.3193}, 2012.

\bibitem{bastug2014think}
E.~Ba{\c{s}}tu{\u{g}}, M.~Bennis, and M.~Debbah, ``{Think before reacting:
  Proactive caching in {5G} small cell networks},'' \emph{Wiley, submitted},
  2015.

\bibitem{dutta2015predictive}
S.~Dutta, A.~Narang, S.~Bhattacherjee, A.~S. Das, and D.~Krishnaswamy,
  ``Predictive caching framework for mobile wireless networks,'' in \emph{{IEEE
  International Conference on Mobile Data Management (MDM)}}, 2015, pp.
  179--184.

\bibitem{xu2005survey}
R.~Xu, D.~Wunsch \emph{et~al.}, ``{Survey of clustering algorithms},''
  \emph{IEEE Transactions on Neural Networks}, vol.~16, no.~3, pp. 645--678,
  2005.

\bibitem{murthy1998automatic}
S.~K. Murthy, ``{Automatic construction of decision trees from data: A
  multi-disciplinary survey},'' \emph{Kluwer Data mining and knowledge
  discovery}, vol.~2, no.~4, pp. 345--389, 1998.

\bibitem{ramsay2006functional}
J.~O. Ramsay, \emph{Functional data analysis}.\hskip 1em plus 0.5em minus
  0.4em\relax Wiley Online Library, 2006.

\bibitem{ramsay1991some}
J.~O. Ramsay and C.~Dalzell, ``Some tools for functional data analysis,''
  \emph{JSTOR Journal of the Royal Statistical Society. Series B
  (Methodological)}, pp. 539--572, 1991.

\bibitem{mozer2000predicting}
M.~C. Mozer, R.~Wolniewicz, D.~B. Grimes, E.~Johnson, and H.~Kaushansky,
  ``Predicting subscriber dissatisfaction and improving retention in the
  wireless telecommunications industry,'' \emph{IEEE Transactions on Neural
  Networks}, vol.~11, no.~3, pp. 690--696, 2000.

\bibitem{kaaniche2010mobility}
H.~Kaaniche and F.~Kamoun, ``{Mobility prediction in wireless ad hoc networks
  using neural networks},'' \emph{Journal of Telecommunications}, vol.~2,
  no.~1, pp. 95--101, 2010.

\bibitem{chen2015rate}
C.~Chen, X.~Zhu, G.~de~Veciana, A.~C. Bovik, and R.~W. Heath, ``{Rate
  adaptation and admission control for video transmission with subjective
  quality constraints},'' \emph{IEEE Journal of Selected Topics in Signal
  Processing}, vol.~9, no.~1, pp. 22--36, 2015.

\bibitem{chen2013markov}
C.~Chen, R.~W. Heath, A.~C. Bovik, and G.~de~Veciana, ``{A Markov decision
  model for adaptive scheduling of stored scalable videos},'' \emph{IEEE
  Transactions on Circuits and Systems for Video Technology}, vol.~23, no.~6,
  pp. 1081--1095, 2013.

\bibitem{bianchi2013networked}
D.~Bianchi, A.~Ferrara, and M.~Di~Benedetto, ``{Networked model predictive
  traffic control with time varying optimization horizon: The Grenoble South
  Ring case study},'' in \emph{{IEEE European Control Conference (ECC)}}, 2013,
  pp. 4039--4044.

\bibitem{witheephanich2014min}
K.~Witheephanich, J.~M. Esca{\~n}o, D.~Mu{\~n}oz de~la Pe{\~n}a, and M.~J.
  Hayes, ``A min--max model predictive control approach to robust power
  management in ambulatory wireless sensor networks,'' \emph{IEEE Systems
  Journal}, vol.~8, no.~4, pp. 1060--1073, 2014.

\bibitem{boyd2004convex}
S.~Boyd and L.~Vandenberghe, \emph{{Convex optimization}}.\hskip 1em plus 0.5em
  minus 0.4em\relax Cambridge university press, 2004.

\bibitem{schrijver1998theory}
A.~Schrijver, \emph{{Theory of linear and integer programming}}.\hskip 1em plus
  0.5em minus 0.4em\relax John Wiley \& Sons, 1998.

\bibitem{qin2003survey}
S.~J. Qin and T.~A. Badgwell, ``{A survey of industrial model predictive
  control technology},'' \emph{Elsevier Control engineering practice}, vol.~11,
  no.~7, pp. 733--764, 2003.

\bibitem{puterman2014markov}
M.~L. Puterman, \emph{{Markov decision processes: discrete stochastic dynamic
  programming}}.\hskip 1em plus 0.5em minus 0.4em\relax John Wiley \& Sons,
  2014.

\bibitem{sutton1998reinforcement}
R.~S. Sutton and A.~G. Barto, \emph{{Reinforcement learning: An
  introduction}}.\hskip 1em plus 0.5em minus 0.4em\relax MIT press Cambridge,
  1998, vol.~1, no.~1.

\bibitem{fu2010systematic}
F.~Fu and M.~van~der Schaar, ``{A systematic framework for dynamically
  optimizing multi-user wireless video transmission},'' \emph{IEEE Journal on
  Selected Areas in Communications}, vol.~28, no.~3, pp. 308--320, 2010.

\bibitem{hossain20155g}
E.~Hossain and M.~Hasan, ``{5G cellular: key enabling technologies and research
  challenges},'' \emph{IEEE Instrumentation \& Measurement Magazine}, vol.~18,
  no.~3, pp. 11--21, 2015.

\bibitem{giordano2002mobile}
S.~Giordano \emph{et~al.}, ``Mobile ad hoc networks,'' \emph{Handbook of
  wireless networks and mobile computing}, pp. 325--346, 2002.

\bibitem{asadi2014survey}
A.~Asadi, Q.~Wang, and V.~Mancuso, ``{A Survey on Device-to-Device
  Communication in Cellular Networks},'' \emph{IEEE Communications Surveys \&
  Tutorials}, vol.~16, no.~4, pp. 1801--1819, Fourthquarter 2014.

\bibitem{alfuqaha2015internet}
A.~Al-Fuqaha, M.~Guizani, M.~Mohammadi, M.~Aledhari, and M.~Ayyash, ``Internet
  of things: A survey on enabling technologies, protocols, and applications,''
  \emph{IEEE Communications Surveys \& Tutorials}, vol.~17, no.~4, pp.
  2347--2376, 2015.

\bibitem{zanella2014internet}
A.~Zanella, N.~Bui, A.~Castellani, L.~Vangelista, and M.~Zorzi, ``{Internet of
  Things} for smart cities,'' \emph{IEEE Internet of Things Journal}, vol.~1,
  no.~1, pp. 22--32, 2014.

\bibitem{xu2014internet}
L.~D. Xu, W.~He, and S.~Li, ``{Internet of Things in Industries: A Survey},''
  \emph{{IEEE Transactions on Industrial Informatics}}, vol.~10, no.~4, pp.
  2233--2243, Nov 2014.

\bibitem{zimmermann1980osi}
H.~Zimmermann, ``{OSI reference model--The ISO model of architecture for open
  systems interconnection},'' \emph{IEEE Transactions on communications},
  vol.~28, no.~4, pp. 425--432, 1980.

\bibitem{NGMN}
\BIBentryALTinterwordspacing
NGMN. {{Next Generation Mobile Networks}}. [Online]. Available:
  \url{{http://www.ngmn.de/publications/all-downloads/article/ngmn-5g-white-paper.html}}
\BIBentrySTDinterwordspacing

\bibitem{malanchini2016wireless}
I.~Malanchini, S.~Valentin, and O.~Aydin, ``{Wireless resource sharing for
  multiple operators: Generalization, fairness, and the value of prediction},''
  \emph{Elsevier Computer Networks}, vol. 100, pp. 110--123, 2016.

\bibitem{fettweis2014tactile}
G.~P. Fettweis, ``{The tactile internet: applications and challenges},''
  \emph{IEEE Vehicular Technology Magazine}, vol.~9, no.~1, pp. 64--70, 2014.

\bibitem{suryaprakash2016reliability}
V.~Suryaprakash and I.~Malanchini, ``{Reliability in future radio access
  networks: from linguistic to quantitative definitions},'' in \emph{{IEEE/ACM
  International Symposium on Quality of Service (IWQoS)}}, 2016.

\bibitem{singer2015sharing}
N.~Singer, ``{Sharing data, but not happily},''
  {http://www.nytimes.com/2015/06/05/technology/consumers-conflicted-over-data-mining-policies-report-finds.html?\_r=0},
  2015, the New York Times, [Online; accessed 5-November-2016].

\bibitem{wan2014context}
J.~Wan, D.~Zhang, S.~Zhao, L.~Yang, and J.~Lloret, ``{Context-aware vehicular
  cyber-physical systems with cloud support: architecture, challenges, and
  solutions},'' \emph{IEEE Communications Magazine}, vol.~52, no.~8, pp.
  106--113, 2014.

\end{thebibliography}

\end{document}